\documentclass[10pt]{article}
\usepackage{amsmath}
\usepackage{amsfonts}

\advance \textheight by \topskip
\advance \textheight by 1pt
\makeatother
\def\bea{\begin{eqnarray}}
\def\ena{\end{eqnarray}}
\def\no{\nonumber}
\def\ep{\epsilon}
\newtheorem{prop}{Proposition}
\newtheorem{theorem}{Theorem}
\newtheorem{defn}{Definition}
\newtheorem{lemma}{Lemma}
\newtheorem{cor}{Corollary}

\def\pf{{\it Proof.} \quad}
\title{
On Form Factors of SU(2) Invariant Thirring Model
}

\author{
Atsushi Nakayashiki\thanks{
Faculty of Mathematics,
Kyushu University,
Ropponmatsu 4-2-1, Fukuoka 810-8560, Japan, \quad 
e-mail: atsushi@rc.kyushu-u.ac.jp}
\quad and \quad
Yoshihiro Takeyama\thanks{
Research Institute for Mathmatical Sciences, Kyoto University,
Kyoto 606-8502, Japan, \quad
e-mail: ninihuni@kurims.kyoto-u.ac.jp} 
\thanks{Research Fellow of the Japan Society for the Promotion of Science.}
}

\date{}
\begin{document}
\maketitle
\begin{abstract}
Integral formulae for form factors of a large family of 
charged local operators in $SU(2)$ invariant Thirring model are given
extending Smirnov's construction of form factors of chargeless local operators
in the sine-Gordon model.
New abelian symmetry acting on this family of local operators is found.
It creates Lukyanov's operators which are not in the above family of local
operators in general.
\end{abstract}

\section{Introduction}
We study the problem to determine all local
operators in the $SU(2)$-invariant Thirring model ($SU(2)$ ITM).
In \cite{smir2} F. Smirnov has constructed sufficient number of 
charge zero local operators of the quantum sine-Gordon model (SGM) by
giving their form factors and 
conjectured that they exhausted all local operators of the theory.
In this paper we shall give a similar construction 
of form factors of local operators with any charge (or weight) for 
$SU(2)$ ITM.

One of the important problems in the study of integrable 
quantum field theories is to determine all local operators in the theory.
The form factor bootstrap approach is one of the most appropriate methods
to study this problem.
In this approach the problem is reduced to determining sets of functions
which satisfy certain axioms \cite{smirbook}, the axioms for locality.
After the pioneering work of Karowski et al.\cite{KW} and 
Smirnov et al. \cite{KS,smirbook} there was an important
progress in constructing form factors of local operators in the last 
decade \cite{JM1,luk1,smir2}.
Recently there is a progress in determining the form factors of
several specific operators \cite{BK1,BK2,luk2,luk3}.
Nevertheless to calculate the character of the space of local operators
of SGM or $SU(2)$ ITM for example is not yet carried out.
There are some results on this problem for models with 
diagonal S-matrices \cite{Koubek,BBS}.
The problem is not yet solved for models with non-diagonal S-matrices.

Thus it is important to study a general structure of the operators of 
\cite{smir2}.
By this reason we are interested in a more general situation,
the description of the charged local operators.
The construction of Smirnov \cite{smir2} 
is applicable to the non-zero charged case without much change.
Unexpectedly we have found that the form factors of some local operators
are not obtained by this construction in the charged case.
The operator $\Lambda_{-1}(y)$ 
introduced by Lukyanov in \cite{luk1}
gives such an example (see \S6 and \S7 for more precise statement).
To improve this drawback we have found a new symmetry.
It produces a family of local operators from the ones
constructed by Smirnov's way.
We conjecture that this larger family of local operators contains
all operators defined in \cite{luk1}.

Let us explain the construction given in this paper briefly.
For $SU(2)$ ITM the axioms for locality
imply that the $n$-particle form factor satisfies the $SU(2)$ qKZ equation
of level zero ({\it cf.} \cite{npt});
\begin{eqnarray}
&&
f(\beta_1,\cdots,\beta_j-2\pi i,\cdots,\beta_n)=(-1)^{n/2}
K_j(\beta_1,\cdots,\beta_n)f(\beta_1,\cdots,\beta_n),
\label{qkz}
\\
&&
K_j(\beta_1,\cdots,\beta_n)=
S_{j,j-1}(\beta_j-\beta_{j-1}-2\pi i)\cdots 
S_{j,1}(\beta_j-\beta_{1}-2\pi i)
\nonumber
\\
&&
\qquad\qquad\qquad\qquad
{}\times 
S_{j,n}(\beta_j-\beta_{n})\cdots S_{j,j+1}(\beta_j-\beta_{j+1}),
\no
\end{eqnarray}
where $\beta_j$'s are rapidities of particles, $S(\beta)$ is the S-matrix
of $SU(2)$ ITM, $S_{ij}(\beta)$ is $S(\beta)$ acting on $i$-th and $j$-th
tensor components, $f$ takes the value in $V^{\otimes n}$, 
$V\simeq \mathbb{C}^2$ is the vector representation of $SU(2)$.
Thus to describe solutions of qKZ equation (\ref{qkz}) 
becomes a first step to determine sets of form factors.
The integral formulae for solutions of (\ref{qkz}) were studied in \cite{npt}.
The solution space of (\ref{qkz}) was determined in \cite{tar}
as the space of certain polynomials which we call cycles.
More precisely, in \cite{tar}, a set of cycles which generate the solution
space is given and all relations among them are
described. We remark that, these relations, 
the deformed Riemann bilinear relations and
some linear dependence relations of cycles, 
are originally discovered by Smirnov \cite{bilinear}.
Thus we know the complete description of the solutions of (\ref{qkz}).

The remaining condition for form factors to define local operators 
is on certain residues which relates $n$-particle form factor to 
$(n-2)$-particle form factor.
The strategy of the construction is as follows.
To each solution of the $m$-particle qKZ equation, which is regular at
$\beta_m=\beta_{m-1}+\pi i$ and $S$-symmetric ({\it cf.} \S2.2 (II)), 
we shall construct a set of $S$-symmetric solutions $f_n$,
$n=m+2,m+4,...$ of $n$-particle qKZ equation satisfying the residue
condition. In this way we construct a set of form factors $\{f_n\}$
which satisfy $f_n=0$ $(n<m)$ for given $m$ and $f_m$.
The operator given by such form factor $\{f_n\}$ is called 
$m$-minimal and $f_m$ is called the initial form factor.
The charge of such operator is the $SU(2)$ weight of the vector function $f_n$
which should be same for all $n$.
To summarize we construct, for given $m$ and an initial form factor $f_m$,
a set of form factors of an $m$-minimal local operator with the charge equal
to the $SU(2)$ weight of $f_m$.

The formula of the cycle of $f_n$, $n\geq m+2$, 
at weight zero is given by Smirnov \cite{smir2} 
using certain determinant.
As a matter of fact these formulae of cycles are valid 
at non-zero weight cases with small change.
Nevertheless there are some subtleties.
In order to give the initial function $f_m$ arbitrary,
assuming the $S$-symmetry and the regularity condition at 
$\beta_m=\beta_{m-1}+\pi i$, we need to take a different type 
of integral formula from the zero weight case of \cite{smir2}.
This fact is found in \cite{tar} for the solutions of (\ref{qkz}).
By this reason the non-zero weight case is technically more involved.

The construction of the paper is as follows.
In section 2 we shall introduce the $SU(2)$ ITM and 
describe the axiom of locality for form factors.
We review the description of the solutions of the $SU(2)$ qKZ
equation at level zero in section 3. 
In section 4 the formula of cycles for form factors of 
charged local operators are presented.
The proof is given in section 5.
In section 6 we introduce new symmetry.
The cycles for Lukyanov's operators $\Lambda_{-1}(y)$ and $T(y)$ 
are given as special cases of section 6 in section 7.
In section 8 one time integrated formulae for form factors of chargeless
local operators are given. 
The formulae by Smirnov for the form factors of 
the energy momentum tensor \cite{smir1,smirbook} are originally given
as a special case of this form.
A list of cycles for important local operators is given in
appendix A.

\section{SU(2) ITM}
\par
\noindent
{\bf 2.1.} Model.
We mainly follow the notations in \cite{smir1,smirbook}.
Let $V=\mathbb{C}v_{+}\oplus \mathbb{C}v_{-}$ be the two dimensional
vector space describing the one particle states created by the
Faddeev-Zamolodchikov creation operator, where $v_{+}$ and $v_{-}$ 
correspond to the kink and the anti-kink respectively.
We denote by $\beta$ the rapidity of the particles.
It parametrizes the energy $p_0$ and the momentum $p_1$ by
\bea
&&
p_0=M\text{cosh}\beta,
\qquad
p_1=M\text{sinh}\beta,
\no
\ena
where $M$ is the mass of the particles.
The S-matrix is considered as the operator 
$S(\beta)\in {\rm End}(V^{\otimes 2})$ given by
\begin{eqnarray}
&&
S(\beta)=S_0(\beta)\frac{\beta-\pi i P}{\beta-\pi i},
\nonumber
\end{eqnarray}
where $P$ is the permutation operator and the scalar 
function $S_0(\beta)$ is given by
\begin{eqnarray}
&&
S_0(\beta)=
\frac{
\Gamma(\frac{1}{2}+\frac{\beta}{2\pi i})\Gamma(-\frac{\beta}{2\pi i})
}
{
\Gamma(\frac{1}{2}-\frac{\beta}{2\pi i})\Gamma(\frac{\beta}{2\pi i})
}.
\nonumber
\end{eqnarray}
This S-matrix satisfies the following unitarity and crossing symmetry 
relations:
\begin{eqnarray}
&&
S(\beta)S(-\beta)=1,
\quad
S(\pi i -\beta)=C_1 {}^{t_1}S(\beta) C_1,
\nonumber
\end{eqnarray}
where ${}^{t_1}S(\beta)$ is the transpose with respect to
the first component, $C_1=C\otimes 1$ and $C$ is the two by two matrix
given by
$$
C=
\left[\begin{array}{cc}
0&1\\
-1&0\\
\end{array}\right].
$$
Notice that $C$ is not symmetric as opposed to the usual case.
\vskip3mm

\noindent
{\bf 2.2.} Axioms for locality.
\par
\noindent
Consider a set of functions
$f(\beta_1,\cdots,\beta_n)$ with values in $V^{\otimes n}$ for all 
non-negative even integers $n$.
The operator whose form factors are $\{f_n\}$ is a local
operator if and only if the following conditions are satisfied
\cite{smir1,smirbook}.
\vskip3mm

\noindent
(I). $P_{i, i+1}S_{i, i+1}(\beta_i-\beta_{i+1})f(\beta_1,\cdots,\beta_n)
=f(\cdots,\beta_{i+1},\beta_i,\cdots)$.
\vskip2mm
\noindent
(II). $P_{n-1, n}\cdots P_{1, 2}f(\beta_1-2\pi i,\beta_2,\cdots,\beta_n)=
(-1)^{\frac{n}{2}}f(\beta_2,\cdots,\beta_n,\beta_1)$.
\vskip2mm
\noindent
(III). The functions $f(\beta_1,\cdots,\beta_n)$ are analytic in $\beta_n$ 
except the simple poles at $\beta_n=\beta_j+\pi i$ in the strip 
$0<{\rm Im} \beta_n<2\pi$ for real $\beta_1,...,\beta_{n-1}$.
The residues at $\beta_n=\beta_{n-1}+\pi i$ should be given by
\bea
&&2\pi i \text{res}_{\beta_n=\beta_{n-1}+\pi i} 
f(\beta_1,\cdots,\beta_n) 
\no \\
&& \!\!\!\! 
{}=\big(I-(-1)^{\frac{n}{2}-1}
S_{n-1, n-2}(\beta_{n-1}-\beta_{n-2})\cdots 
S_{n-1, 1}(\beta_{n-1}-\beta_{1}))
f(\beta_1,\cdots,\beta_{n-2}\big)\otimes \mathbf{e}_0,
\no 
\ena
where $\mathbf{e}_0=v_{+}\otimes v_{-}-v_{-}\otimes v_{+}$, 
$P_{ij}$ and $S_{ij}(\beta)$ are the operators 
acting on the $i$-th and $j$-th components as $P$ and $S(\beta)$ respectively.
\vskip3mm

\noindent
The model has the SU(2)-invariance.
Let us briefly describe the corresponding properties of form factors.
We denote by $E$, $F$, $H$ the standard basis of $sl_2$:
\bea
&&
[E,F]=H,
\quad
[H,E]=2E,
\quad
[H,F]=-2F.
\no
\ena
The vector space $V$ is considered as the vector representation 
of $sl_2$ by
\bea
&&
v_{+}=\left[\begin{array}{c}1\\0\end{array}\right],
\quad
v_{-}=\left[\begin{array}{c}0\\1\end{array}\right],
\quad
E=\sigma^{+},
\quad
F=\sigma^{-},
\quad
H=2\sigma^3,
\no
\ena
where $\sigma^a (a=3, \pm)$ 
is the Pauli matrices
$$
\sigma^+=
\left[\begin{array}{cc}0&1\\0&0\end{array}\right],
\quad
\sigma^-=
\left[\begin{array}{cc}0&0\\1&0\end{array}\right],
\quad
\sigma^3=
\left[\begin{array}{cc}1&0\\0&-1\end{array}\right].
$$
Then $V^{\otimes n}$ becomes a representation of $sl_2$ by
\bea
&&
E=\Sigma^{+},
\quad
F=\Sigma^{-},
\quad
H=2\Sigma^3,
\no
\ena
where $\Sigma^{a}=\sum_{j=1}^n\sigma_j^{a}$
and $\sigma_j^{a}$ acts on $j$-th tensor component.
The operators $S(\beta)$ and $P$ commute with the action of $sl_2$ on 
$V^{\otimes 2}$.
This is the SU(2) invariance of the model.
The vector $\mathbf{e}_0$ in (III) is the
$sl_2$ singlet.
Thus if a set of functions $\{f(\beta_1,\cdots,\beta_n)\}$ satisfies
(I), (II), (III), so does
$\{Xf(\beta_1,\cdots,\beta_n)\}$ for any $X\in sl_2$.

In this paper we consider even particle form factors only.
By this reason $n$ is always assumed to be even in this paper.

\section{Solutions of qKZ equation}
\medskip

\noindent
{\bf 3.1} qKZ equation.

If the function $\psi(\beta_1,\cdots,\beta_n)$ satisfies the axioms (I) 
and (II) for locality then
\begin{eqnarray}
&&
\psi(\beta_1,\cdots,\beta_j-2\pi i,\cdots,\beta_n)=(-1)^{n/2}
K_j(\beta_1,\cdots,\beta_n)\psi(\beta_1,\cdots,\beta_n),
\label{qkz1}
\\
&&
K_j(\beta_1,\cdots,\beta_n)=
S_{j,j-1}(\beta_j-\beta_{j-1}-2\pi i)\cdots 
S_{j,1}(\beta_j-\beta_{1}-2\pi i)
\nonumber
\\
&&
\qquad\qquad\qquad\qquad
{}\times 
S_{j,n}(\beta_j-\beta_{n})\cdots S_{j,j+1}(\beta_j-\beta_{j+1}).
\label{qkzop1}
\end{eqnarray}
This is nothing but the rational qKZ equation of level zero 
({\it cf.}\cite{npt}).
The integral formulae for solutions of this equation have been studied 
in \cite{npt,smir1,smirbook,tar}.
Let us recall the results in these papers.

Denote by $\widehat{K}_j(\beta_1,\cdots,\beta_n)$ the operator
obtained from $K_j(\beta_1,\cdots,\beta_n)$ by replacing $S(\beta)$
by $\widehat{S}(\beta):=S(\beta)/S_0(\beta)$.
We first consider the equation
\begin{eqnarray}
&&
\psi(\beta_1,\cdots,\beta_j-2\pi i,\cdots,\beta_n)=
\widehat{K}_j(\beta_1,\cdots,\beta_n)\psi(\beta_1,\cdots,\beta_n).
\label{qkz2}
\end{eqnarray}
\vskip5mm

\noindent
{\bf 3.2} Deformed cocycles.

For a subset $M=\{m_1,\cdots,m_\ell\}\subset \{1,\cdots,n\}$, 
$m_{1} < \cdots < m_{\ell}$, let 
$g_M$ and $w_M$ be functions defined by
\begin{eqnarray}
&&
g_M(\alpha_1\cdots \alpha_\ell)
\, :=\,
\prod_{a=1}^\ell
\biggl(
\frac{1}{\alpha_a-\beta_{m_a}}
\prod_{j=1}^{m_a-1}
\frac{\alpha_a-\beta_j+\pi i}{\alpha_a-\beta_j}
\biggr)
\prod_{1\leq a<b \leq \ell}(\alpha_a-\alpha_b+\pi i),
\nonumber
\\
&&
w_M\, :=\, \text{Asym}\, g_M,
\nonumber
\end{eqnarray}
where anti-symmetrization of a function 
$f(\alpha_1,\cdots,\alpha_\ell)$ is defined by
\begin{eqnarray}
&&
\text{Asym}\, f:=\sum_{\sigma \in S_n} (\hbox{sgn}\, \sigma ) 
\cdot f(\alpha_{\sigma(1)},\cdots,\alpha_{\sigma(\ell)}).
\nonumber
\end{eqnarray}
\vskip5mm

\noindent
{\bf 3.3.} Deformed cycles.

Let ${\cal C}$ be the space of $2\pi i$ periodic functions
of $\beta_1$,...,$\beta_n$.
Define the vector spaces $\hat{{\cal F}}_q$, $\bar{{\cal F}}_q$ and 
${\cal F}_q$ by
\begin{eqnarray}
&&
\hat{{\cal F}}_q:=\bigoplus_{k=0}^{n}{\cal C}
\frac{A^k}{\prod_{j=1}^n(1-AB_j^{-1})}
\supset
\bar{{\cal F}}_q:=\bigoplus_{k=0}^{n-1}{\cal C}
\frac{A^k}{\prod_{j=1}^n(1-AB_j^{-1})}
\supset
{\cal F}_q:=\bigoplus_{k=1}^{n-1}{\cal C}
\frac{A^k}{\prod_{j=1}^n(1-AB_j^{-1})},
\nonumber
\end{eqnarray}
where $A=\exp(-\alpha)$ and $B_j=\exp(-\beta_j)$.
Then the vector space $\bar{{\cal F}}_q^{\otimes \ell}$ 
($\wedge^\ell \bar{{\cal F}}_q$) can be
identified with the space of functions of the form
\begin{eqnarray}
&&
\frac{f(\alpha_1,\cdots,\alpha_\ell;\beta_1,\cdots,\beta_n)}
{\prod_{a=1}^{\ell}\prod_{j=1}^n(1-A_aB_j^{-1})}
\label{poly}
\end{eqnarray}
where $f$ is a (n anti-symmetric) polynomial of $A_1$,...,$A_\ell$ of 
order
less than $n$ in each variable with the coefficients in ${\cal C}$.
The spaces $\hat{{\cal F}}_q^{\otimes \ell}$, 
$\wedge^\ell \hat{{\cal F}}_q$, ${\cal F}_q^{\otimes \ell}$ and
$\wedge^\ell {\cal F}_q$ are similarly understood.

We call the elements of $\hat{{\cal F}}_q^{\otimes \ell}$ cycles.
We sometimes use the term cycle for the elements of 
$\hat{{\cal F}}_q^{\otimes \ell}$ multiplied by the denominator 
of (\ref{poly}).

\vskip5mm

\noindent
{\bf 3.4} Solutions.

Let $\phi(\alpha)$ be the function
\begin{eqnarray}
&&
\phi(\alpha)\,
=
\,
\phi(\alpha ; \beta_{1}, \cdots , \beta_{n}):=
\prod_{j=1}^n\,
\frac{\Gamma(\frac{\alpha-\beta_j+\pi i}{-2\pi i})}{\Gamma(\frac{\alpha-\beta_j}
{-2\pi i})}\;.
\nonumber
\end{eqnarray}
For a subset $M=\{m_1,\cdots,m_\ell\}\subset \{1,\cdots,n\}$, 
$m_{1} < \cdots < m_{\ell}$ we denote by $v_M$ the vector in 
$V^{\otimes n}$ defined by
\begin{eqnarray}
&&
v_M \, :=\, v_{\epsilon_1}\otimes\cdots\otimes v_{\epsilon_n},
\nonumber
\end{eqnarray}
where $M=\{i \, \vert \, \epsilon_i=-\,\}$. Sometimes $v_M$ is
denoted by $v_{\ep_1,\cdots,\ep_n}$ for the sake of convenience.

For any $W\in \hat{{\cal F}}_q^{\otimes \ell}$ define the 
$V^{\otimes n}$ valued function $\psi_W$ by
\begin{eqnarray}
&&
\psi_W(\beta_1,\cdots,\beta_n)
:=
\sum_{\sharp M=\ell}v_M\,
\int_{C^\ell}
\prod_{a=1}^\ell d\alpha_a
\prod_{a=1}^\ell \phi(\alpha_a)
w_MW.
\label{sol1}
\end{eqnarray}
The contour $C$ is a simple curve
going from $-\infty$ to $\infty$, separating sets
$\cup_{j=1}^n(\beta_j-\pi i-2\pi i\mathbb{Z}_{\leq 0})$
and
$\cup_{j=1}^n(\beta_j-2\pi i\mathbb{Z}_{\geq 0})$ (Figure 1).
\vskip5mm
%%%%%%%%%%%%%%%%%%%%%%%%%%%%%%
%WinTpicVersion3.08
\unitlength 0.1in
\begin{center}
\begin{picture}( 49.8000, 14.5000)( 13.5000,-23.1000)
% DOT 2 0 3 0
% 2 2390 1000 2390 1000
% 
\special{pn 8}%
\special{sh 1}%
\special{ar 2390 1000 10 10 0  6.28318530717959E+0000}%
\special{sh 1}%
\special{ar 2390 1000 10 10 0  6.28318530717959E+0000}%
% DOT 2 0 3 0
% 2 2400 1820 2400 1820
% 
\special{pn 8}%
\special{sh 1}%
\special{ar 2400 1820 10 10 0  6.28318530717959E+0000}%
\special{sh 1}%
\special{ar 2400 1820 10 10 0  6.28318530717959E+0000}%
% CIRCLE 2 0 3 0
% 4 2400 1410 2400 1410 2400 1410 2400 1410
% 
\special{pn 8}%
\special{ar 2400 1410 0 0  0.0000000 6.2831853}%
% CIRCLE 2 0 3 0
% 4 2390 1410 2420 1440 2420 1440 2420 1440
% 
\special{pn 8}%
\special{ar 2390 1410 42 42  0.0000000 6.2831853}%
% CIRCLE 2 0 3 0
% 4 2420 2200 2420 2200 2400 2220 2400 2220
% 
\special{pn 8}%
\special{ar 2420 2200 0 0  0.0000000 6.2831853}%
% CIRCLE 2 0 3 0
% 4 2410 2210 2430 2240 2430 2240 2430 2240
% 
\special{pn 8}%
\special{ar 2410 2210 36 36  0.0000000 6.2831853}%
% STR 2 0 3 0
% 3 2500 930 2500 1030 2 0
% $\beta_1+\pi i$
\put(25.0000,-10.3000){\makebox(0,0)[lb]{$\beta_1+\pi i$}}%
% STR 2 0 3 0
% 3 2550 1380 2550 1480 2 0
% $\beta_1$
\put(25.5000,-14.8000){\makebox(0,0)[lb]{$\beta_1$}}%
% STR 2 0 3 0
% 3 2580 1760 2580 1860 2 0
% $\beta_1-\pi i$
\put(25.8000,-18.6000){\makebox(0,0)[lb]{$\beta_1-\pi i$}}%
% STR 2 0 3 0
% 3 2590 2180 2590 2280 2 0
% $\beta_1-2\pi i$
\put(25.9000,-22.8000){\makebox(0,0)[lb]{$\beta_1-2\pi i$}}%
% DOT 2 0 3 0
% 2 4210 1010 4200 1010
% 
\special{pn 8}%
\special{sh 1}%
\special{ar 4210 1010 10 10 0  6.28318530717959E+0000}%
\special{sh 1}%
\special{ar 4200 1010 10 10 0  6.28318530717959E+0000}%
% DOT 2 0 3 0
% 2 4210 1800 4210 1790
% 
\special{pn 8}%
\special{sh 1}%
\special{ar 4210 1800 10 10 0  6.28318530717959E+0000}%
\special{sh 1}%
\special{ar 4210 1790 10 10 0  6.28318530717959E+0000}%
% CIRCLE 2 0 3 0
% 4 4210 1400 4210 1400 4200 1400 4210 1400
% 
\special{pn 8}%
\special{ar 4210 1400 0 0  6.2831853 6.2831853}%
\special{ar 4210 1400 0 0  0.0000000 3.1415927}%
% CIRCLE 2 0 3 0
% 4 4200 1410 4220 1450 4220 1450 4220 1450
% 
\special{pn 8}%
\special{ar 4200 1410 46 46  0.0000000 6.2831853}%
% CIRCLE 2 0 3 0
% 4 4210 2210 4210 2210 4210 2210 4220 2200
% 
\special{pn 8}%
\special{ar 4210 2210 0 0  5.4977871 6.2831853}%
% CIRCLE 2 0 3 0
% 4 4190 2200 4210 2230 4220 2230 4220 2230
% 
\special{pn 8}%
\special{ar 4190 2200 36 36  0.0000000 6.2831853}%
% STR 2 0 3 0
% 3 4330 950 4330 1050 2 0
% $\beta_n+\pi i$
\put(43.3000,-10.5000){\makebox(0,0)[lb]{$\beta_n+\pi i$}}%
% STR 2 0 3 0
% 3 4360 1340 4360 1440 2 0
% $\beta_n$
\put(43.6000,-14.4000){\makebox(0,0)[lb]{$\beta_n$}}%
% STR 2 0 3 0
% 3 4360 1760 4360 1860 2 0
% $\beta_n-\pi i$
\put(43.6000,-18.6000){\makebox(0,0)[lb]{$\beta_n-\pi i$}}%
% STR 2 0 3 0
% 3 4350 2140 4350 2240 2 0
% $\beta_n-2\pi i$
\put(43.5000,-22.4000){\makebox(0,0)[lb]{$\beta_n-2\pi i$}}%
% SPLINE 2 0 3 0
% 40 1350 1180 2230 1160 2230 1160 4480 1160 4480 1160 5220 1230 5310 1460 5310 1460 5100 1550 5010 1550 4710 1580 3690 1560 3680 1560 2660 1560 2510 1590 2460 1590 2460 1590 2180 1750 2180 1750 2290 2010 2290 2010 2580 2020 2580 2020 3370 2020 3370 2020 3900 2000 3900 2000 4390 1980 4390 1980 4950 1980 4950 1980 5520 1950 5520 1950 5800 1900 5800 1900 5900 1740 5970 1470 6110 1230 6270 1230 6260 1230
% 
\special{pn 8}%
\special{pa 1350 1180}%
\special{pa 1382 1180}%
\special{pa 1414 1178}%
\special{pa 1448 1176}%
\special{pa 1480 1174}%
\special{pa 1512 1174}%
\special{pa 1544 1172}%
\special{pa 1576 1170}%
\special{pa 1608 1170}%
\special{pa 1640 1168}%
\special{pa 1672 1168}%
\special{pa 1704 1166}%
\special{pa 1736 1166}%
\special{pa 1768 1164}%
\special{pa 1800 1164}%
\special{pa 1832 1162}%
\special{pa 1864 1162}%
\special{pa 1896 1160}%
\special{pa 1928 1160}%
\special{pa 1960 1160}%
\special{pa 1992 1160}%
\special{pa 2024 1160}%
\special{pa 2056 1160}%
\special{pa 2088 1160}%
\special{pa 2120 1160}%
\special{pa 2152 1160}%
\special{pa 2182 1160}%
\special{pa 2214 1160}%
\special{pa 2246 1160}%
\special{pa 2278 1162}%
\special{pa 2310 1162}%
\special{pa 2340 1164}%
\special{pa 2372 1164}%
\special{pa 2404 1166}%
\special{pa 2434 1168}%
\special{pa 2466 1170}%
\special{pa 2498 1170}%
\special{pa 2528 1172}%
\special{pa 2560 1174}%
\special{pa 2590 1176}%
\special{pa 2622 1178}%
\special{pa 2654 1180}%
\special{pa 2684 1182}%
\special{pa 2716 1184}%
\special{pa 2746 1186}%
\special{pa 2778 1188}%
\special{pa 2810 1190}%
\special{pa 2840 1192}%
\special{pa 2872 1194}%
\special{pa 2902 1198}%
\special{pa 2934 1200}%
\special{pa 2964 1202}%
\special{pa 2996 1204}%
\special{pa 3028 1206}%
\special{pa 3058 1208}%
\special{pa 3090 1210}%
\special{pa 3120 1212}%
\special{pa 3152 1214}%
\special{pa 3184 1216}%
\special{pa 3214 1218}%
\special{pa 3246 1220}%
\special{pa 3278 1222}%
\special{pa 3310 1222}%
\special{pa 3340 1224}%
\special{pa 3372 1226}%
\special{pa 3404 1228}%
\special{pa 3436 1228}%
\special{pa 3466 1230}%
\special{pa 3498 1230}%
\special{pa 3530 1230}%
\special{pa 3562 1232}%
\special{pa 3594 1232}%
\special{pa 3626 1232}%
\special{pa 3658 1232}%
\special{pa 3690 1232}%
\special{pa 3722 1232}%
\special{pa 3754 1232}%
\special{pa 3786 1232}%
\special{pa 3820 1230}%
\special{pa 3852 1230}%
\special{pa 3884 1228}%
\special{pa 3916 1226}%
\special{pa 3950 1224}%
\special{pa 3982 1222}%
\special{pa 4016 1220}%
\special{pa 4048 1218}%
\special{pa 4082 1216}%
\special{pa 4114 1212}%
\special{pa 4148 1210}%
\special{pa 4182 1206}%
\special{pa 4214 1202}%
\special{pa 4248 1198}%
\special{pa 4282 1192}%
\special{pa 4316 1188}%
\special{pa 4350 1184}%
\special{pa 4384 1178}%
\special{pa 4418 1172}%
\special{pa 4452 1166}%
\special{pa 4486 1160}%
\special{pa 4522 1152}%
\special{pa 4556 1146}%
\special{pa 4590 1138}%
\special{pa 4626 1132}%
\special{pa 4660 1126}%
\special{pa 4694 1120}%
\special{pa 4728 1114}%
\special{pa 4762 1108}%
\special{pa 4796 1104}%
\special{pa 4830 1102}%
\special{pa 4864 1100}%
\special{pa 4896 1100}%
\special{pa 4928 1102}%
\special{pa 4960 1104}%
\special{pa 4992 1108}%
\special{pa 5022 1116}%
\special{pa 5054 1124}%
\special{pa 5082 1136}%
\special{pa 5112 1148}%
\special{pa 5140 1164}%
\special{pa 5168 1184}%
\special{pa 5194 1204}%
\special{pa 5220 1230}%
\special{pa 5244 1256}%
\special{pa 5266 1286}%
\special{pa 5286 1318}%
\special{pa 5302 1350}%
\special{pa 5314 1382}%
\special{pa 5318 1412}%
\special{pa 5318 1440}%
\special{pa 5308 1466}%
\special{pa 5290 1488}%
\special{pa 5266 1506}%
\special{pa 5236 1522}%
\special{pa 5204 1534}%
\special{pa 5168 1542}%
\special{pa 5132 1548}%
\special{pa 5098 1550}%
\special{pa 5064 1552}%
\special{pa 5032 1550}%
\special{pa 5002 1550}%
\special{pa 4970 1552}%
\special{pa 4938 1554}%
\special{pa 4908 1556}%
\special{pa 4876 1560}%
\special{pa 4844 1564}%
\special{pa 4812 1568}%
\special{pa 4780 1572}%
\special{pa 4748 1576}%
\special{pa 4716 1580}%
\special{pa 4682 1584}%
\special{pa 4650 1586}%
\special{pa 4618 1588}%
\special{pa 4586 1590}%
\special{pa 4554 1592}%
\special{pa 4522 1592}%
\special{pa 4490 1592}%
\special{pa 4458 1594}%
\special{pa 4426 1592}%
\special{pa 4394 1592}%
\special{pa 4362 1592}%
\special{pa 4330 1590}%
\special{pa 4298 1590}%
\special{pa 4266 1588}%
\special{pa 4234 1586}%
\special{pa 4202 1584}%
\special{pa 4170 1582}%
\special{pa 4138 1580}%
\special{pa 4106 1578}%
\special{pa 4074 1576}%
\special{pa 4042 1574}%
\special{pa 4010 1572}%
\special{pa 3978 1570}%
\special{pa 3946 1568}%
\special{pa 3914 1568}%
\special{pa 3884 1566}%
\special{pa 3852 1564}%
\special{pa 3820 1562}%
\special{pa 3788 1562}%
\special{pa 3756 1562}%
\special{pa 3724 1560}%
\special{pa 3692 1560}%
\special{pa 3660 1560}%
\special{pa 3628 1560}%
\special{pa 3596 1558}%
\special{pa 3562 1558}%
\special{pa 3530 1556}%
\special{pa 3498 1554}%
\special{pa 3466 1552}%
\special{pa 3434 1550}%
\special{pa 3402 1548}%
\special{pa 3370 1544}%
\special{pa 3338 1542}%
\special{pa 3306 1540}%
\special{pa 3274 1536}%
\special{pa 3242 1534}%
\special{pa 3208 1532}%
\special{pa 3176 1530}%
\special{pa 3144 1528}%
\special{pa 3112 1526}%
\special{pa 3080 1526}%
\special{pa 3048 1524}%
\special{pa 3016 1524}%
\special{pa 2984 1524}%
\special{pa 2952 1524}%
\special{pa 2920 1526}%
\special{pa 2888 1528}%
\special{pa 2856 1530}%
\special{pa 2826 1532}%
\special{pa 2794 1536}%
\special{pa 2762 1540}%
\special{pa 2730 1546}%
\special{pa 2698 1552}%
\special{pa 2668 1558}%
\special{pa 2636 1566}%
\special{pa 2604 1574}%
\special{pa 2572 1582}%
\special{pa 2542 1588}%
\special{pa 2510 1590}%
\special{pa 2478 1590}%
\special{pa 2446 1592}%
\special{pa 2412 1594}%
\special{pa 2376 1602}%
\special{pa 2342 1612}%
\special{pa 2310 1624}%
\special{pa 2278 1640}%
\special{pa 2250 1660}%
\special{pa 2224 1680}%
\special{pa 2204 1706}%
\special{pa 2188 1732}%
\special{pa 2178 1762}%
\special{pa 2174 1794}%
\special{pa 2176 1826}%
\special{pa 2184 1860}%
\special{pa 2196 1894}%
\special{pa 2212 1926}%
\special{pa 2230 1954}%
\special{pa 2254 1980}%
\special{pa 2278 2002}%
\special{pa 2306 2018}%
\special{pa 2334 2030}%
\special{pa 2364 2038}%
\special{pa 2396 2040}%
\special{pa 2428 2040}%
\special{pa 2462 2038}%
\special{pa 2496 2034}%
\special{pa 2528 2028}%
\special{pa 2562 2024}%
\special{pa 2596 2018}%
\special{pa 2630 2014}%
\special{pa 2664 2010}%
\special{pa 2696 2008}%
\special{pa 2730 2006}%
\special{pa 2762 2004}%
\special{pa 2794 2002}%
\special{pa 2828 2002}%
\special{pa 2860 2002}%
\special{pa 2892 2002}%
\special{pa 2924 2002}%
\special{pa 2956 2002}%
\special{pa 2986 2004}%
\special{pa 3018 2004}%
\special{pa 3050 2006}%
\special{pa 3082 2008}%
\special{pa 3112 2010}%
\special{pa 3144 2012}%
\special{pa 3176 2012}%
\special{pa 3206 2014}%
\special{pa 3238 2016}%
\special{pa 3270 2018}%
\special{pa 3300 2018}%
\special{pa 3332 2020}%
\special{pa 3364 2020}%
\special{pa 3396 2020}%
\special{pa 3428 2020}%
\special{pa 3460 2020}%
\special{pa 3490 2020}%
\special{pa 3522 2020}%
\special{pa 3554 2018}%
\special{pa 3586 2018}%
\special{pa 3618 2016}%
\special{pa 3650 2014}%
\special{pa 3682 2014}%
\special{pa 3716 2012}%
\special{pa 3748 2010}%
\special{pa 3780 2008}%
\special{pa 3812 2006}%
\special{pa 3844 2004}%
\special{pa 3876 2002}%
\special{pa 3908 2000}%
\special{pa 3940 1998}%
\special{pa 3972 1996}%
\special{pa 4004 1994}%
\special{pa 4036 1992}%
\special{pa 4068 1992}%
\special{pa 4100 1990}%
\special{pa 4132 1988}%
\special{pa 4164 1986}%
\special{pa 4194 1986}%
\special{pa 4226 1984}%
\special{pa 4258 1984}%
\special{pa 4290 1982}%
\special{pa 4322 1982}%
\special{pa 4354 1980}%
\special{pa 4388 1980}%
\special{pa 4420 1980}%
\special{pa 4452 1980}%
\special{pa 4484 1980}%
\special{pa 4516 1980}%
\special{pa 4548 1982}%
\special{pa 4580 1982}%
\special{pa 4612 1982}%
\special{pa 4644 1982}%
\special{pa 4676 1982}%
\special{pa 4708 1982}%
\special{pa 4740 1984}%
\special{pa 4772 1984}%
\special{pa 4804 1984}%
\special{pa 4836 1984}%
\special{pa 4868 1982}%
\special{pa 4900 1982}%
\special{pa 4932 1982}%
\special{pa 4962 1980}%
\special{pa 4994 1978}%
\special{pa 5026 1976}%
\special{pa 5056 1976}%
\special{pa 5088 1974}%
\special{pa 5118 1970}%
\special{pa 5150 1968}%
\special{pa 5180 1966}%
\special{pa 5212 1964}%
\special{pa 5244 1962}%
\special{pa 5276 1960}%
\special{pa 5308 1958}%
\special{pa 5340 1956}%
\special{pa 5372 1954}%
\special{pa 5404 1952}%
\special{pa 5438 1952}%
\special{pa 5472 1952}%
\special{pa 5506 1950}%
\special{pa 5540 1950}%
\special{pa 5574 1950}%
\special{pa 5608 1950}%
\special{pa 5644 1948}%
\special{pa 5676 1946}%
\special{pa 5708 1940}%
\special{pa 5738 1932}%
\special{pa 5766 1922}%
\special{pa 5792 1906}%
\special{pa 5816 1888}%
\special{pa 5836 1864}%
\special{pa 5854 1838}%
\special{pa 5870 1808}%
\special{pa 5884 1778}%
\special{pa 5898 1748}%
\special{pa 5910 1718}%
\special{pa 5922 1686}%
\special{pa 5932 1656}%
\special{pa 5940 1626}%
\special{pa 5948 1596}%
\special{pa 5956 1566}%
\special{pa 5962 1534}%
\special{pa 5966 1502}%
\special{pa 5970 1468}%
\special{pa 5974 1434}%
\special{pa 5976 1400}%
\special{pa 5982 1364}%
\special{pa 5990 1332}%
\special{pa 6002 1302}%
\special{pa 6020 1276}%
\special{pa 6048 1254}%
\special{pa 6082 1238}%
\special{pa 6128 1228}%
\special{pa 6182 1224}%
\special{pa 6232 1224}%
\special{pa 6268 1228}%
\special{pa 6280 1230}%
\special{pa 6260 1230}%
\special{sp}%
% VECTOR 2 0 3 0
% 2 1830 1160 1910 1160
% 
\special{pn 8}%
\special{pa 1830 1160}%
\special{pa 1910 1160}%
\special{fp}%
\special{sh 1}%
\special{pa 1910 1160}%
\special{pa 1844 1140}%
\special{pa 1858 1160}%
\special{pa 1844 1180}%
\special{pa 1910 1160}%
\special{fp}%
% VECTOR 2 0 3 0
% 2 4460 1970 4540 1970
% 
\special{pn 8}%
\special{pa 4460 1970}%
\special{pa 4540 1970}%
\special{fp}%
\special{sh 1}%
\special{pa 4540 1970}%
\special{pa 4474 1950}%
\special{pa 4488 1970}%
\special{pa 4474 1990}%
\special{pa 4540 1970}%
\special{fp}%
% STR 2 0 3 0
% 3 1500 970 1500 1070 2 0
% $C$
\put(15.0000,-10.7000){\makebox(0,0)[lb]{$C$}}%
% VECTOR 2 0 3 0
% 2 6280 1240 6330 1240
% 
\special{pn 8}%
\special{pa 6280 1240}%
\special{pa 6330 1240}%
\special{fp}%
\special{sh 1}%
\special{pa 6330 1240}%
\special{pa 6264 1220}%
\special{pa 6278 1240}%
\special{pa 6264 1260}%
\special{pa 6330 1240}%
\special{fp}%
% STR 2 0 3 0
% 3 3200 960 3200 1060 2 0
% $\cdots$
\put(32.0000,-10.6000){\makebox(0,0)[lb]{$\cdots$}}%
% STR 2 0 3 0
% 3 3190 1390 3190 1490 2 0
% $\cdots$
\put(31.9000,-14.9000){\makebox(0,0)[lb]{$\cdots$}}%
% STR 2 0 3 0
% 3 3210 1760 3210 1860 2 0
% $\cdots$
\put(32.1000,-18.6000){\makebox(0,0)[lb]{$\cdots$}}%
% STR 2 0 3 0
% 3 3220 2150 3220 2250 2 0
% $\cdots$
\put(32.2000,-22.5000){\makebox(0,0)[lb]{$\cdots$}}%
% STR 2 0 3 0
% 3 3230 2380 3230 2480 2 0
% {\bf Figure 1}
\put(32.3000,-24.8000){\makebox(0,0)[lb]{{\bf Figure 1}}}%
\end{picture}%
\end{center}
%%%%%%%%%%%%%%%%%%%%%%%%%%%%%%
\vskip5mm
\noindent
In \cite{npt} it is proved that the integral (\ref{sol1}) converges.
Let $(V^{\otimes n})_\lambda$ and $(V^{\otimes n})_\lambda^{sing}$
be the spaces defined by
\begin{eqnarray}
&&
(V^{\otimes n})_\lambda :=
\{\, v\in V^{\otimes n}\, \vert \, \Sigma^3 v=\lambda v\},
\nonumber
\\
&&
(V^{\otimes n})_\lambda^{sing}
:=
\{\, v\in (V^{\otimes n})_\lambda \, \vert \, \Sigma^+ v=0\}.
\nonumber
\end{eqnarray}

The following Theorem was proved in \cite{npt}.

\begin{theorem}
Suppose that $n\geq 2\ell$.
Then for any $W\in \wedge^\ell \hat{{\cal F}}_q$
the function $\psi_W$ takes the value in 
$(V^{\otimes n})_{n-2\ell}^{sing}$ and 
is a solution of the qKZ equation (\ref{qkz2}). 
If $W$ is symmetric with respect to $\beta_1$,...,$\beta_n$, then it
satisfies the equations
\bea
&&
\psi_{W}(\cdots, \beta_{j+1}, \beta_{j}, \cdots )
=
P_{j, j+1}\widehat{S}_{j, j+1}(\beta_{j}-\beta_{j+1})
\psi_{W}(\cdots , \beta_{j}, \beta_{j+1}, \cdots ),
\label{eqI'}
\\
&&
P_{n-1, n}\cdots P_{1, 2}
\psi_{W}(\beta_{1}-2\pi i, \beta_{2}, \cdots , \beta_{n})
=
\psi_{W}(\beta_{2}, \cdots , \beta_{n}, \beta_{1}).
\label{eqII'}
\ena
\end{theorem}
\vskip3mm

The assignment $W \mapsto \psi_W$ defines a map
\begin{eqnarray}
&&
I_\ell: 
\wedge^\ell \bar{{\cal F}}_q 
\longrightarrow
{\cal C}\otimes_{\mathbb{C}}(V^{\otimes n})_{n-2\ell}^{sing}.
\nonumber
\end{eqnarray}
This map has a kernel \cite{tar}.
Let us recall it.
Let $\Theta(A)$ be the function
\begin{eqnarray}
&&
\Theta(A):=\prod_{j=1}^n
\frac{1+AB_j^{-1}}{1-AB_j^{-1}}.
\nonumber
\end{eqnarray}
Then we define $\Xi^{(1)}(A)\in \bar{{\cal F}}_q$ and
$\Xi^{(2)}(A_1,A_2)\in \wedge^2\bar{{\cal F}}_q$ by
\begin{eqnarray}
&&
\Xi^{(1)}(A):=\Theta(A)-1,
\nonumber
\\
&&
\Xi^{(2)}(A_1,A_2):=
(\Theta(A_1)\Theta(A_2)-1)
\frac{A_1A_2^{-1}-1}{A_1A_2^{-1}+1}
+
\Theta(A_1)-\Theta(A_2).
\nonumber
\end{eqnarray}

The following theorem was proved in \cite{tar}.

\begin{theorem}
The map $I_\ell$ is surjective and its kernel is
\begin{eqnarray}
&&
{\rm Ker} \, I_\ell=
\Xi^{(1)}\wedge^{\ell-1}\bar{{\cal F}}_q
+\Xi^{(2)}\wedge^{\ell-2}\bar{{\cal F}}_q.
\nonumber
\end{eqnarray}
\end{theorem}
\vskip3mm

By the theorem we have
\begin{eqnarray}
&&
\frac{\wedge^\ell\bar{{\cal F}}_q}
{\Xi^{(1)}\wedge^{\ell-1}\bar{{\cal F}}_q
+\Xi^{(2)}\wedge^{\ell-2}\bar{{\cal F}}_q}
\simeq
{\cal C}
\otimes_{\mathbb{C}}
(V^{\otimes n})_{n-2\ell}^{sing}.
\label{solsp1}
\end{eqnarray}
Let us simplify the LHS.
Notice that
\begin{eqnarray}
&&
\Xi^{(1)}(A)
=
2\frac{\bar{e}_1(B)A+\bar{e}_3(B)A^3+\cdots+\bar{e}_{n-1}(B)A^{n-1}}
{\prod_{j=1}^n(1-AB_j^{-1})},
\label{rel1}
\end{eqnarray}
where 
$$
\bar{e}_k(B):=\sum_{i_1<\cdots<i_k}B_{i_1}^{-1}\cdots B_{i_k}^{-1}.
$$
We solve (\ref{rel1}) in $A^{n-1}$ and substitute it
in LHS of (\ref{solsp1}).
In this way, if we set
\begin{eqnarray}
&&
\bar{{\cal F}}_q^{red}=\oplus_{k=0}^{n-2}{\cal C}
\frac{A^k}{\prod_{j=1}^n(1-AB_j^{-1})},
\nonumber
\end{eqnarray}
then
\begin{eqnarray}
&&
\hbox{LHS of (\ref{solsp1})}
\simeq
\frac{\wedge^\ell\bar{{\cal F}}_q^{red}}
{\Xi^{(2)}_{red}\wedge^{\ell-2}\bar{{\cal F}}_q^{red}}.
\nonumber
\end{eqnarray}
Here $\Xi^{(2)}_{red}$ is obtained from $\Xi^{(2)}$
by substituting 
\begin{eqnarray}
&&
A^{n-1}=\frac{-1}{\bar{e}_{n-1}(B)}
(\bar{e}_1(B)A+\bar{e}_3(B)A^3+\cdots+\bar{e}_{n-3}(B)A^{n-3}).
\label{xi1}
\end{eqnarray}
To summarize, the solution space (\ref{solsp1}) is isomorphic to
\begin{eqnarray}
&&
\frac{\wedge^\ell\bar{{\cal F}}_q^{red}}
{\Xi^{(2)}_{red}\wedge^{\ell-2}\bar{{\cal F}}_q^{red}},
\label{solsp2}
\end{eqnarray}
as a vector space over ${\cal C}$.
This space is generated, over ${\cal C}$,
by functions
\begin{eqnarray}
&&
\frac{\text{Asym}\,(A^{r_1}_1\cdots A^{r_\ell}_\ell)}
{\prod_{a=1}^\ell\prod_{j=1}^n(1-A_aB_j^{-1})},
\quad
0\leq r_1<\cdots<r_\ell\leq n-2.
\nonumber
\end{eqnarray}
\vskip2mm

\noindent
{\bf Remark.} For $W\in \hat{{\cal F}}_q^{\otimes \ell}$ we have
$\text{Asym}(W)\in \wedge^\ell \hat{{\cal F}}_q$ and
$\psi_W=\frac{1}{\ell !}\psi_{\text{Asym}(W)}$.
Thus we sometimes specify $W\in \hat{{\cal F}}_q^{\otimes \ell}$
to specify the element $\text{Asym}(W)$ in (\ref{solsp1}).
\vskip5mm

\section{Form factors}
\par
\noindent
{\bf 4.1.} Double gamma function.

To connect the solutions of (\ref{qkz2}) to (\ref{qkz1}),
we shall introduce the double gamma function
$\Gamma_{2}(x | \omega_{1}, \omega_{2})$ 
following \cite{JM}.

We assume that ${\rm Re}\omega_{j}>0$. 
$\Gamma_{2}(x| \omega_{1}, \omega_{2})^{-1}$ is an entire function of $x$. 
$\Gamma_{2}(x| \omega_{1}, \omega_{2})$ is meromorphic with poles at
\bea
x \in \omega_{1}\Bbb{Z}_{\le 0}+\omega_{2}\Bbb{Z}_{\le 0}
\ena
and symmetric with respect to $\omega_{1}, \omega_{2}$.
The following formula holds:
\bea
\frac
{\Gamma_{2}(x+\omega_{1}| \omega_{1}, \omega_{2})}
{\Gamma_{2}(x| \omega_{1}, \omega_{2})}
=\frac{1}{\Gamma_{1}(x| \omega_{2})},
\ena
where 
\bea
\Gamma_{1}(x| \omega):=\frac{\omega^{\frac{x}{\omega}-\frac{1}{2}}}{\sqrt{2\pi}}
\Gamma( \frac{x}{\omega}).
\label{gamma1}
\ena

With this function, we define
\bea
\zeta(\beta):=\frac
{\Gamma_{2}(-i\beta +3\pi)\Gamma_{2}(i\beta +\pi)}
{\Gamma_{2}(-i\beta +2\pi)\Gamma_{2}(i\beta )}, \quad
\Gamma_{2}(x)=\Gamma_{2}(x|2\pi , 2\pi).
\ena
This function satisfies the following equations
\bea
\zeta(\beta -2\pi i)=\zeta(-\beta), \quad
\zeta(\beta)\zeta(\beta -\pi i)=
\frac{1}{\Gamma_{1}(-i\beta +\pi)\Gamma_{1}(i\beta )},
\label{zetarel}
\ena
where $\Gamma_{1}(x)=\Gamma_{1}(x| 2\pi)$.
\vskip3mm

\noindent
{\bf 4.2.} Solution of (I) and (II).

Given a solution $\psi$ of the equations 
(\ref{eqI'}) and (\ref{eqII'}),
define 
\bea
f:=e^{\frac{n}{4}\sum_{j=1}^{n}\beta_{j}}
\prod_{1 \le j<j' \le n}\zeta(\beta_{j}-\beta_{j'})\psi .
\ena
Then it can be easily checked that $f$ satisfies 
(I) and (II) in the axioms for locality.
In particular $f$ is a solution of the qKZ equation (\ref{qkz1}).
\vskip3mm

\noindent
{\bf 4.3.} Minimal form factors.

\begin{defn}
An operator ${\cal O}$ is called $m$-minimal if 
its $n$-particle form factors vanish for all 
$n\leq m-1$.
\end{defn}
\vskip3mm

To each $m$ even and each solution of the equations (I), (II) with
$n=m$, we shall construct a set of form factors
$\{f(\beta_1,\cdots,\beta_n)\}$ of an $m$-minimal operator 
such that the initial form factor $f(\beta_1,\cdots,\beta_m)$ is the given one.

For 
\bea
W=\frac{P(A_1,\cdots,A_\ell)}
{\prod_{a=1}^\ell\prod_{j=1}^n(1-A_aB_j^{-1})}
\in \widehat{{\cal F}}_{q}^{\otimes \ell},
\no
\ena
we set
\bea
&&
\Psi_P:=\psi_W.
\nonumber
\ena
Define $E_n(t\vert B)$ and $E_n^{odd}(t\vert B)$ by
\bea
&&
E_n(t\vert B):=
\exp{(\sum_{k=-\infty}^{\infty}t_{k}\sum_{j=1}^{n}B_{j}^{k})},
\quad
E_n^{odd}(t\vert B):=
\exp{(\sum_{k=-\infty}^{\infty}t_{2k-1}\sum_{j=1}^{n}B_{j}^{2k-1})}.
\no
\ena

Let 
\bea
&&
P_{m(\pm)}(A_{1}, \cdots , A_{r}|B_{1}, \cdots , B_{m})=
E_m(t\vert B)(\prod_{j=1}^m B_j)^s
\prod_{1 \le j<j' \le m}(B_{j}^{\pm 1}+B_{j'}^{\pm 1}) 
\prod_{a=1}^{r}A_{a}^{k_{a}},
\quad
r\geq 0,
\label{allr}
\\
&&
P_{m(\pm)}(B_{1}, \cdots , B_{m})=
E_m^{odd}(t\vert B)(\prod_{j=1}^m B_j)^s,
\quad
r=0,
\label{rzero}
\ena
be the cycles for initial form factors, where $0\leq k_a\leq m$ for all $a$
and $0\leq s\leq \frac{m}{2}$.
We consider both (\ref{allr}) and (\ref{rzero}) for $r=0$.
In the case of $r=0$, $\Psi_{P_{m(\pm)}}$ is understood as 
(notice that $M=\phi$)
\bea
&&
\Psi_{P_{m(\pm)}}=P_{m(\pm)}v_{+}\otimes\cdots\otimes v_{+}.
\no
\ena
Set
\bea
&&
P_{n(\pm)}(A_{1}, \cdots , A_{\ell_n}|B_{1}, \cdots , B_{n})=
c_{n}^{m, r, s}
E_n(t\vert B)(\prod_{j=1}^n B_j)^s
\prod_{a=1}^{r}A_{a}^{k_{a}}
\prod_{a=r+1}^{\ell_n}A_{a}^{n+1-2s+2r-2a}
\no \\
&&
\qquad\qquad\qquad\qquad\qquad\qquad\qquad
\times
D_{n}^{\pm}(A_{1}, \cdots , A_{r}|B_{1}, \cdots , B_{n}),
\label{defnP}
\ena
where
\bea
&& \quad
c_{n}^{m, r, s}:=
(-1)^{\frac{1}{2}(n-m)s+\frac{1}{2}m(m-1)}
\left( \frac{-2\pi}{\zeta(-\pi i)} \right)^{\frac{n-m}{2}}
(-2\pi i)^{\frac{(m-n)(2n+2r-m)}{4}}, 
\label{defc} 
\\
&& \quad
D_{n}^{\pm}(A_{1}, \cdots , A_{r}|B_{1}, \cdots , B_{n})
:=\!
\frac{1}{\prod_{1 \le j<j' \le n}(B_{j}^{\pm 1}-B_{j'}^{\pm 1})}\!
\left|
\begin{array}{ccc}
1 & \cdots & 1 \\
B_{1}^{\pm 2} & \cdots & B_{n}^{\pm 2} \\
\vdots & \vdots & \vdots \\
B_{1}^{\pm(n+m-2)} & \cdots & B_{n}^{\pm(n+m-2)} \\
H_{1}^{\pm}(B_{1}) & \cdots & H_{1}^{\pm}(B_{n}) \\
\vdots & \vdots & \vdots \\
H_{\ell_n -r}^{\pm}(B_{1}) & \cdots & H_{\ell_n -r}^{\pm}(B_{n}) 
\end{array}
\!\!\right| , 
\label{cycle1}
\\
&& \quad
D_{n}^{\pm}(B_{1}, \cdots , B_{n})
:=(-1)^{\frac{1}{8}(n-m)(n-m+2)}
\exp{(-\sum_{k=-\infty}^{\infty}t_{2k}\sum_{j=1}^nB_j^{2k})},
\label{cycle11}
\\
&& \quad
H_{p}^{\pm}(B):=\exp{(-2\sum_{k=-\infty}^{\infty}t_{2k}B^{2k})}B^{\pm(2p-1)}
\prod_{a=1}^{r}(1-A_{a}^{2}B^{-2}). 
\label{cycle2}
\ena
Here $n$ is even, $n\geq m$, $\ell_n=r+(n-m)/2$, (\ref{cycle1}) is for 
$r \geq 0$, and 
(\ref{cycle11}) is for the initial form factor (\ref{rzero}). 
We set $P_n=0$ for $n<m$.

\begin{theorem}\label{mainth}
The set of functions
\bea
&&
f_{P_{n(\pm)}}=
e^{\frac{n}{4}\sum_{j=1}^{n}\beta_{j}}
\prod_{1 \le j<j' \le n}\zeta(\beta_{j}-\beta_{j'})\Psi_{P_{n(\pm)}}
\label{minimalff}
\ena
satisfies (I), (II), (III) and defines an $m$-minimal local operator.
Each form factor of (\ref{minimalff}) is in $(V^{\otimes n})_{m-2r}^{sing}$.
\end{theorem}
\vskip3mm

We will prove this theorem in Section \ref{pfsec}.

On the space of local operators defined by the form factors in 
Theorem \ref{mainth}, two kinds of abelian symmetries manifest.

One corresponds to $t_k$ with $k$ odd.
It describes the action of local integrals of motion.
The other is $t_k$ for $k$ even.
It describes some non-trivial symmetry.
It is interesting to study wether one can define this abelian
symmetry without using the explicit form of form factors.
\vskip3mm

\noindent
{\bf 4.4.} Spins of local operators.

\begin{defn}
Let $f_{{\cal O}}(\beta_1,\cdots,\beta_n)$ be the form factor 
of a local operator ${\cal O}$.
If
\bea
f_{{\cal O}}(\beta_1+\theta,\cdots,\beta_n+\theta)
=\exp(\theta S({\cal O}))f_{{\cal O}}(\beta_1,\cdots,\beta_n)
\nonumber
\ena
for all $n$ and a parameter $\theta$, 
then $S({\cal O})$ is called the spin of ${\cal O}$.
\end{defn}

Let us calculate the spin of the operator ${\cal O}$ whose $n$-particle
form factor is given by ({\ref{minimalff}) with all $t_k=0$.
By shifting $\beta_j$'s and $\alpha_a$'s simultaneously, 
one easily finds that
\bea
&&
f_{P_{n(\pm)}}(\beta_1+\theta,\cdots,\beta_n+\theta)\vert_{\forall t_j=0}
=\exp(\theta S_{s,m}^\pm(k_1,\cdots,k_r))
f_{P_n}(\beta_1,\cdots,\beta_n)\vert_{\forall t_j=0},
\nonumber
\ena
where
\bea
S_{s,m}^\pm(k_1,\cdots,k_r)
&=&
-\sum_{a=1}^r\, k_a+\frac{m^2}{4}-ms
\mp\frac{1}{2}m(m-1),
\quad \hbox{ for (\ref{defnP}) with (\ref{cycle1})},
\no
\\
&=&
\frac{m^2}{4}-ms, \quad \hbox{ for (\ref{defnP}) with (\ref{cycle11})}.
\no
\ena
Thus we have

\begin{prop}
The spin of the local operator whose form factor is 
$f_{P_{n(\pm)}}\vert_{\forall t_k=0}$ is $S_{s,m}^{\pm}(k_1,\cdots,k_r)$.
\end{prop}

The multiplication of
\bea
&&
I_{2k-1}=\sum_{j=1}^n B_j^{-(2k-1)},
\nonumber
\ena
to $n$-particle form factor for all $n$ simultaneously
increases the spin of the operator by $2k-1$.
As we mentioned $I_{2k-1}$ corresponds to the local integrals of motion
of spin $2k-1$.
The differentiation with respect to $t_{2k}$ increases the
spin by $-2k$.
\vskip3mm

\noindent
{\bf 4.5.} $su(2)$ current.

Let us describe the form factors of $su(2)$ currents
as a special case of $f_{P_{n(\pm)}}$ of Theorem \ref{mainth}.

The form factors of the $su(2)$ currents were determined in
\cite{KS,smir1}.
The form factor $f^{+}_\sigma$ of the current $j^{+}_\sigma$, $\sigma=\pm$,
is ({\it cf.} \cite{npt})
\bea
&&
f^{+}_\sigma(\beta_1,\cdots,\beta_{2n})
=C^{j^{+}_\sigma}
\exp(\frac{n}{2}\sum_{j=1}^{2n}\beta_j)
\prod_{1\le j<j' \le 2n}\zeta(\beta_j-\beta_{j'})
\Psi_{P_\sigma}(\beta_1,\cdots,\beta_{2n}),
\label{current}
\ena
where $\ell=n-1$ in the formula of $\Psi_{P_\sigma}$
and
\bea
&&
P_{+}=(-1)^{n-1}c^{2,0,1}_{2n}\prod_{j=1}^{2n} B_j\prod_{a=1}^{n-1}A_a^{2a-1},
\quad
P_{-}=(-1)^{n-1}c^{2,0,0}_{2n}\prod_{a=1}^{n-1}A_a^{2a+1}.
\no
\ena
The number $C^{j^{+}_\sigma}$ is a constant which does not depend on 
$n$ and $\sigma$.
The cycle $P_{+}$ is $P_{2n(+)}\vert_{\forall t_j=0}$ with
$m=2$, $r=0$, $s=1$ and $P_{-}$ is 
$P_{2n(+)}\vert_{\forall t_j=0}$ with $m=2$, $r=0$, $s=0$.
In both cases we use (\ref{defnP}) and (\ref{cycle11}). 
In particular
\bea
&&
S(j^{+}_\sigma)=-\sigma.
\no
\ena

The two particle form factors are without integral and  
given by 
\bea
&&
f^{+}_\sigma(\beta_1,\beta_{2})
=C^{j^{+}_\sigma}
\exp(\frac{-\sigma}{2}(\beta_1+\beta_2))
v_{+}\otimes v_{+}.
\no
\ena

\vskip3mm

\noindent
{\bf 4.6.} Energy momentum tensor.

The $n$-particle form factor $f_{\mu\nu}$ of the energy momentum tensor 
$T_{\mu\nu}$ was determined in \cite{smirbook}.
It is given by
\bea
&&
f_{\mu\nu}(\beta_1,\cdots,\beta_{2n})
=
C^{T_{\mu\nu}}
\exp(\frac{n}{2}\sum_{j=1}^{2n}\beta_j)
\prod_{1 \le j<j' \le 2n}\zeta(\beta_j-\beta_{j'})\Psi_{P_{\mu\nu}},
\no
\ena
where $P_{\mu\nu}$ is 
\bea
&&
P_{\mu\nu}=
c^{2,1,0}_{2n}
\biggl(
\sum_{j=1}^{2n}B_j^{-1}-(-1)^{\nu}\sum_{j=1}^{2n}B_j
\biggr)
\biggl(
(-1)^{\mu+n}\prod_{j=1}^{2n}B_j\prod_{a=1}^{n-1}A_a^{2a-1}+
\prod_{a=1}^{n-1}A_a^{2a+1}
\biggr).
\no
\ena
The number $C^{T_{\mu\nu}}$ is a constant given by (\ref{EMconst})
which is independent of $n$, $\mu$, $\nu$.
The following proposition is proved in \cite{smirbook}.
Since there is a misprint in \cite{smirbook}, we shall give a proof of it.
It will also serve to understand the role of the relation (\ref{xi1})
between cycles in proving the properties of form factors.
\begin{prop}$\quad$
{\bf 1.} $f_{\mu\nu}=f_{\nu\mu}. $
$\qquad$
{\bf 2.} $\partial_0 f_{0\nu}-\partial_1 f_{1\nu}=0$.
\end{prop}
\vskip3mm

\noindent
{\it Proof.}
1. 
For two polynomials $P$ and $Q$ of $A_1$, ..., $A_{n-1}$, 
we write $P\cong Q$ if $\text{Asym}(P)=\text{Asym}(Q)$ in (\ref{solsp1}).
Precisely speaking we consider $\text{Asym}(P)$ as an element of 
(\ref{solsp1}) after dividing it by the denominator of (\ref{poly}).
Notice that 
\bea
&&
\bar{e}_{2n-1}(B)=\prod_{j=1}^{2n} B_j^{-1} \sum_{j=1}^{2n}B_j
\no
\ena 
in the relation (\ref{xi1}), where $n$ is replaced by $2n$.
Using (\ref{xi1}) we can rewrite $\text{Asym}(\prod_{a=1}^{n-1}A_a^{2a+1})$
in terms of $\text{Asym}(\prod_{a=1}^{n-1}A_a^{2a-1})$ and get
\bea
&&
P_{\mu\nu}\cong
c^{2,1,1}_{2n}
\biggl(
\sum_{j=1}^{2n}B_j^{-1}-(-1)^{\nu}\sum_{j=1}^{2n}B_j
\biggr)
\biggl(
\sum_{j=1}^{2n}B_j^{-1}-(-1)^{\mu}\sum_{j=1}^{2n}B_j
\biggr)
\frac{\prod_{j=1}^{2n}B_j}{\sum_{j=1}^{2n}B_j}
\prod_{a=1}^{n-1}A_a^{2a-1},
\label{symmetric}
\ena
where we use $c^{2,1,1}_{2n}=(-1)^{n-1}c^{2,1,0}_{2n}$.
Thus 1 is proved.
\par
\noindent
2. Since $\partial_0$ and $\partial_1$ act on $2n$-particle form factors 
by the multiplication of $-iM\sum_{j=1}^{2n}\text{cosh}(\beta_j)$ and
$-iM\sum_{j=1}^{2n}\text{sinh}(\beta_j)$ respectively,
the claim is equivalent to
\bea
&&
(\sum_{j=1}^{2n}B_j^{-1}+\sum_{j=1}^{2n}B_j)f_{0\nu}
=(\sum_{j=1}^{2n}B_j^{-1}-\sum_{j=1}^{2n}B_j)f_{1\nu}.
\no
\ena
This follows from the equation for cycles
\bea
&&
(\sum_{j=1}^{2n}B_j^{-1}+\sum_{j=1}^{2n}B_j)P_{0\nu}
\cong
(\sum_{j=1}^{2n}B_j^{-1}-\sum_{j=1}^{2n}B_j)P_{1\nu}.
\no
\ena
Let us prove this equation.
Since, by the expression (\ref{symmetric}),
\bea
&&
\frac{1}{2}(P_{0\nu}-P_{1\nu})
\cong
-c^{2,1,1}_{2n}
(\sum_{j=1}^{2n}B_j^{-1}-(-1)^{\nu}\sum_{j=1}^{2n}B_j)
\prod_{j=1}^{2n}B_j\prod_{a=1}^{n-1}A_a^{2a-1},
\no
\\
&&
\frac{1}{2}(P_{0\nu}+P_{1\nu})
\cong
c^{2,1,1}_{2n}(\sum_{j=1}^{2n}B_j^{-1}-(-1)^{\nu}\sum_{j=1}^{2n}B_j)
\frac{\sum_{j=1}^{2n}B_j^{-1}}{\sum_{j=1}^{2n}B_j}
\prod_{j=1}^{2n}B_j\prod_{a=1}^{n-1}A_a^{2a-1},
\no
\ena
we have
\bea
&&
\sum_{j=1}^{2n}B_j(P_{0\nu}+P_{1\nu})
\cong
-\sum_{j=1}^{2n}B_j^{-1}(P_{0\nu}-P_{1\nu}).
\ena
Thus 2 is proved. q.e.d
\vskip3mm

Using (\ref{xi1}) we have
\bea
&&
P_{z}:=c^{2,1,0}_{2n}(\sum_{j=1}^{2n}B_j^{-1})\prod_{a=1}^{n-1}A_a^{2a+1}
\cong
\frac{1}{4}(P_{00}+P_{11}+2P_{01}),
\no
\\
&&
P_{\bar{z}}:=
c^{2,1,1}_{2n} \prod_{j=1}^{2n}B_j(\sum_{j=1}^{2n}B_j)
\prod_{a=1}^{n-1}A_a^{2a-1} 
\cong
\frac{1}{4}(P_{00}+P_{11}-2P_{01}),
\no
\ena
where we have defined $P_{z}$ and $P_{\bar{z}}$ by these equations.
These are special cases of our cycles,
$P_z$ is described by
$P_{2n(-)}$ with 
$r=1$, $m=2$, $s=0$, $k_1=0$ and 
$P_{\bar{z}}$ by
$P_{2n(+)}$ with 
$r=1$, $m=2$, $s=1$, $k_1=0$.

The form factors $f_{P_{z}}$, $f_{P_{\bar{z}}}$ describe
the operators $T_z$ and $T_{\bar{z}}$ which correspond
to holomorphic and antiholomorphic part of the energy momentum tensor
in the CFT limit. 
It follows from the description of the cycles above that the spins of 
$T_z$ and $T_{\bar{z}}$ are $2$ and $-2$ respectively.

Let us introduce the coordinates
\bea
&&
z=x_0+x_1,
\qquad
\bar{z}=x_0-x_1.
\no
\ena
Define the trace $\Theta$ of the energy momentum tensor by
\bea
&&
\Theta=\frac{1}{4}(T_{11}-T_{00}).
\no
\ena
Then the conservation laws of $T_{\mu\nu}$ are written as
\bea
&&
\partial_{\bar{z}}T_z=\partial_z\Theta,
\qquad
\partial_zT_{\bar{z}}=\partial_{\bar{z}}\Theta,
\no
\ena
which are in turn equivalent to the relations of cycles:
\bea
&&
(\sum_{j=1}^{2n}B_j) P_z
\cong
(\sum_{j=1}^{2n}B_j^{-1}) P_{tr},
\qquad
(\sum_{j=1}^{2n}B_j^{-1})P_{\bar{z}}
\cong
(\sum_{j=1}^{2n}B_j) P_{tr},
\no
\ena
where $P_{tr}$ is the cycle which describes $\Theta$:
\bea
&&
P_{tr}=c^{2,1,0}_{2n}(\sum_{j=1}^{2n}B_j)\prod_{a=1}^{n-1}A_a^{2a+1}.
\no
\ena
These relations among cycles can be directly checked using
(\ref{xi1}) as in the previous proposition.

The two particle form factors $f_{\mu\nu}(\beta_1,\beta_2)$,
$f_{P_z}(\beta_1,\beta_2)$, $f_{P_{\bar{z}}}(\beta_1,\beta_2)$
are given by
\bea
f_{\mu\nu}(\beta_1,\beta_2)&=&
4\pi iC^{T_{\mu\nu}}\zeta(\beta_1-\beta_{2})
\frac{e^{\frac{1}{2}(\beta_1+\beta_2)}}{e^{\beta_1}+e^{\beta_2}}
\sum_{i=1}^2(e^{\beta_i}-(-1)^\mu e^{-\beta_i})
\sum_{j=1}^2(e^{\beta_j}-(-1)^\nu e^{-\beta_j})
\no
\\
&\times& \frac{1}{\beta_2-\beta_1-\pi i}
(v_{+}\otimes v_{-}-v_{-}\otimes v_{+}),
\no
\\
f_{P_z}(\beta_1,\beta_2)&=&
4\pi iC^{T_{\mu\nu}}\zeta(\beta_1-\beta_{2})
e^{\frac{1}{2}(\beta_1+\beta_2)}(e^{\beta_1}+e^{\beta_2})
\frac{1}{\beta_2-\beta_1-\pi i}(v_{+}\otimes v_{-}-v_{-}\otimes v_{+}),
\no
\\
f_{P_{\bar{z}}}(\beta_1,\beta_2)&=&
4\pi iC^{T_{\mu\nu}}\zeta(\beta_1-\beta_{2})
e^{-\frac{1}{2}(\beta_1+\beta_2)}(e^{-\beta_1}+e^{-\beta_2})
\frac{1}{\beta_2-\beta_1-\pi i}(v_{+}\otimes v_{-}-v_{-}\otimes v_{+}).
\no
\ena
Notice that the formulae for two particle form factors of $T_{\mu\nu}$ 
are given by one fold integral. In the case of weight zero, 
which is the case for $T_{\mu\nu}$, this integral can be calculated 
by some general reason \cite{npt}. 
The above formulae for two particle form factors can be calculated
in such a way.
The general one time integrated formulae for $n$-particle form factors 
of weight zero are given in \S 6.

The constant $C^{T_{\mu\nu}}$ is determined by the condition
\bea
&&
P_\mu\vert \beta>_{\pm}=M\frac{\text{e}^\beta+(-1)^\mu \text{e}^{-\beta}}{2}
\vert \beta>_{\pm},
\quad
P_\mu=\int_{-\infty}^{\infty}T_{\mu0}(x_0,x_1)dx_1,
\no
\ena
where $\vert \beta>_{\pm}$ is 
the one particle states of the kink ($+$) and the anti-kink ($-$) 
with the rapidity $\beta$. 
Explicitly it is given by
\bea
&&
C^{T_{\mu\nu}}=-\frac{M^2}{32\pi^2 \zeta(-\pi i)}.
\label{EMconst}
\ena
\vskip5mm

\section{Proof of Theorem \ref{mainth}}\label{pfsec}

\noindent
Fix non-negative integers $r$ and $m$ such that $m-2r\geq 0$ and
$m$ is even.
For an even integer $n$ set $\ell_n=r+(n-m)/2$.
Consider the set of polynomials 
$P_n(A_1,\cdots,A_{\ell_n}\vert B_1,\cdots,B_n)$
in $A_j$'s with the coefficients in the polynomials of $B_j^{\pm1}$'s
satisfying the following conditions.
\begin{description}
\item[(i).] $0\leq \text{deg}_{A_a}\, P_n \leq n$ for $1\leq a\leq \ell_n$.
\item[(ii).] There exists a set of polynomials 
$\overline{P}_{n}=$
$\overline{P}_{n}(A_1,\cdots,A_{\ell_n}\vert B_1,\cdots,B_{n-2}| B)$ such that
\bea
&&
\text{Asym}\left\{ P_n(A_{1}, \cdots , A_{\ell_{n}}|B_{1}, \cdots , B_{n-2}, 
B, -B) \right\} 
=
\text{Asym}\left\{ \prod_{a=1}^{\ell_n -1}(1-A_{a}^{2}B^{-2}) 
\overline{P}_{n}
\right\}, 
\label{cycletoshow1}
\\
&&
\overline{P}_{n}(A_{1}, \cdots \! , A_{\ell_n \! -1}, \pm B|B_{1}, \cdots \! , B_{n-2}|B)=
\pm B^{n-1}d_{n}P_{n-2}(A_{1}, \cdots \! , A_{\ell_n \! -1}|B_{1}, \cdots \! , B_{n-2}),
\label{cycletoshow2}
\ena
where $d_{n}$ is the constant given by (\ref{defd}).
\end{description}

We shall prove the following two things.
\begin{description}
\item[(a).] If $\{P_n\}$ satisfies (i) and (ii), then $\{f_{P_n}\}$
defined by (\ref{minimalff}), satisfies (III) in the axioms for 
locality.

\item[(b).] The $P_{n(\pm)}$'s of (\ref{defnP}) satisfy (i) and (ii).
\end{description}
\vskip3mm

\noindent
{\it Proof of (a).}
In the following we set $\ell=\ell_n$ for the sake of simplicity.
We calculate the residue of $f_{P_{n}}$ at $\beta_{n}=\beta_{n-1}+\pi i$ explicitly.
Note that the function $\prod_{j<j'}\zeta(\beta_{j}-\beta_{j'})$ is regular at 
$\beta_{n}=\beta_{n-1}+\pi i$.
Hence, it suffices to calculate the residue of $\Psi_{P_{n}}$.

Recall the definition of $\Psi_{P_{n}}$:
\bea
\Psi_{P_{n}}(\beta_{1}, \cdots , \beta_{n})=
\sum_{\# M=\ell}v_{M} \int_{C^{\ell}} \prod_{a=1}^{\ell}d\alpha_{a}
\prod_{a=1}^{\ell}\phi(\alpha_{a}) w_{M}(\alpha_{1}, \cdots , \alpha_{\ell})
\frac{P_{n}(A_{1}, \cdots , A_{\ell})}
{\prod_{a=1}^{\ell}\prod_{j=1}^{n}(1-A_{a}B_{j}^{-1})},
\label{sol00}
\ena
where $A_{a}=e^{-\alpha_{a}}, B_{j}=e^{-\beta_{j}}$. 
We set
\bea
&&
\varphi(\alpha)=\varphi(\alpha ; \beta_{1}, \cdots , \beta_{n})
:=(2\pi i)^{n}
\frac{\phi(\alpha ; \beta_{1}, \cdots , \beta_{n})}
{\prod_{j=1}^{n}(1-AB_{j}^{-1})} \no \\
&& \qquad {}=
\prod_{j=1}^{n} \left\{
e^{\frac{\alpha -\beta_{j}}{2}}
\Gamma\left( \frac{\alpha -\beta_{j}+2\pi i}{2\pi i} \right)
\Gamma\left( \frac{\alpha -\beta_{j}+\pi i}{-2\pi i} \right) \right\},
\ena
where $A=e^{-\alpha}$.
Using this function, we rewrite (\ref{sol00}) as 
\bea
&&
\Psi_{P_{n}}(\beta_{1}, \cdots , \beta_{n})=
(2\pi i)^{-n\ell}\sum_{\# M=\ell}v_{M}I_{M}, \no \\
&&
I_{M}:=\int_{C^{\ell}}\prod_{a=1}^{\ell}d\alpha_{a} 
\prod_{a=1}^{\ell}\varphi(\alpha_{a}) w_{M}(\alpha_{1}, \cdots , \alpha_{\ell})
P_{n}(A_{1}, \cdots , A_{\ell}). 
\label{sol01}
\ena
The singularity at $\beta_{n}=\beta_{n-1}+\pi i$ comes from pinches of the contour $C$
by poles of the integrand of $I_{M}$.

\begin{prop} \label{simple}
The solution $\Psi_{P_{n}}$ has a simple pole at $\beta_{n}=\beta_{n-1}+\pi i$.
\end{prop}

\pf
Note that the function $w_{M}$ is anti-symmetric 
with respect to $\alpha_{1}, \cdots , \alpha_{\ell}$.
Hence we have
\bea
I_{M}=\int_{C^{\ell}}\prod_{a=1}^{\ell}d\alpha_{a}
\prod_{a=1}^{\ell}\varphi(\alpha_{a})
g_{M}(\alpha_{1}, \cdots , \alpha_{\ell})
\prod_{1\le a<b \le \ell}(A_{a}-A_{b}) L(A_{1}, \cdots , A_{\ell}), 
\ena 
where $L(A_{1}, \cdots , A_{\ell})$ is the polynomial satisfying
\bea
\prod_{1\le a<b \le \ell}(A_{a}-A_{b}) L(A_{1}, \cdots , A_{\ell})
=\text{Asym}\left\{P_{n}(A_{1}, \cdots , A_{\ell})\right\}.
\ena

In the limit $\beta_{n} \to \beta_{n-1} + \pi i$, 
the contour $C$ may be pinched at 
$\alpha_{a}=\beta_{n-1}-\pi i, \beta_{n-1}$, and $\beta_{n-1}+\pi i$ 
(Figure 2). 
\vskip5mm
%%%%%%%%%%%%%%%%%%%%%%%%%%%%%%%%%%%
%WinTpicVersion3.08
\unitlength 0.1in
\begin{center}
\begin{picture}( 49.7500, 14.8000)( 13.5500,-23.4000)
% DOT 2 0 3 0
% 2 2390 1000 2390 1000
% 
\special{pn 8}%
\special{sh 1}%
\special{ar 2390 1000 10 10 0  6.28318530717959E+0000}%
\special{sh 1}%
\special{ar 2390 1000 10 10 0  6.28318530717959E+0000}%
% DOT 2 0 3 0
% 2 2400 1820 2400 1820
% 
\special{pn 8}%
\special{sh 1}%
\special{ar 2400 1820 10 10 0  6.28318530717959E+0000}%
\special{sh 1}%
\special{ar 2400 1820 10 10 0  6.28318530717959E+0000}%
% CIRCLE 2 0 3 0
% 4 2400 1410 2400 1410 2400 1410 2400 1410
% 
\special{pn 8}%
\special{ar 2400 1410 0 0  0.0000000 6.2831853}%
% CIRCLE 2 0 3 0
% 4 2390 1410 2420 1440 2420 1440 2420 1440
% 
\special{pn 8}%
\special{ar 2390 1410 42 42  0.0000000 6.2831853}%
% CIRCLE 2 0 3 0
% 4 2420 2200 2420 2200 2400 2220 2400 2220
% 
\special{pn 8}%
\special{ar 2420 2200 0 0  0.0000000 6.2831853}%
% CIRCLE 2 0 3 0
% 4 2410 2210 2430 2240 2430 2240 2430 2240
% 
\special{pn 8}%
\special{ar 2410 2210 36 36  0.0000000 6.2831853}%
% STR 2 0 3 0
% 3 2500 930 2500 1030 2 0
% $\beta_{n-1}+\pi i$
\put(25.0000,-10.3000){\makebox(0,0)[lb]{$\beta_{n-1}+\pi i$}}%
% STR 2 0 3 0
% 3 2550 1380 2550 1480 2 0
% $\beta_{n-1}$
\put(25.5000,-14.8000){\makebox(0,0)[lb]{$\beta_{n-1}$}}%
% STR 2 0 3 0
% 3 2580 1760 2580 1860 2 0
% $\beta_{n-1}-\pi i$
\put(25.8000,-18.6000){\makebox(0,0)[lb]{$\beta_{n-1}-\pi i$}}%
% STR 2 0 3 0
% 3 2590 2180 2590 2280 2 0
% $\beta_{n-1}-2\pi i$
\put(25.9000,-22.8000){\makebox(0,0)[lb]{$\beta_{n-1}-2\pi i$}}%
% DOT 2 0 3 0
% 2 4210 1010 4200 1010
% 
\special{pn 8}%
\special{sh 1}%
\special{ar 4210 1010 10 10 0  6.28318530717959E+0000}%
\special{sh 1}%
\special{ar 4200 1010 10 10 0  6.28318530717959E+0000}%
% DOT 2 0 3 0
% 2 4210 1800 4210 1790
% 
\special{pn 8}%
\special{sh 1}%
\special{ar 4210 1800 10 10 0  6.28318530717959E+0000}%
\special{sh 1}%
\special{ar 4210 1790 10 10 0  6.28318530717959E+0000}%
% CIRCLE 2 0 3 0
% 4 4210 1400 4210 1400 4200 1400 4210 1400
% 
\special{pn 8}%
\special{ar 4210 1400 0 0  6.2831853 6.2831853}%
\special{ar 4210 1400 0 0  0.0000000 3.1415927}%
% CIRCLE 2 0 3 0
% 4 4200 1410 4220 1450 4220 1450 4220 1450
% 
\special{pn 8}%
\special{ar 4200 1410 46 46  0.0000000 6.2831853}%
% CIRCLE 2 0 3 0
% 4 4210 2210 4210 2210 4210 2210 4220 2200
% 
\special{pn 8}%
\special{ar 4210 2210 0 0  5.4977871 6.2831853}%
% CIRCLE 2 0 3 0
% 4 4190 2200 4210 2230 4220 2230 4220 2230
% 
\special{pn 8}%
\special{ar 4190 2200 36 36  0.0000000 6.2831853}%
% STR 2 0 3 0
% 3 4330 950 4330 1050 2 0
% $\beta_n+\pi i$
\put(43.3000,-10.5000){\makebox(0,0)[lb]{$\beta_n+\pi i$}}%
% STR 2 0 3 0
% 3 4360 1340 4360 1440 2 0
% $\beta_n$
\put(43.6000,-14.4000){\makebox(0,0)[lb]{$\beta_n$}}%
% STR 2 0 3 0
% 3 4360 1760 4360 1860 2 0
% $\beta_n-\pi i$
\put(43.6000,-18.6000){\makebox(0,0)[lb]{$\beta_n-\pi i$}}%
% STR 2 0 3 0
% 3 4350 2140 4350 2240 2 0
% $\beta_n-2\pi i$
\put(43.5000,-22.4000){\makebox(0,0)[lb]{$\beta_n-2\pi i$}}%
% SPLINE 2 0 3 0
% 40 1355 1180 2235 1165 2235 1165 4485 1178 4485 1178 5224 1252 5313 1482 5313 1482 5103 1571 5013 1571 4713 1599 3693 1573 3683 1573 2663 1567 2513 1597 2463 1596 2463 1596 2182 1755 2182 1755 2290 2015 2290 2015 2580 2027 2580 2027 3370 2031 3370 2031 3900 2014 3900 2014 4390 1997 4390 1997 4950 2000 4950 2000 5520 1973 5520 1973 5801 1925 5801 1925 5902 1766 5973 1496 6114 1257 6274 1258 6264 1258
% 
\special{pn 8}%
\special{pa 1356 1180}%
\special{pa 1388 1180}%
\special{pa 1420 1178}%
\special{pa 1452 1176}%
\special{pa 1484 1176}%
\special{pa 1516 1174}%
\special{pa 1548 1174}%
\special{pa 1580 1172}%
\special{pa 1614 1172}%
\special{pa 1646 1170}%
\special{pa 1678 1170}%
\special{pa 1710 1168}%
\special{pa 1742 1168}%
\special{pa 1774 1166}%
\special{pa 1806 1166}%
\special{pa 1838 1166}%
\special{pa 1870 1164}%
\special{pa 1902 1164}%
\special{pa 1934 1164}%
\special{pa 1966 1164}%
\special{pa 1998 1164}%
\special{pa 2030 1164}%
\special{pa 2062 1164}%
\special{pa 2092 1164}%
\special{pa 2124 1164}%
\special{pa 2156 1164}%
\special{pa 2188 1164}%
\special{pa 2220 1166}%
\special{pa 2252 1166}%
\special{pa 2282 1166}%
\special{pa 2314 1168}%
\special{pa 2346 1170}%
\special{pa 2378 1170}%
\special{pa 2408 1172}%
\special{pa 2440 1174}%
\special{pa 2472 1176}%
\special{pa 2502 1178}%
\special{pa 2534 1180}%
\special{pa 2564 1182}%
\special{pa 2596 1184}%
\special{pa 2628 1186}%
\special{pa 2658 1188}%
\special{pa 2690 1190}%
\special{pa 2720 1192}%
\special{pa 2752 1194}%
\special{pa 2782 1196}%
\special{pa 2814 1198}%
\special{pa 2846 1202}%
\special{pa 2876 1204}%
\special{pa 2908 1206}%
\special{pa 2938 1208}%
\special{pa 2970 1210}%
\special{pa 3002 1214}%
\special{pa 3032 1216}%
\special{pa 3064 1218}%
\special{pa 3094 1220}%
\special{pa 3126 1222}%
\special{pa 3158 1224}%
\special{pa 3188 1226}%
\special{pa 3220 1228}%
\special{pa 3252 1230}%
\special{pa 3282 1232}%
\special{pa 3314 1234}%
\special{pa 3346 1236}%
\special{pa 3378 1238}%
\special{pa 3408 1238}%
\special{pa 3440 1240}%
\special{pa 3472 1242}%
\special{pa 3504 1242}%
\special{pa 3536 1244}%
\special{pa 3568 1244}%
\special{pa 3600 1244}%
\special{pa 3632 1246}%
\special{pa 3664 1246}%
\special{pa 3696 1246}%
\special{pa 3728 1246}%
\special{pa 3760 1246}%
\special{pa 3792 1246}%
\special{pa 3824 1244}%
\special{pa 3856 1244}%
\special{pa 3890 1242}%
\special{pa 3922 1242}%
\special{pa 3954 1240}%
\special{pa 3988 1238}%
\special{pa 4020 1236}%
\special{pa 4054 1234}%
\special{pa 4086 1232}%
\special{pa 4120 1228}%
\special{pa 4152 1226}%
\special{pa 4186 1222}%
\special{pa 4220 1218}%
\special{pa 4254 1214}%
\special{pa 4286 1210}%
\special{pa 4320 1206}%
\special{pa 4354 1200}%
\special{pa 4388 1196}%
\special{pa 4424 1190}%
\special{pa 4458 1184}%
\special{pa 4492 1178}%
\special{pa 4526 1170}%
\special{pa 4560 1164}%
\special{pa 4596 1158}%
\special{pa 4630 1150}%
\special{pa 4664 1144}%
\special{pa 4700 1138}%
\special{pa 4734 1132}%
\special{pa 4768 1128}%
\special{pa 4802 1124}%
\special{pa 4836 1122}%
\special{pa 4868 1120}%
\special{pa 4902 1120}%
\special{pa 4934 1122}%
\special{pa 4966 1124}%
\special{pa 4998 1130}%
\special{pa 5028 1136}%
\special{pa 5058 1146}%
\special{pa 5088 1156}%
\special{pa 5116 1170}%
\special{pa 5144 1186}%
\special{pa 5172 1206}%
\special{pa 5198 1228}%
\special{pa 5224 1252}%
\special{pa 5248 1280}%
\special{pa 5270 1310}%
\special{pa 5290 1342}%
\special{pa 5306 1374}%
\special{pa 5316 1404}%
\special{pa 5322 1436}%
\special{pa 5320 1464}%
\special{pa 5310 1488}%
\special{pa 5292 1510}%
\special{pa 5268 1528}%
\special{pa 5238 1544}%
\special{pa 5204 1554}%
\special{pa 5170 1564}%
\special{pa 5134 1568}%
\special{pa 5098 1572}%
\special{pa 5066 1572}%
\special{pa 5034 1572}%
\special{pa 5004 1572}%
\special{pa 4972 1572}%
\special{pa 4940 1574}%
\special{pa 4908 1578}%
\special{pa 4876 1580}%
\special{pa 4844 1584}%
\special{pa 4812 1588}%
\special{pa 4780 1592}%
\special{pa 4748 1596}%
\special{pa 4716 1600}%
\special{pa 4684 1602}%
\special{pa 4652 1604}%
\special{pa 4620 1606}%
\special{pa 4588 1608}%
\special{pa 4556 1610}%
\special{pa 4524 1610}%
\special{pa 4492 1610}%
\special{pa 4460 1610}%
\special{pa 4428 1610}%
\special{pa 4396 1608}%
\special{pa 4364 1608}%
\special{pa 4332 1606}%
\special{pa 4300 1604}%
\special{pa 4268 1604}%
\special{pa 4236 1602}%
\special{pa 4204 1600}%
\special{pa 4172 1598}%
\special{pa 4140 1596}%
\special{pa 4108 1592}%
\special{pa 4076 1590}%
\special{pa 4044 1588}%
\special{pa 4012 1586}%
\special{pa 3980 1584}%
\special{pa 3948 1582}%
\special{pa 3916 1580}%
\special{pa 3884 1578}%
\special{pa 3852 1578}%
\special{pa 3820 1576}%
\special{pa 3788 1574}%
\special{pa 3756 1574}%
\special{pa 3724 1574}%
\special{pa 3692 1574}%
\special{pa 3660 1574}%
\special{pa 3628 1572}%
\special{pa 3596 1572}%
\special{pa 3564 1570}%
\special{pa 3532 1568}%
\special{pa 3500 1566}%
\special{pa 3468 1564}%
\special{pa 3436 1562}%
\special{pa 3404 1558}%
\special{pa 3372 1556}%
\special{pa 3340 1552}%
\special{pa 3308 1550}%
\special{pa 3274 1548}%
\special{pa 3242 1544}%
\special{pa 3210 1542}%
\special{pa 3178 1540}%
\special{pa 3146 1538}%
\special{pa 3114 1536}%
\special{pa 3082 1534}%
\special{pa 3050 1532}%
\special{pa 3018 1532}%
\special{pa 2986 1532}%
\special{pa 2954 1532}%
\special{pa 2922 1532}%
\special{pa 2890 1534}%
\special{pa 2858 1536}%
\special{pa 2826 1540}%
\special{pa 2794 1542}%
\special{pa 2764 1548}%
\special{pa 2732 1552}%
\special{pa 2700 1560}%
\special{pa 2668 1566}%
\special{pa 2638 1574}%
\special{pa 2606 1582}%
\special{pa 2574 1590}%
\special{pa 2544 1596}%
\special{pa 2512 1598}%
\special{pa 2480 1596}%
\special{pa 2448 1596}%
\special{pa 2412 1600}%
\special{pa 2378 1606}%
\special{pa 2344 1616}%
\special{pa 2310 1630}%
\special{pa 2280 1646}%
\special{pa 2250 1664}%
\special{pa 2226 1686}%
\special{pa 2204 1710}%
\special{pa 2190 1738}%
\special{pa 2180 1768}%
\special{pa 2176 1800}%
\special{pa 2178 1834}%
\special{pa 2186 1866}%
\special{pa 2196 1900}%
\special{pa 2212 1932}%
\special{pa 2232 1960}%
\special{pa 2254 1986}%
\special{pa 2280 2008}%
\special{pa 2308 2024}%
\special{pa 2336 2036}%
\special{pa 2366 2042}%
\special{pa 2398 2046}%
\special{pa 2430 2046}%
\special{pa 2464 2044}%
\special{pa 2496 2040}%
\special{pa 2530 2034}%
\special{pa 2564 2030}%
\special{pa 2598 2026}%
\special{pa 2632 2022}%
\special{pa 2666 2018}%
\special{pa 2698 2014}%
\special{pa 2732 2012}%
\special{pa 2764 2012}%
\special{pa 2796 2010}%
\special{pa 2830 2010}%
\special{pa 2862 2010}%
\special{pa 2894 2010}%
\special{pa 2926 2010}%
\special{pa 2956 2012}%
\special{pa 2988 2012}%
\special{pa 3020 2014}%
\special{pa 3052 2016}%
\special{pa 3084 2018}%
\special{pa 3114 2020}%
\special{pa 3146 2022}%
\special{pa 3178 2024}%
\special{pa 3208 2024}%
\special{pa 3240 2026}%
\special{pa 3272 2028}%
\special{pa 3302 2030}%
\special{pa 3334 2030}%
\special{pa 3366 2032}%
\special{pa 3398 2032}%
\special{pa 3430 2032}%
\special{pa 3460 2032}%
\special{pa 3492 2032}%
\special{pa 3524 2032}%
\special{pa 3556 2030}%
\special{pa 3588 2030}%
\special{pa 3620 2028}%
\special{pa 3652 2028}%
\special{pa 3684 2026}%
\special{pa 3716 2024}%
\special{pa 3748 2022}%
\special{pa 3782 2022}%
\special{pa 3814 2020}%
\special{pa 3846 2018}%
\special{pa 3878 2016}%
\special{pa 3910 2014}%
\special{pa 3942 2012}%
\special{pa 3974 2010}%
\special{pa 4006 2008}%
\special{pa 4038 2008}%
\special{pa 4070 2006}%
\special{pa 4102 2004}%
\special{pa 4134 2004}%
\special{pa 4166 2002}%
\special{pa 4196 2000}%
\special{pa 4228 2000}%
\special{pa 4260 2000}%
\special{pa 4292 1998}%
\special{pa 4324 1998}%
\special{pa 4356 1998}%
\special{pa 4390 1998}%
\special{pa 4422 1998}%
\special{pa 4454 1998}%
\special{pa 4486 1998}%
\special{pa 4518 1998}%
\special{pa 4550 1998}%
\special{pa 4582 2000}%
\special{pa 4614 2000}%
\special{pa 4646 2000}%
\special{pa 4678 2002}%
\special{pa 4710 2002}%
\special{pa 4742 2002}%
\special{pa 4774 2002}%
\special{pa 4806 2002}%
\special{pa 4838 2002}%
\special{pa 4870 2002}%
\special{pa 4902 2002}%
\special{pa 4934 2002}%
\special{pa 4964 2000}%
\special{pa 4996 1998}%
\special{pa 5028 1998}%
\special{pa 5058 1996}%
\special{pa 5090 1994}%
\special{pa 5120 1992}%
\special{pa 5152 1990}%
\special{pa 5182 1988}%
\special{pa 5214 1986}%
\special{pa 5246 1984}%
\special{pa 5278 1982}%
\special{pa 5310 1980}%
\special{pa 5342 1978}%
\special{pa 5374 1976}%
\special{pa 5406 1976}%
\special{pa 5440 1974}%
\special{pa 5474 1974}%
\special{pa 5508 1974}%
\special{pa 5542 1974}%
\special{pa 5576 1974}%
\special{pa 5612 1974}%
\special{pa 5646 1972}%
\special{pa 5678 1970}%
\special{pa 5710 1964}%
\special{pa 5740 1956}%
\special{pa 5768 1946}%
\special{pa 5794 1930}%
\special{pa 5818 1912}%
\special{pa 5838 1888}%
\special{pa 5856 1862}%
\special{pa 5872 1832}%
\special{pa 5888 1802}%
\special{pa 5900 1772}%
\special{pa 5912 1742}%
\special{pa 5924 1712}%
\special{pa 5934 1682}%
\special{pa 5944 1652}%
\special{pa 5952 1622}%
\special{pa 5960 1590}%
\special{pa 5966 1558}%
\special{pa 5970 1526}%
\special{pa 5974 1492}%
\special{pa 5976 1458}%
\special{pa 5980 1424}%
\special{pa 5986 1390}%
\special{pa 5994 1356}%
\special{pa 6006 1328}%
\special{pa 6026 1300}%
\special{pa 6054 1280}%
\special{pa 6090 1264}%
\special{pa 6136 1254}%
\special{pa 6190 1252}%
\special{pa 6238 1252}%
\special{pa 6274 1256}%
\special{pa 6284 1258}%
\special{pa 6264 1258}%
\special{sp}%
% VECTOR 2 0 3 0
% 2 1830 1160 1910 1160
% 
\special{pn 8}%
\special{pa 1830 1160}%
\special{pa 1910 1160}%
\special{fp}%
\special{sh 1}%
\special{pa 1910 1160}%
\special{pa 1844 1140}%
\special{pa 1858 1160}%
\special{pa 1844 1180}%
\special{pa 1910 1160}%
\special{fp}%
% VECTOR 2 0 3 0
% 2 4460 1970 4540 1970
% 
\special{pn 8}%
\special{pa 4460 1970}%
\special{pa 4540 1970}%
\special{fp}%
\special{sh 1}%
\special{pa 4540 1970}%
\special{pa 4474 1950}%
\special{pa 4488 1970}%
\special{pa 4474 1990}%
\special{pa 4540 1970}%
\special{fp}%
% STR 2 0 3 0
% 3 1500 970 1500 1070 2 0
% $C$
\put(15.0000,-10.7000){\makebox(0,0)[lb]{$C$}}%
% VECTOR 2 0 3 0
% 2 6280 1240 6330 1240
% 
\special{pn 8}%
\special{pa 6280 1240}%
\special{pa 6330 1240}%
\special{fp}%
\special{sh 1}%
\special{pa 6330 1240}%
\special{pa 6264 1220}%
\special{pa 6278 1240}%
\special{pa 6264 1260}%
\special{pa 6330 1240}%
\special{fp}%
% VECTOR 2 0 3 0
% 2 3300 1250 3340 1250
% 
\special{pn 8}%
\special{pa 3300 1250}%
\special{pa 3340 1250}%
\special{fp}%
\special{sh 1}%
\special{pa 3340 1250}%
\special{pa 3274 1230}%
\special{pa 3288 1250}%
\special{pa 3274 1270}%
\special{pa 3340 1250}%
\special{fp}%
% LINE 2 2 3 0
% 2 2390 1010 4140 1370
% 
\special{pn 8}%
\special{pa 2390 1010}%
\special{pa 4140 1370}%
\special{dt 0.045}%
% LINE 2 2 3 0
% 2 2430 1430 4200 1800
% 
\special{pn 8}%
\special{pa 2430 1430}%
\special{pa 4200 1800}%
\special{dt 0.045}%
% LINE 2 2 3 0
% 2 2410 1830 4170 2180
% 
\special{pn 8}%
\special{pa 2410 1830}%
\special{pa 4170 2180}%
\special{dt 0.045}%
% STR 2 0 3 0
% 3 3560 2410 3560 2510 2 0
% {\bf Figure 2}
\put(35.6000,-25.1000){\makebox(0,0)[lb]{{\bf Figure 2}}}%
\end{picture}%
\end{center}
%%%%%%%%%%%%%%%%%%%%%%%%%%%%%%%%%%%
\vskip5mm

\noindent
In order to avoid these pinches, 
we deform $C$ by taking residues at these points. 
Then we shall prove that, 
once we take the residue with respect to one integration variable 
$\alpha_{a}$, 
the remaining integrand has no poles that may pinch the contour 
for other integration variables. 
The proof is given by dividing the cases.
\vskip3mm

\noindent
(I). $M \subset \{1, \cdots , n-2 \}$.
\par
\noindent
We shall prove that $I_M$ has a simple pole at $\beta_n=\beta_{n-1}+\pi i$
in the case $M \subset \{1, \cdots , n-2 \}$.
Since there are no poles at $\alpha_{a}=\beta_{n-1},\beta_n$ for any $a$,
the contour $C$ is pinched only by the poles at
$\alpha_{a}=\beta_{n-1}-\pi i$ and $\alpha_{a}=\beta_{n}-2\pi i$.
To avoid this pinch, we deform the contour for $\alpha_{a}$ 
by taking the residue at $\alpha_{a}=\beta_{n-1}-\pi i$:
\bea
\int_{C}d\alpha_{a}=\int_{C'}d\alpha_{a}
+2\pi i{\rm Res}_{\alpha_{a}=\beta_{n-1}-\pi i}.
\label{pf1}
\ena
The integration contour $C'$ is such that 
it satisfies the same conditions as $C$ for 
$\beta_j+\pi i \mathbb{Z}$, $j\neq n-1,n$ and 
it goes between $\beta_{n-1}$ and $\beta_{n-1}-\pi i$,
between $\beta_{n}-\pi i$ and $\beta_{n}-2\pi i$ (Figure 3).
\vskip5mm
%%%%%%%%%%%%%%%%%%%%%%%%%%%%%%%%%%%
%WinTpicVersion3.08
\begin{center}
\unitlength 0.1in
\begin{picture}( 34.3000, 16.7000)( 20.1000,-24.4000)
% DOT 2 0 3 0
% 2 2810 1010 2810 1010
% 
\special{pn 8}%
\special{sh 1}%
\special{ar 2810 1010 10 10 0  6.28318530717959E+0000}%
\special{sh 1}%
\special{ar 2810 1010 10 10 0  6.28318530717959E+0000}%
% DOT 2 0 3 0
% 2 2800 2210 2810 2210
% 
\special{pn 8}%
\special{sh 1}%
\special{ar 2800 2210 10 10 0  6.28318530717959E+0000}%
\special{sh 1}%
\special{ar 2810 2210 10 10 0  6.28318530717959E+0000}%
% DOT 2 0 3 0
% 2 3980 1020 3980 1020
% 
\special{pn 8}%
\special{sh 1}%
\special{ar 3980 1020 10 10 0  6.28318530717959E+0000}%
\special{sh 1}%
\special{ar 3980 1020 10 10 0  6.28318530717959E+0000}%
% DOT 2 0 3 0
% 2 4010 1600 4010 1600
% 
\special{pn 8}%
\special{sh 1}%
\special{ar 4010 1600 10 10 0  6.28318530717959E+0000}%
\special{sh 1}%
\special{ar 4010 1600 10 10 0  6.28318530717959E+0000}%
% DOT 2 0 3 0
% 2 4000 2210 4000 2210
% 
\special{pn 8}%
\special{sh 1}%
\special{ar 4000 2210 10 10 0  6.28318530717959E+0000}%
\special{sh 1}%
\special{ar 4000 2210 10 10 0  6.28318530717959E+0000}%
% STR 2 0 3 0
% 3 2870 950 2870 1050 2 0
% $\beta_{n-1}$
\put(28.7000,-10.5000){\makebox(0,0)[lb]{$\beta_{n-1}$}}%
% STR 2 0 3 0
% 3 4030 940 4030 1040 2 0
% $\beta_n$
\put(40.3000,-10.4000){\makebox(0,0)[lb]{$\beta_n$}}%
% STR 2 0 3 0
% 3 3060 1600 3060 1700 2 0
% $\beta_{n-1}-\pi i$
\put(30.6000,-17.0000){\makebox(0,0)[lb]{$\beta_{n-1}-\pi i$}}%
% STR 2 0 3 0
% 3 4090 1620 4090 1720 2 0
% $\beta_{n}-\pi i$
\put(40.9000,-17.2000){\makebox(0,0)[lb]{$\beta_{n}-\pi i$}}%
% STR 2 0 3 0
% 3 2900 2190 2900 2290 2 0
% $\beta_{n-1}-2\pi i$
\put(29.0000,-22.9000){\makebox(0,0)[lb]{$\beta_{n-1}-2\pi i$}}%
% STR 2 0 3 0
% 3 4140 2220 4140 2320 2 0
% $\beta_n-2\pi i$
\put(41.4000,-23.2000){\makebox(0,0)[lb]{$\beta_n-2\pi i$}}%
% DOT 2 0 3 0
% 2 2820 1620 2820 1620
% 
\special{pn 8}%
\special{sh 1}%
\special{ar 2820 1620 10 10 0  6.28318530717959E+0000}%
\special{sh 1}%
\special{ar 2820 1620 10 10 0  6.28318530717959E+0000}%
% CIRCLE 2 0 3 0
% 4 2810 1620 2860 1820 2830 1620 2830 1620
% 
\special{pn 8}%
\special{ar 2810 1620 206 206  0.0000000 6.2831853}%
% VECTOR 2 0 3 0
% 2 2720 1440 2660 1460
% 
\special{pn 8}%
\special{pa 2720 1440}%
\special{pa 2660 1460}%
\special{fp}%
\special{sh 1}%
\special{pa 2660 1460}%
\special{pa 2730 1458}%
\special{pa 2712 1444}%
\special{pa 2718 1420}%
\special{pa 2660 1460}%
\special{fp}%
% VECTOR 2 0 3 0
% 2 2010 1020 2590 1020
% 
\special{pn 8}%
\special{pa 2010 1020}%
\special{pa 2590 1020}%
\special{fp}%
\special{sh 1}%
\special{pa 2590 1020}%
\special{pa 2524 1000}%
\special{pa 2538 1020}%
\special{pa 2524 1040}%
\special{pa 2590 1020}%
\special{fp}%
% SPLINE 2 0 3 0
% 13 2600 1030 2730 1110 2910 1180 3220 1340 3490 1470 3870 1680 4140 1750 4340 1630 4570 1400 4690 1210 4920 1030 5190 970 5440 970
% 
\special{pn 8}%
\special{pa 2600 1030}%
\special{pa 2628 1048}%
\special{pa 2654 1066}%
\special{pa 2682 1084}%
\special{pa 2708 1100}%
\special{pa 2738 1114}%
\special{pa 2766 1126}%
\special{pa 2796 1138}%
\special{pa 2826 1150}%
\special{pa 2856 1160}%
\special{pa 2886 1172}%
\special{pa 2916 1182}%
\special{pa 2946 1196}%
\special{pa 2974 1208}%
\special{pa 3004 1222}%
\special{pa 3032 1238}%
\special{pa 3060 1252}%
\special{pa 3088 1268}%
\special{pa 3116 1284}%
\special{pa 3144 1300}%
\special{pa 3172 1314}%
\special{pa 3200 1330}%
\special{pa 3228 1344}%
\special{pa 3258 1360}%
\special{pa 3286 1374}%
\special{pa 3316 1388}%
\special{pa 3344 1400}%
\special{pa 3374 1414}%
\special{pa 3402 1428}%
\special{pa 3432 1442}%
\special{pa 3460 1456}%
\special{pa 3488 1470}%
\special{pa 3516 1484}%
\special{pa 3544 1498}%
\special{pa 3572 1514}%
\special{pa 3600 1530}%
\special{pa 3626 1544}%
\special{pa 3654 1560}%
\special{pa 3680 1576}%
\special{pa 3708 1592}%
\special{pa 3736 1608}%
\special{pa 3764 1624}%
\special{pa 3792 1640}%
\special{pa 3822 1656}%
\special{pa 3852 1670}%
\special{pa 3882 1686}%
\special{pa 3912 1700}%
\special{pa 3944 1714}%
\special{pa 3974 1728}%
\special{pa 4006 1738}%
\special{pa 4038 1746}%
\special{pa 4068 1752}%
\special{pa 4100 1754}%
\special{pa 4130 1752}%
\special{pa 4160 1746}%
\special{pa 4188 1736}%
\special{pa 4216 1722}%
\special{pa 4244 1706}%
\special{pa 4272 1686}%
\special{pa 4298 1666}%
\special{pa 4324 1644}%
\special{pa 4350 1622}%
\special{pa 4376 1602}%
\special{pa 4400 1580}%
\special{pa 4424 1558}%
\special{pa 4448 1536}%
\special{pa 4472 1514}%
\special{pa 4494 1492}%
\special{pa 4516 1470}%
\special{pa 4536 1446}%
\special{pa 4556 1422}%
\special{pa 4574 1396}%
\special{pa 4592 1370}%
\special{pa 4608 1342}%
\special{pa 4624 1314}%
\special{pa 4640 1286}%
\special{pa 4658 1260}%
\special{pa 4674 1232}%
\special{pa 4694 1206}%
\special{pa 4714 1182}%
\special{pa 4736 1158}%
\special{pa 4760 1136}%
\special{pa 4784 1114}%
\special{pa 4810 1094}%
\special{pa 4836 1076}%
\special{pa 4864 1058}%
\special{pa 4892 1044}%
\special{pa 4922 1030}%
\special{pa 4952 1018}%
\special{pa 4982 1008}%
\special{pa 5012 998}%
\special{pa 5044 990}%
\special{pa 5076 984}%
\special{pa 5108 980}%
\special{pa 5140 976}%
\special{pa 5172 972}%
\special{pa 5204 970}%
\special{pa 5236 968}%
\special{pa 5268 968}%
\special{pa 5300 968}%
\special{pa 5332 968}%
\special{pa 5364 968}%
\special{pa 5396 970}%
\special{pa 5428 970}%
\special{pa 5440 970}%
\special{sp}%
% STR 2 0 3 0
% 3 2080 840 2080 940 2 0
% $C'$
\put(20.8000,-9.4000){\makebox(0,0)[lb]{$C'$}}%
% STR 2 0 3 0
% 3 3230 2510 3230 2610 2 0
% {\bf Figure 3}
\put(32.3000,-26.1000){\makebox(0,0)[lb]{{\bf Figure 3}}}%
% VECTOR 2 0 3 0
% 2 4550 1410 4610 1340
% 
\special{pn 8}%
\special{pa 4550 1410}%
\special{pa 4610 1340}%
\special{fp}%
\special{sh 1}%
\special{pa 4610 1340}%
\special{pa 4552 1378}%
\special{pa 4576 1380}%
\special{pa 4582 1404}%
\special{pa 4610 1340}%
\special{fp}%
\end{picture}%
\end{center}
%%%%%%%%%%%%%%%%%%%%%%%%%%%%%%%%%%%
\vskip5mm

\noindent
The residue in (\ref{pf1}) is given by
\bea
{\rm Res}_{\alpha_{a}=\beta_{n-1}-\pi i} 
&=&
e^{\frac{1}{2}\sum_{j=1}^{n-2}(\beta_{n-1}-\beta_{j}-\pi i)}
\Gamma(\frac{1}{2})(-2\pi i)
\Gamma(\frac{\beta_{n-1}-\beta_{n}+\pi i}{2\pi i})
\Gamma(\frac{\beta_{n-1}-\beta_{n}}{-2\pi i})
\varphi'(\beta_{n-1}-\pi i) 
\no \\
&\times&
\int_{C^{\ell-1}}\prod_{a'\neq a}^{\ell}d\alpha_{a'}
\prod_{a'\neq a}^{\ell}\varphi(\alpha_{a'})
\Big\{
g_{M}
\prod_{1\le a'<b' \le \ell}(A_{a'}-A_{b'}) L
\Big\}\Big\vert_{\alpha_a=\beta_{n-1}-\pi i},
\label{pf2}
\ena
where
\bea
\varphi'(\alpha):=\varphi(\alpha ; \beta_{1}, \cdots, \beta_{n-2}).
\label{varphi'}
\ena
In (\ref{pf2})
$\prod_{1\le a'<b' \le \ell}(A_{a'}-A_{b'})\vert_{\alpha_a=\beta_{n-1}-\pi i}$
contains the divisor
\bea
&&
\prod_{b\neq a}(e^{-\alpha_b}-e^{-\beta_{n-1}})
\no
\ena
which vanishes at $\alpha_b=\beta_{n-1}-\pi i$.
Thus the integration contour $C^{\ell-1}$ in (\ref{pf2}) can be replaced
by $C^{'\ell-1}$. 
Then ${\rm Res}_{\alpha_{1}=\beta_{n-1}-\pi i}$ has a simple pole at 
$\beta_{n}=\beta_{n-1}+\pi i$ which comes from 
$\Gamma(\frac{\beta_{n-1}-\beta_{n}+\pi i}{2\pi i})$.
Making the decomposition (\ref{pf1})
in all variables $\alpha_a$ we find that 
$I_M$ has a simple pole at $\beta_n=\beta_{n-1}+\pi i$.
\vskip3mm

\noindent
(II). $M=M'\cup \{n-1\}$, $M' \subset \{1, \cdots , n-2 \}$.
\par
\noindent
Notice that there are no poles at $\alpha_a=\beta_n$.
The pinches of the integration contour in the limit 
$\beta_n\rightarrow \beta_{n-1}+\pi i$ can occur at 
$\alpha_a=\beta_{n-1},\beta_{n-1}-\pi i$.
We decompose the integarl as (Figure 4)
\bea
\int_{C}d\alpha_{a}=
\int_{C'}d\alpha_{a}
-2\pi i{\rm Res}_{\alpha_{a}=\beta_{n-1}}
+2\pi i{\rm Res}_{\alpha_{a}=\beta_{n-1}-\pi i}.
\no
\ena
\vskip5mm
%%%%%%%%%%%%%%%%%%%%%%%%%
%WinTpicVersion3.08
\unitlength 0.1in
\begin{center}
\begin{picture}( 35.6000, 16.7000)( 14.0000,-24.4000)
% DOT 2 0 3 0
% 2 2800 2210 2810 2210
% 
\special{pn 8}%
\special{sh 1}%
\special{ar 2800 2210 10 10 0  6.28318530717959E+0000}%
\special{sh 1}%
\special{ar 2810 2210 10 10 0  6.28318530717959E+0000}%
% DOT 2 0 3 0
% 2 3980 1020 3980 1020
% 
\special{pn 8}%
\special{sh 1}%
\special{ar 3980 1020 10 10 0  6.28318530717959E+0000}%
\special{sh 1}%
\special{ar 3980 1020 10 10 0  6.28318530717959E+0000}%
% DOT 2 0 3 0
% 2 4010 1600 4010 1600
% 
\special{pn 8}%
\special{sh 1}%
\special{ar 4010 1600 10 10 0  6.28318530717959E+0000}%
\special{sh 1}%
\special{ar 4010 1600 10 10 0  6.28318530717959E+0000}%
% DOT 2 0 3 0
% 2 4000 2210 4000 2210
% 
\special{pn 8}%
\special{sh 1}%
\special{ar 4000 2210 10 10 0  6.28318530717959E+0000}%
\special{sh 1}%
\special{ar 4000 2210 10 10 0  6.28318530717959E+0000}%
% STR 2 0 3 0
% 3 4030 940 4030 1040 2 0
% $\beta_n$
\put(40.3000,-10.4000){\makebox(0,0)[lb]{$\beta_n$}}%
% STR 2 0 3 0
% 3 3060 1600 3060 1700 2 0
% $\beta_{n-1}-\pi i$
\put(30.6000,-17.0000){\makebox(0,0)[lb]{$\beta_{n-1}-\pi i$}}%
% STR 2 0 3 0
% 3 4090 1620 4090 1720 2 0
% $\beta_{n}-\pi i$
\put(40.9000,-17.2000){\makebox(0,0)[lb]{$\beta_{n}-\pi i$}}%
% STR 2 0 3 0
% 3 2900 2190 2900 2290 2 0
% $\beta_{n-1}-2\pi i$
\put(29.0000,-22.9000){\makebox(0,0)[lb]{$\beta_{n-1}-2\pi i$}}%
% STR 2 0 3 0
% 3 4140 2220 4140 2320 2 0
% $\beta_n-2\pi i$
\put(41.4000,-23.2000){\makebox(0,0)[lb]{$\beta_n-2\pi i$}}%
% DOT 2 0 3 0
% 2 2820 1620 2820 1620
% 
\special{pn 8}%
\special{sh 1}%
\special{ar 2820 1620 10 10 0  6.28318530717959E+0000}%
\special{sh 1}%
\special{ar 2820 1620 10 10 0  6.28318530717959E+0000}%
% CIRCLE 2 0 3 0
% 4 2810 1620 2860 1820 2830 1620 2830 1620
% 
\special{pn 8}%
\special{ar 2810 1620 206 206  0.0000000 6.2831853}%
% VECTOR 2 0 3 0
% 2 2720 1440 2660 1460
% 
\special{pn 8}%
\special{pa 2720 1440}%
\special{pa 2660 1460}%
\special{fp}%
\special{sh 1}%
\special{pa 2660 1460}%
\special{pa 2730 1458}%
\special{pa 2712 1444}%
\special{pa 2718 1420}%
\special{pa 2660 1460}%
\special{fp}%
% STR 2 0 3 0
% 3 2080 840 2080 940 2 0
% $C'$
\put(20.8000,-9.4000){\makebox(0,0)[lb]{$C'$}}%
% STR 2 0 3 0
% 3 3230 2510 3230 2610 2 0
% {\bf Figure 4}
\put(32.3000,-26.1000){\makebox(0,0)[lb]{{\bf Figure 4}}}%
% DOT 2 0 3 0
% 2 2810 1010 2810 1010
% 
\special{pn 8}%
\special{sh 1}%
\special{ar 2810 1010 10 10 0  6.28318530717959E+0000}%
\special{sh 1}%
\special{ar 2810 1010 10 10 0  6.28318530717959E+0000}%
% CIRCLE 2 0 3 0
% 4 2810 1020 2790 1220 2790 1220 2790 1220
% 
\special{pn 8}%
\special{ar 2810 1020 202 202  0.0000000 6.2831853}%
% STR 2 0 3 0
% 3 3050 960 3050 1060 2 0
% $\beta_{n-1}$
\put(30.5000,-10.6000){\makebox(0,0)[lb]{$\beta_{n-1}$}}%
% VECTOR 2 0 3 0
% 2 1400 1010 2000 1010
% 
\special{pn 8}%
\special{pa 1400 1010}%
\special{pa 2000 1010}%
\special{fp}%
\special{sh 1}%
\special{pa 2000 1010}%
\special{pa 1934 990}%
\special{pa 1948 1010}%
\special{pa 1934 1030}%
\special{pa 2000 1010}%
\special{fp}%
% SPLINE 2 0 3 0
% 17 2010 1020 2200 1040 2280 1110 2370 1210 2550 1280 2800 1330 3150 1330 3320 1420 3500 1590 3940 1750 4170 1760 4370 1630 4530 1310 4620 1120 4670 1010 4800 950 4960 940
% 
\special{pn 8}%
\special{pa 2010 1020}%
\special{pa 2044 1018}%
\special{pa 2078 1018}%
\special{pa 2112 1018}%
\special{pa 2142 1022}%
\special{pa 2172 1028}%
\special{pa 2202 1040}%
\special{pa 2228 1058}%
\special{pa 2252 1080}%
\special{pa 2274 1104}%
\special{pa 2296 1128}%
\special{pa 2316 1154}%
\special{pa 2336 1178}%
\special{pa 2358 1200}%
\special{pa 2384 1220}%
\special{pa 2410 1236}%
\special{pa 2440 1248}%
\special{pa 2470 1258}%
\special{pa 2502 1268}%
\special{pa 2534 1276}%
\special{pa 2566 1284}%
\special{pa 2598 1292}%
\special{pa 2628 1300}%
\special{pa 2660 1308}%
\special{pa 2690 1316}%
\special{pa 2722 1322}%
\special{pa 2754 1326}%
\special{pa 2784 1330}%
\special{pa 2816 1332}%
\special{pa 2848 1330}%
\special{pa 2880 1330}%
\special{pa 2914 1328}%
\special{pa 2946 1324}%
\special{pa 2978 1322}%
\special{pa 3010 1320}%
\special{pa 3042 1320}%
\special{pa 3074 1320}%
\special{pa 3106 1322}%
\special{pa 3138 1328}%
\special{pa 3168 1336}%
\special{pa 3198 1346}%
\special{pa 3226 1360}%
\special{pa 3256 1374}%
\special{pa 3282 1392}%
\special{pa 3310 1412}%
\special{pa 3334 1432}%
\special{pa 3358 1454}%
\special{pa 3382 1476}%
\special{pa 3404 1500}%
\special{pa 3426 1522}%
\special{pa 3450 1544}%
\special{pa 3472 1566}%
\special{pa 3496 1586}%
\special{pa 3520 1606}%
\special{pa 3546 1624}%
\special{pa 3574 1640}%
\special{pa 3602 1654}%
\special{pa 3630 1666}%
\special{pa 3660 1678}%
\special{pa 3690 1690}%
\special{pa 3722 1700}%
\special{pa 3752 1710}%
\special{pa 3784 1718}%
\special{pa 3816 1726}%
\special{pa 3848 1732}%
\special{pa 3882 1740}%
\special{pa 3914 1746}%
\special{pa 3946 1752}%
\special{pa 3980 1758}%
\special{pa 4012 1762}%
\special{pa 4044 1766}%
\special{pa 4076 1768}%
\special{pa 4108 1768}%
\special{pa 4140 1766}%
\special{pa 4170 1760}%
\special{pa 4200 1752}%
\special{pa 4230 1740}%
\special{pa 4260 1724}%
\special{pa 4286 1708}%
\special{pa 4314 1688}%
\special{pa 4338 1666}%
\special{pa 4360 1642}%
\special{pa 4382 1616}%
\special{pa 4400 1590}%
\special{pa 4418 1564}%
\special{pa 4432 1536}%
\special{pa 4446 1506}%
\special{pa 4460 1478}%
\special{pa 4472 1448}%
\special{pa 4484 1418}%
\special{pa 4498 1388}%
\special{pa 4510 1358}%
\special{pa 4522 1330}%
\special{pa 4536 1300}%
\special{pa 4550 1272}%
\special{pa 4564 1244}%
\special{pa 4578 1216}%
\special{pa 4592 1186}%
\special{pa 4606 1158}%
\special{pa 4618 1128}%
\special{pa 4630 1096}%
\special{pa 4640 1064}%
\special{pa 4654 1036}%
\special{pa 4670 1010}%
\special{pa 4694 990}%
\special{pa 4722 974}%
\special{pa 4752 962}%
\special{pa 4786 954}%
\special{pa 4818 948}%
\special{pa 4850 944}%
\special{pa 4882 942}%
\special{pa 4914 940}%
\special{pa 4946 940}%
\special{pa 4960 940}%
\special{sp}%
% VECTOR 2 0 3 0
% 2 2790 820 2910 820
% 
\special{pn 8}%
\special{pa 2790 820}%
\special{pa 2910 820}%
\special{fp}%
\special{sh 1}%
\special{pa 2910 820}%
\special{pa 2844 800}%
\special{pa 2858 820}%
\special{pa 2844 840}%
\special{pa 2910 820}%
\special{fp}%
% VECTOR 2 0 3 0
% 2 3910 1740 3930 1740
% 
\special{pn 8}%
\special{pa 3910 1740}%
\special{pa 3930 1740}%
\special{fp}%
\special{sh 1}%
\special{pa 3930 1740}%
\special{pa 3864 1720}%
\special{pa 3878 1740}%
\special{pa 3864 1760}%
\special{pa 3930 1740}%
\special{fp}%
\end{picture}%
\end{center}
%%%%%%%%%%%%%%%%%%%%%%%%%
\vskip5mm

\noindent
It can be easily checked that there do not occur pinches of 
the integration contour in the limit 
$\beta_n\rightarrow \beta_{n-1}+\pi i$ for both
${\rm Res}_{\alpha_{a}=\beta_{n-1}}$ and 
${\rm Res}_{\alpha_{a}=\beta_{n-1}-\pi i}$.
\vskip3mm

\noindent
(III). $M=M'\cup \{n\}$, $M' \subset \{1, \cdots , n-2 \}$.
\par
\noindent
The pinches of integration contour in the limit 
$\beta_n\rightarrow \beta_{n-1}+\pi i$ can occur at
$\alpha_a=\beta_{n-1},\beta_{n-1}\pm \pi i$.
We decompose the integral as
\bea
\int_{C}d\alpha_{a}=
\int_{C''}d\alpha_{a}
+2\pi i{\rm Res}_{\alpha_{a}=\beta_{n-1}+\pi i}
-2\pi i{\rm Res}_{\alpha_{a}=\beta_{n-1}}
+2\pi i{\rm Res}_{\alpha_{a}=\beta_{n-1}-\pi i}.
\label{decomp}
\ena
The integration contour $C''$ separates the same sets as $C$ for
$\beta_j+\pi i \mathbb{Z}$, $j\neq n-1$ and 
separates 
$\{\beta_{n-1}\pm \pi i\}\cup \{\beta_{n-1}-2\pi i\mathbb{Z}_{\geq 1}\}$
and $\{\beta_{n-1}\}\cup \{\beta_{n-1}+\pi i+2\pi i\mathbb{Z}_{\geq 1}\}$
(Figure 5).
\vskip5mm
%%%%%%%%%%%%%%%%%%%%%%%%%%%%%%%%%%
%WinTpicVersion3.08
\unitlength 0.1in
\begin{center}
\begin{picture}( 49.6000, 27.1000)( 14.8000,-32.1000)
% DOT 2 0 3 0
% 2 2800 610 2800 610
% 
\special{pn 8}%
\special{sh 1}%
\special{ar 2800 610 10 10 0  6.28318530717959E+0000}%
\special{sh 1}%
\special{ar 2800 610 10 10 0  6.28318530717959E+0000}%
% DOT 2 0 3 0
% 2 2810 1210 2810 1210
% 
\special{pn 8}%
\special{sh 1}%
\special{ar 2810 1210 10 10 0  6.28318530717959E+0000}%
\special{sh 1}%
\special{ar 2810 1210 10 10 0  6.28318530717959E+0000}%
% DOT 2 0 3 0
% 2 2810 1810 2810 1810
% 
\special{pn 8}%
\special{sh 1}%
\special{ar 2810 1810 10 10 0  6.28318530717959E+0000}%
\special{sh 1}%
\special{ar 2810 1810 10 10 0  6.28318530717959E+0000}%
% DOT 2 0 3 0
% 2 2800 2410 2800 2410
% 
\special{pn 8}%
\special{sh 1}%
\special{ar 2800 2410 10 10 0  6.28318530717959E+0000}%
\special{sh 1}%
\special{ar 2800 2410 10 10 0  6.28318530717959E+0000}%
% DOT 2 0 3 0
% 2 2810 3020 2810 3020
% 
\special{pn 8}%
\special{sh 1}%
\special{ar 2810 3020 10 10 0  6.28318530717959E+0000}%
\special{sh 1}%
\special{ar 2810 3020 10 10 0  6.28318530717959E+0000}%
% CIRCLE 2 0 3 0
% 4 2810 1200 2830 1370 2830 1370 2830 1370
% 
\special{pn 8}%
\special{ar 2810 1200 172 172  0.0000000 6.2831853}%
% CIRCLE 2 0 3 0
% 4 2810 1800 2830 1960 2830 1960 2830 1960
% 
\special{pn 8}%
\special{ar 2810 1800 162 162  0.0000000 6.2831853}%
% CIRCLE 2 0 3 0
% 4 2800 2420 2810 2570 2810 2570 2810 2570
% 
\special{pn 8}%
\special{ar 2800 2420 150 150  0.0000000 6.2831853}%
% DOT 2 0 3 0
% 2 4590 610 4590 610
% 
\special{pn 8}%
\special{sh 1}%
\special{ar 4590 610 10 10 0  6.28318530717959E+0000}%
\special{sh 1}%
\special{ar 4590 610 10 10 0  6.28318530717959E+0000}%
% DOT 2 0 3 0
% 2 4610 1210 4620 1200
% 
\special{pn 8}%
\special{sh 1}%
\special{ar 4610 1210 10 10 0  6.28318530717959E+0000}%
\special{sh 1}%
\special{ar 4620 1200 10 10 0  6.28318530717959E+0000}%
% DOT 2 0 3 0
% 2 4610 1810 4610 1810
% 
\special{pn 8}%
\special{sh 1}%
\special{ar 4610 1810 10 10 0  6.28318530717959E+0000}%
\special{sh 1}%
\special{ar 4610 1810 10 10 0  6.28318530717959E+0000}%
% DOT 2 0 3 0
% 2 4600 2400 4600 2400
% 
\special{pn 8}%
\special{sh 1}%
\special{ar 4600 2400 10 10 0  6.28318530717959E+0000}%
\special{sh 1}%
\special{ar 4600 2400 10 10 0  6.28318530717959E+0000}%
% DOT 2 0 3 0
% 2 4590 3010 4590 3010
% 
\special{pn 8}%
\special{sh 1}%
\special{ar 4590 3010 10 10 0  6.28318530717959E+0000}%
\special{sh 1}%
\special{ar 4590 3010 10 10 0  6.28318530717959E+0000}%
% STR 2 0 3 0
% 3 2880 570 2880 670 2 0
% $\beta_{n-1}+2\pi i$
\put(28.8000,-6.7000){\makebox(0,0)[lb]{$\beta_{n-1}+2\pi i$}}%
% STR 2 0 3 0
% 3 3030 1190 3030 1290 2 0
% $\beta_{n-1}+\pi i$
\put(30.3000,-12.9000){\makebox(0,0)[lb]{$\beta_{n-1}+\pi i$}}%
% STR 2 0 3 0
% 3 3030 1800 3030 1900 2 0
% $\beta_{n-1}$
\put(30.3000,-19.0000){\makebox(0,0)[lb]{$\beta_{n-1}$}}%
% STR 2 0 3 0
% 3 3040 2420 3040 2520 2 0
% $\beta_{n-1}-\pi i$
\put(30.4000,-25.2000){\makebox(0,0)[lb]{$\beta_{n-1}-\pi i$}}%
% STR 2 0 3 0
% 3 2890 2990 2890 3090 2 0
% $\beta_{n-1}-2\pi i$
\put(28.9000,-30.9000){\makebox(0,0)[lb]{$\beta_{n-1}-2\pi i$}}%
% STR 2 0 3 0
% 3 4710 580 4710 680 2 0
% $\beta_n+2\pi i$
\put(47.1000,-6.8000){\makebox(0,0)[lb]{$\beta_n+2\pi i$}}%
% STR 2 0 3 0
% 3 4750 1210 4750 1310 2 0
% $\beta_n+\pi i$
\put(47.5000,-13.1000){\makebox(0,0)[lb]{$\beta_n+\pi i$}}%
% STR 2 0 3 0
% 3 4700 1800 4700 1900 2 0
% $\beta_n$
\put(47.0000,-19.0000){\makebox(0,0)[lb]{$\beta_n$}}%
% STR 2 0 3 0
% 3 4780 2440 4780 2540 2 0
% $\beta_n-\pi i$
\put(47.8000,-25.4000){\makebox(0,0)[lb]{$\beta_n-\pi i$}}%
% STR 2 0 3 0
% 3 4720 2990 4720 3090 2 0
% $\beta_n-2\pi i$
\put(47.2000,-30.9000){\makebox(0,0)[lb]{$\beta_n-2\pi i$}}%
% VECTOR 2 0 3 0
% 2 1480 1500 1640 1500
% 
\special{pn 8}%
\special{pa 1480 1500}%
\special{pa 1640 1500}%
\special{fp}%
\special{sh 1}%
\special{pa 1640 1500}%
\special{pa 1574 1480}%
\special{pa 1588 1500}%
\special{pa 1574 1520}%
\special{pa 1640 1500}%
\special{fp}%
% STR 2 0 3 0
% 3 1570 1160 1570 1260 2 0
% $C''$
\put(15.7000,-12.6000){\makebox(0,0)[lb]{$C''$}}%
% STR 2 0 3 0
% 3 3800 3280 3800 3380 2 0
% {\bf Figure 5}
\put(38.0000,-33.8000){\makebox(0,0)[lb]{{\bf Figure 5}}}%
% VECTOR 2 0 3 0
% 2 2860 1030 2820 1030
% 
\special{pn 8}%
\special{pa 2860 1030}%
\special{pa 2820 1030}%
\special{fp}%
\special{sh 1}%
\special{pa 2820 1030}%
\special{pa 2888 1050}%
\special{pa 2874 1030}%
\special{pa 2888 1010}%
\special{pa 2820 1030}%
\special{fp}%
% VECTOR 2 0 3 0
% 2 2660 1820 2650 1760
% 
\special{pn 8}%
\special{pa 2660 1820}%
\special{pa 2650 1760}%
\special{fp}%
\special{sh 1}%
\special{pa 2650 1760}%
\special{pa 2642 1830}%
\special{pa 2660 1814}%
\special{pa 2682 1822}%
\special{pa 2650 1760}%
\special{fp}%
% VECTOR 2 0 3 0
% 2 2810 2260 2760 2270
% 
\special{pn 8}%
\special{pa 2810 2260}%
\special{pa 2760 2270}%
\special{fp}%
\special{sh 1}%
\special{pa 2760 2270}%
\special{pa 2830 2278}%
\special{pa 2812 2260}%
\special{pa 2822 2238}%
\special{pa 2760 2270}%
\special{fp}%
% LINE 2 0 3 0
% 2 1640 1510 2220 1510
% 
\special{pn 8}%
\special{pa 1640 1510}%
\special{pa 2220 1510}%
\special{fp}%
% SPLINE 2 0 3 0
% 9 2220 1510 2350 1270 2450 1080 2580 970 2770 920 2970 930 3090 980 3170 1050 3320 1120
% 
\special{pn 8}%
\special{pa 2220 1510}%
\special{pa 2236 1482}%
\special{pa 2252 1456}%
\special{pa 2268 1428}%
\special{pa 2284 1400}%
\special{pa 2300 1372}%
\special{pa 2314 1344}%
\special{pa 2330 1314}%
\special{pa 2344 1286}%
\special{pa 2356 1256}%
\special{pa 2370 1228}%
\special{pa 2384 1198}%
\special{pa 2398 1170}%
\special{pa 2412 1140}%
\special{pa 2428 1114}%
\special{pa 2446 1086}%
\special{pa 2466 1062}%
\special{pa 2488 1038}%
\special{pa 2512 1016}%
\special{pa 2538 996}%
\special{pa 2564 980}%
\special{pa 2594 964}%
\special{pa 2622 952}%
\special{pa 2654 942}%
\special{pa 2684 934}%
\special{pa 2716 928}%
\special{pa 2748 922}%
\special{pa 2780 920}%
\special{pa 2812 918}%
\special{pa 2844 918}%
\special{pa 2876 918}%
\special{pa 2908 920}%
\special{pa 2940 924}%
\special{pa 2972 930}%
\special{pa 3004 938}%
\special{pa 3034 950}%
\special{pa 3062 964}%
\special{pa 3090 980}%
\special{pa 3114 1000}%
\special{pa 3138 1022}%
\special{pa 3162 1044}%
\special{pa 3188 1062}%
\special{pa 3214 1078}%
\special{pa 3244 1092}%
\special{pa 3274 1104}%
\special{pa 3304 1116}%
\special{pa 3320 1120}%
\special{sp}%
% SPLINE 2 0 3 0
% 11 3310 1120 3800 1210 4120 1330 4420 1460 4700 1640 4820 1800 4850 1960 4780 2050 4640 2080 4430 1990 4320 1900
% 
\special{pn 8}%
\special{pa 3310 1120}%
\special{pa 3342 1124}%
\special{pa 3374 1128}%
\special{pa 3406 1134}%
\special{pa 3438 1138}%
\special{pa 3470 1142}%
\special{pa 3502 1146}%
\special{pa 3534 1152}%
\special{pa 3564 1158}%
\special{pa 3596 1162}%
\special{pa 3628 1168}%
\special{pa 3660 1176}%
\special{pa 3690 1182}%
\special{pa 3722 1190}%
\special{pa 3752 1198}%
\special{pa 3784 1206}%
\special{pa 3814 1214}%
\special{pa 3844 1224}%
\special{pa 3874 1234}%
\special{pa 3904 1244}%
\special{pa 3934 1256}%
\special{pa 3964 1268}%
\special{pa 3994 1280}%
\special{pa 4024 1292}%
\special{pa 4054 1304}%
\special{pa 4084 1316}%
\special{pa 4114 1328}%
\special{pa 4142 1340}%
\special{pa 4172 1352}%
\special{pa 4202 1364}%
\special{pa 4232 1376}%
\special{pa 4260 1388}%
\special{pa 4290 1402}%
\special{pa 4320 1414}%
\special{pa 4348 1428}%
\special{pa 4378 1440}%
\special{pa 4406 1454}%
\special{pa 4436 1468}%
\special{pa 4464 1482}%
\special{pa 4494 1498}%
\special{pa 4522 1514}%
\special{pa 4550 1530}%
\special{pa 4578 1546}%
\special{pa 4604 1564}%
\special{pa 4630 1582}%
\special{pa 4656 1602}%
\special{pa 4680 1622}%
\special{pa 4704 1644}%
\special{pa 4726 1666}%
\special{pa 4746 1690}%
\special{pa 4766 1714}%
\special{pa 4786 1740}%
\special{pa 4802 1768}%
\special{pa 4818 1796}%
\special{pa 4834 1828}%
\special{pa 4844 1860}%
\special{pa 4852 1892}%
\special{pa 4856 1924}%
\special{pa 4852 1954}%
\special{pa 4842 1984}%
\special{pa 4824 2012}%
\special{pa 4800 2036}%
\special{pa 4774 2054}%
\special{pa 4744 2068}%
\special{pa 4714 2078}%
\special{pa 4680 2082}%
\special{pa 4648 2082}%
\special{pa 4616 2078}%
\special{pa 4584 2070}%
\special{pa 4554 2060}%
\special{pa 4524 2048}%
\special{pa 4496 2032}%
\special{pa 4468 2016}%
\special{pa 4440 1998}%
\special{pa 4414 1978}%
\special{pa 4390 1958}%
\special{pa 4364 1938}%
\special{pa 4340 1918}%
\special{pa 4320 1900}%
\special{sp}%
% SPLINE 2 0 3 0
% 14 4320 1920 4240 1830 4010 1680 3590 1540 3180 1490 2690 1490 2590 1510 2500 1570 2470 1810 2520 1970 2740 2070 2960 2110 3170 2200 3370 2270
% 
\special{pn 8}%
\special{pa 4320 1920}%
\special{pa 4300 1896}%
\special{pa 4280 1872}%
\special{pa 4258 1848}%
\special{pa 4236 1826}%
\special{pa 4212 1804}%
\special{pa 4186 1784}%
\special{pa 4160 1766}%
\special{pa 4134 1748}%
\special{pa 4106 1730}%
\special{pa 4078 1714}%
\special{pa 4050 1700}%
\special{pa 4020 1686}%
\special{pa 3992 1672}%
\special{pa 3962 1658}%
\special{pa 3932 1646}%
\special{pa 3902 1634}%
\special{pa 3872 1622}%
\special{pa 3842 1610}%
\special{pa 3812 1600}%
\special{pa 3782 1590}%
\special{pa 3752 1580}%
\special{pa 3720 1572}%
\special{pa 3690 1564}%
\special{pa 3658 1556}%
\special{pa 3628 1548}%
\special{pa 3596 1542}%
\special{pa 3564 1536}%
\special{pa 3534 1530}%
\special{pa 3502 1524}%
\special{pa 3470 1520}%
\special{pa 3438 1516}%
\special{pa 3406 1512}%
\special{pa 3376 1508}%
\special{pa 3344 1504}%
\special{pa 3312 1500}%
\special{pa 3280 1498}%
\special{pa 3248 1496}%
\special{pa 3216 1492}%
\special{pa 3184 1490}%
\special{pa 3152 1488}%
\special{pa 3120 1486}%
\special{pa 3088 1484}%
\special{pa 3054 1484}%
\special{pa 3022 1482}%
\special{pa 2990 1480}%
\special{pa 2958 1480}%
\special{pa 2926 1480}%
\special{pa 2894 1480}%
\special{pa 2862 1480}%
\special{pa 2830 1480}%
\special{pa 2798 1482}%
\special{pa 2766 1484}%
\special{pa 2734 1486}%
\special{pa 2704 1490}%
\special{pa 2672 1492}%
\special{pa 2640 1498}%
\special{pa 2608 1504}%
\special{pa 2578 1516}%
\special{pa 2548 1530}%
\special{pa 2522 1548}%
\special{pa 2500 1572}%
\special{pa 2484 1596}%
\special{pa 2472 1626}%
\special{pa 2466 1656}%
\special{pa 2462 1688}%
\special{pa 2462 1722}%
\special{pa 2464 1758}%
\special{pa 2468 1792}%
\special{pa 2472 1826}%
\special{pa 2478 1860}%
\special{pa 2486 1894}%
\special{pa 2496 1924}%
\special{pa 2508 1952}%
\special{pa 2524 1976}%
\special{pa 2546 1998}%
\special{pa 2572 2016}%
\special{pa 2600 2030}%
\special{pa 2630 2042}%
\special{pa 2662 2054}%
\special{pa 2696 2062}%
\special{pa 2730 2068}%
\special{pa 2764 2074}%
\special{pa 2796 2080}%
\special{pa 2828 2084}%
\special{pa 2860 2088}%
\special{pa 2890 2094}%
\special{pa 2922 2100}%
\special{pa 2952 2108}%
\special{pa 2982 2118}%
\special{pa 3010 2128}%
\special{pa 3040 2140}%
\special{pa 3070 2154}%
\special{pa 3098 2168}%
\special{pa 3128 2182}%
\special{pa 3158 2194}%
\special{pa 3188 2208}%
\special{pa 3216 2220}%
\special{pa 3248 2230}%
\special{pa 3278 2240}%
\special{pa 3308 2250}%
\special{pa 3338 2260}%
\special{pa 3368 2270}%
\special{pa 3370 2270}%
\special{sp}%
% SPLINE 2 0 3 0
% 7 3360 2270 3880 2440 4240 2590 4720 2660 5050 2480 5340 2110 5440 1960
% 
\special{pn 8}%
\special{pa 3360 2270}%
\special{pa 3392 2280}%
\special{pa 3422 2288}%
\special{pa 3454 2298}%
\special{pa 3484 2306}%
\special{pa 3514 2316}%
\special{pa 3546 2326}%
\special{pa 3576 2334}%
\special{pa 3606 2344}%
\special{pa 3638 2354}%
\special{pa 3668 2364}%
\special{pa 3698 2374}%
\special{pa 3728 2384}%
\special{pa 3758 2394}%
\special{pa 3788 2406}%
\special{pa 3818 2416}%
\special{pa 3848 2428}%
\special{pa 3878 2440}%
\special{pa 3908 2452}%
\special{pa 3936 2464}%
\special{pa 3966 2476}%
\special{pa 3994 2488}%
\special{pa 4024 2500}%
\special{pa 4054 2514}%
\special{pa 4082 2526}%
\special{pa 4112 2538}%
\special{pa 4142 2552}%
\special{pa 4172 2564}%
\special{pa 4202 2576}%
\special{pa 4232 2588}%
\special{pa 4262 2598}%
\special{pa 4294 2610}%
\special{pa 4324 2620}%
\special{pa 4356 2630}%
\special{pa 4388 2638}%
\special{pa 4420 2646}%
\special{pa 4452 2654}%
\special{pa 4484 2660}%
\special{pa 4516 2664}%
\special{pa 4548 2668}%
\special{pa 4580 2670}%
\special{pa 4612 2672}%
\special{pa 4644 2670}%
\special{pa 4676 2668}%
\special{pa 4706 2664}%
\special{pa 4738 2656}%
\special{pa 4768 2648}%
\special{pa 4798 2638}%
\special{pa 4828 2626}%
\special{pa 4858 2614}%
\special{pa 4886 2598}%
\special{pa 4916 2582}%
\special{pa 4942 2564}%
\special{pa 4970 2546}%
\special{pa 4996 2526}%
\special{pa 5022 2506}%
\special{pa 5046 2484}%
\special{pa 5070 2462}%
\special{pa 5092 2440}%
\special{pa 5114 2416}%
\special{pa 5136 2392}%
\special{pa 5156 2368}%
\special{pa 5178 2344}%
\special{pa 5196 2318}%
\special{pa 5216 2292}%
\special{pa 5234 2266}%
\special{pa 5254 2240}%
\special{pa 5272 2214}%
\special{pa 5290 2188}%
\special{pa 5308 2160}%
\special{pa 5324 2134}%
\special{pa 5342 2108}%
\special{pa 5360 2080}%
\special{pa 5378 2054}%
\special{pa 5396 2028}%
\special{pa 5414 2000}%
\special{pa 5432 1974}%
\special{pa 5440 1960}%
\special{sp}%
% SPLINE 2 0 3 0
% 6 5440 1960 5740 1660 5860 1570 6030 1490 6210 1450 6340 1440
% 
\special{pn 8}%
\special{pa 5440 1960}%
\special{pa 5462 1936}%
\special{pa 5484 1914}%
\special{pa 5506 1890}%
\special{pa 5528 1866}%
\special{pa 5550 1844}%
\special{pa 5572 1820}%
\special{pa 5594 1798}%
\special{pa 5618 1774}%
\special{pa 5640 1752}%
\special{pa 5662 1730}%
\special{pa 5686 1708}%
\special{pa 5710 1686}%
\special{pa 5734 1666}%
\special{pa 5758 1646}%
\special{pa 5784 1626}%
\special{pa 5810 1606}%
\special{pa 5836 1588}%
\special{pa 5862 1570}%
\special{pa 5890 1554}%
\special{pa 5918 1538}%
\special{pa 5946 1524}%
\special{pa 5976 1510}%
\special{pa 6006 1500}%
\special{pa 6036 1488}%
\special{pa 6066 1480}%
\special{pa 6098 1472}%
\special{pa 6128 1464}%
\special{pa 6160 1458}%
\special{pa 6192 1452}%
\special{pa 6224 1448}%
\special{pa 6256 1446}%
\special{pa 6288 1444}%
\special{pa 6320 1442}%
\special{pa 6340 1440}%
\special{sp}%
% VECTOR 2 0 3 0
% 2 3960 1650 3870 1610
% 
\special{pn 8}%
\special{pa 3960 1650}%
\special{pa 3870 1610}%
\special{fp}%
\special{sh 1}%
\special{pa 3870 1610}%
\special{pa 3924 1656}%
\special{pa 3920 1632}%
\special{pa 3940 1620}%
\special{pa 3870 1610}%
\special{fp}%
% VECTOR 2 0 3 0
% 2 3250 2240 3340 2260
% 
\special{pn 8}%
\special{pa 3250 2240}%
\special{pa 3340 2260}%
\special{fp}%
\special{sh 1}%
\special{pa 3340 2260}%
\special{pa 3280 2226}%
\special{pa 3288 2248}%
\special{pa 3272 2266}%
\special{pa 3340 2260}%
\special{fp}%
% VECTOR 2 0 3 0
% 2 6320 1460 6440 1440
% 
\special{pn 8}%
\special{pa 6320 1460}%
\special{pa 6440 1440}%
\special{fp}%
\special{sh 1}%
\special{pa 6440 1440}%
\special{pa 6372 1432}%
\special{pa 6388 1450}%
\special{pa 6378 1472}%
\special{pa 6440 1440}%
\special{fp}%
\end{picture}%
\end{center}
%%%%%%%%%%%%%%%%%%%%%%%%%%%%%%%%%
\vskip5mm

\noindent
It follows from the calculation of (I) that, 
for ${\rm Res}_{\alpha_{a}=\beta_{n-1}-\pi i}$,
the pinches of the integration contour in the limit 
$\beta_n\rightarrow \beta_{n-1}+\pi i$ do not occur.

Notice that there are no poles at $\alpha_a=\beta_{n-1}$, $a\neq \ell$
in the integrand of $I_M$.
For ${\rm Res}_{\alpha_{\ell}=\beta_{n-1}}$ it is not difficult to check
that the pinches of the integration contour in the limit 
$\beta_n\rightarrow \beta_{n-1}+\pi i$ do not occur.

Let us consider ${\rm Res}_{\alpha_{a}=\beta_{n-1}+\pi i}$.
The integrand of $I_M$ does not have poles at $\alpha_a=\beta_n$, 
$a\neq \ell$.
Thus it is sufficient to consider the case $a=\ell$.
Since
\bea
&&
\prod_{1 \le a<b \le \ell}(A_a-A_b)\vert_{\alpha_\ell=\beta_{n-1}+\pi i}
=
\prod_{1 \le a<b \le \ell -1}(A_a-A_b)\prod_{b=1}^{\ell-1}(A_b+B_{n-1})
\no
\ena
has zeroes at $\alpha_b=\beta_{n-1}\pm \pi i$, $(\forall b)$,
the pinches of the integration contour do not occur at 
$\alpha_b=\beta_{n-1}\pm \pi i$.
Moreover the integrand of ${\rm Res}_{\alpha_{\ell}=\beta_{n-1}+\pi i}$
does not have poles at $\alpha_b=\beta_{n-1}$, $\forall b$.
Thus 
for ${\rm Res}_{\alpha_{\ell}=\beta_{n-1}+\pi i}$ the pinches of 
the integration contour in the limit 
$\beta_n\rightarrow \beta_{n-1}+\pi i$ do not occur.
In $g_M\vert_{\alpha_{\ell}=\beta_{n-1}+\pi i}$ there is a simple pole at
$\beta_n=\beta_{n-1}+\pi i$. Other factors of the integrand of $I_M$
do not produce poles there. 
Thus as a whole $I_M$ has a simple pole at $\beta_n=\beta_{n-1}+\pi i$.
\vskip3mm

\noindent
(IV). $M=M'\cup \{n-1,n\}$, $M' \subset \{1, \cdots , n-2 \}$.
\par
\noindent
Again we decompose the integral as in (\ref{decomp}).
It follows from the calculation of (I) that there are no poles at
$\alpha_b=\beta_{n-1}\pm \pi i, \beta_{n-1}$, $\forall b$ in the
integrand of ${\rm Res}_{\alpha_{a}=\beta_{n-1}-\pi i}$.

Let us consider ${\rm Res}_{\alpha_{a}=\beta_{n-1}+\pi i}$.
If $a\neq \ell$, there are no poles at $\alpha_a=\beta_n$
in the integrand of $I_M$.
Therefore it is sufficient to consider 
${\rm Res}_{\alpha_{\ell}=\beta_{n-1}+\pi i}$.
In a similar manner to (III) we find that there are no poles at
$\alpha_b=\beta_{n-1}\pm \pi i, \beta_{n-1}$, $\forall b$ in the
integrand of ${\rm Res}_{\alpha_{\ell}=\beta_{n-1}+\pi i}$.

Next consider ${\rm Res}_{\alpha_{a}=\beta_{n-1}}$.
If $a\neq \ell-1,\ell$, then ${\rm Res}_{\alpha_{a}=\beta_{n-1}}=0$.
Consider the case $a=\ell$.
There are no poles at 
$\alpha_b=\beta_n,\beta_{n-1},\beta_{n-1}-\pi i$ 
$(\forall b)$ in the integrand of ${\rm Res}_{\alpha_{\ell}=\beta_{n-1}}$.
Thus the pinches of the integration contour do not occur.

Let us consider ${\rm Res}_{\alpha_{\ell-1}=\beta_{n-1}}$.
There are no poles at $\alpha_b=\beta_n$ $(b\neq \ell)$,
$\alpha_\ell=\beta_{n-1}+\pi i$, and 
$\alpha_b=\beta_{n-1},\beta_{n-1}-\pi i$ $(\forall b)$.
Thus the pinches of the integration contour do not occur.

Thus the proposition is proved.
QED.
\newline

By Proposition \ref{simple}, 
in the calculation of the residue of $\Psi_{P_{n}}$ at 
$\beta_{n}=\beta_{n-1}+\pi i$, 
we can replace $P_{n}$ by $P_{n}|_{\beta_{n}=\beta_{n-1}+\pi i}$. 
Then we apply the assumption (\ref{cycletoshow1}) and
consider the decomposition
\bea
\prod_{a=1}^{\ell-1}(1-A_{a}B_{n-1}^{-2})
\overline{P}_{n}(A_{1}, \cdots , A_{\ell}|B_{1}, \cdots , B_{n-2}|B_{n-1})=
P_{n}^{+}+P_{n}^{-},
\ena
where
\bea
P_{n}^{\pm}=P_{n}^{\pm}(A_{1}, \cdots , A_{\ell}|B_{1}, \cdots , B_{n-2}|B_{n-1})
:=\frac{1\pm A_{\ell}B_{n-1}^{-1}}{2}
\prod_{a=1}^{\ell}(1-A_{a}B_{n-1}^{-2})
\overline{P}_{n}.
\ena
Then we have
\bea
{\rm Res}_{\beta_{n}=\beta_{n-1}+\pi i}\Psi_{P_{n}}=
{\rm Res}_{\beta_{n}=\beta_{n-1}+\pi i}\Psi_{P_{n}^{+}}+
{\rm Res}_{\beta_{n}=\beta_{n-1}+\pi i}\Psi_{P_{n}^{-}}.
\ena

\begin{prop} We have
\bea
&&
{\rm Res}_{\beta_n=\beta_{n-1}+\pi i}\Psi_{P_{n}^{+}}=
U_n\Psi_{\widehat{P}_{n}^{+}}\otimes \mathbf{e}_0,
\label{Res+}
\\
&&
{\rm Res}_{\beta_n=\beta_{n-1}+\pi i}\Psi_{P_{n}^{-}}=
(-1)^{\frac{n}{2}-1}
U_n
S_{n-1, n-2}(\beta_{n-1}-\beta_{n-2}) \cdots S_{n-1, 
1}(\beta_{n-1}-\beta_{1})
\Psi_{\widehat{P}_{n}^{-}}\otimes \mathbf{e}_0,
\label{Res-}
\ena
where
\bea
&&
\widehat{P}_{n}^{\pm}(A_1,\cdots,A_{\ell-1}\vert B_1,\cdots,B_{n-1})
:=
\overline{P}_{n}(A_{1}, \cdots , A_{\ell-1}, \pm B_{n-1}|B_{1}, \cdots , B_{n-2}| B_{n-1}),
\\
&&
U_n=U_n(\beta_1,\cdots,\beta_{n-1}):=
(-1)^{\ell-1}(2\pi i)^{\ell }(-2\pi)^{\frac{n}{2}-1}
e^{\frac{1}{2}\sum_{j=1}^{n-2}(\beta_{n-1}-\beta_{j})} \no \\
&&
\qquad \qquad \qquad \qquad \qquad
{}\times
\prod_{j=1}^{n-2}\left\{
\Gamma_{1}(-i(\beta_{j}-\beta_{n-1})+\pi)
\Gamma_{1}(i(\beta_{j}-\beta_{n-1})) \right\},
\\
&&
\mathbf{e}_0=v_{+}\otimes v_{-}-v_{-}\otimes v_{+}.
\ena
\end{prop}

\pf
First, we calculate ${\rm Res}\Psi_{P_{n}^{+}}$. 
We expand the coefficient $I_{M}$ as follows.
\bea
I_{M}=\sum_{\sigma \in S_{\ell}} ({\rm sgn}\sigma)
\int_{C^{\ell}}\prod_{a=1}^{\ell}d\alpha_{a} \prod_{a=1}^{\ell} \varphi(\alpha_{a}) 
g_{M}(\alpha_{\sigma(1)}, \cdots , \alpha_{\sigma(\ell)})
P_{n}^{+}(e^{-\alpha_{1}}, \cdots , e^{-\alpha_{\ell}}), 
\label{exp}
\ena
where
\bea
P_{n}^{+}(e^{-\alpha_{1}}, \cdots , e^{-\alpha_{\ell}})=
\frac{1+e^{-(\alpha_{\ell}-\beta_{n-1})}}{2}
\prod_{a=1}^{\ell-1}(1-e^{-2(\alpha_{a}-\beta_{n-1})})
\overline{P}_{n}(e^{-\alpha_{1}}, \cdots , e^{-\alpha_{\ell}}).
\ena

Let us consider the poles of the integrand which may pinch the contour $C$.
Because $P_{n}^{+}$ has zeroes at 
$\alpha_{a}=\beta_{n-1} \pm \pi i, (a=1, \cdots , \ell)$, 
the contour $C$ may be pinched by poles at $\alpha_{a}=\beta_{n-1}$ and
$\alpha_{a}=\beta_{n}-\pi i$ (the contour $C$ can be deformed to the contour
in Figure 6).
\vskip5mm
%%%%%%%%%%%%%%%%%%%%%%%%%%%%%%%%%%%%%%
%WinTpicVersion3.08
\unitlength 0.1in
\begin{center}
\begin{picture}( 32.6000, 18.3000)( 16.1000,-27.1000)
% DOT 2 0 3 0
% 2 2390 1000 2390 990
% 
\special{pn 8}%
\special{sh 1}%
\special{ar 2390 1000 10 10 0  6.28318530717959E+0000}%
\special{sh 1}%
\special{ar 2390 990 10 10 0  6.28318530717959E+0000}%
% DOT 2 0 3 0
% 2 2400 1410 2400 1400
% 
\special{pn 8}%
\special{sh 1}%
\special{ar 2400 1410 10 10 0  6.28318530717959E+0000}%
\special{sh 1}%
\special{ar 2400 1400 10 10 0  6.28318530717959E+0000}%
% DOT 2 0 3 0
% 2 2390 1820 2390 1810
% 
\special{pn 8}%
\special{sh 1}%
\special{ar 2390 1820 10 10 0  6.28318530717959E+0000}%
\special{sh 1}%
\special{ar 2390 1810 10 10 0  6.28318530717959E+0000}%
% DOT 2 0 3 0
% 2 2410 2210 2410 2210
% 
\special{pn 8}%
\special{sh 1}%
\special{ar 2410 2210 10 10 0  6.28318530717959E+0000}%
\special{sh 1}%
\special{ar 2410 2210 10 10 0  6.28318530717959E+0000}%
% DOT 2 0 3 0
% 2 2410 2600 2410 2600
% 
\special{pn 8}%
\special{sh 1}%
\special{ar 2410 2600 10 10 0  6.28318530717959E+0000}%
\special{sh 1}%
\special{ar 2410 2600 10 10 0  6.28318530717959E+0000}%
% DOT 2 0 3 0
% 2 3580 990 3580 990
% 
\special{pn 8}%
\special{sh 1}%
\special{ar 3580 990 10 10 0  6.28318530717959E+0000}%
\special{sh 1}%
\special{ar 3580 990 10 10 0  6.28318530717959E+0000}%
% DOT 2 0 3 0
% 2 3600 1410 3600 1400
% 
\special{pn 8}%
\special{sh 1}%
\special{ar 3600 1410 10 10 0  6.28318530717959E+0000}%
\special{sh 1}%
\special{ar 3600 1400 10 10 0  6.28318530717959E+0000}%
% DOT 2 0 3 0
% 2 3610 1820 3610 1820
% 
\special{pn 8}%
\special{sh 1}%
\special{ar 3610 1820 10 10 0  6.28318530717959E+0000}%
\special{sh 1}%
\special{ar 3610 1820 10 10 0  6.28318530717959E+0000}%
% DOT 2 0 3 0
% 2 3620 2210 3620 2210
% 
\special{pn 8}%
\special{sh 1}%
\special{ar 3620 2210 10 10 0  6.28318530717959E+0000}%
\special{sh 1}%
\special{ar 3620 2210 10 10 0  6.28318530717959E+0000}%
% DOT 2 0 3 0
% 2 3620 2600 3620 2600
% 
\special{pn 8}%
\special{sh 1}%
\special{ar 3620 2600 10 10 0  6.28318530717959E+0000}%
\special{sh 1}%
\special{ar 3620 2600 10 10 0  6.28318530717959E+0000}%
% STR 2 0 3 0
% 3 2440 1790 2440 1890 2 0
% $\beta_{n-1}$
\put(24.4000,-18.9000){\makebox(0,0)[lb]{$\beta_{n-1}$}}%
% STR 2 0 3 0
% 3 3670 1780 3670 1880 2 0
% $\beta_n$
\put(36.7000,-18.8000){\makebox(0,0)[lb]{$\beta_n$}}%
% STR 2 0 3 0
% 3 2430 1350 2430 1450 2 0
% $\beta_{n-1}+\pi i$
\put(24.3000,-14.5000){\makebox(0,0)[lb]{$\beta_{n-1}+\pi i$}}%
% STR 2 0 3 0
% 3 2440 950 2440 1050 2 0
% $\beta_{n-1}+2\pi i$
\put(24.4000,-10.5000){\makebox(0,0)[lb]{$\beta_{n-1}+2\pi i$}}%
% STR 2 0 3 0
% 3 3670 1380 3670 1480 2 0
% $\beta_n+\pi i$
\put(36.7000,-14.8000){\makebox(0,0)[lb]{$\beta_n+\pi i$}}%
% STR 2 0 3 0
% 3 3660 960 3660 1060 2 0
% $\beta_n+2\pi i$
\put(36.6000,-10.6000){\makebox(0,0)[lb]{$\beta_n+2\pi i$}}%
% STR 2 0 3 0
% 3 2450 2170 2450 2270 2 0
% $\beta_{n-1}-\pi i$
\put(24.5000,-22.7000){\makebox(0,0)[lb]{$\beta_{n-1}-\pi i$}}%
% STR 2 0 3 0
% 3 2450 2550 2450 2650 2 0
% $\beta_{n-1}-2\pi i$
\put(24.5000,-26.5000){\makebox(0,0)[lb]{$\beta_{n-1}-2\pi i$}}%
% STR 2 0 3 0
% 3 3660 2170 3660 2270 2 0
% $\beta_n-\pi i$
\put(36.6000,-22.7000){\makebox(0,0)[lb]{$\beta_n-\pi i$}}%
% STR 2 0 3 0
% 3 3650 2550 3650 2650 2 0
% $\beta_n-2\pi i$
\put(36.5000,-26.5000){\makebox(0,0)[lb]{$\beta_n-2\pi i$}}%
% VECTOR 2 0 3 0
% 2 1610 1600 2250 1590
% 
\special{pn 8}%
\special{pa 1610 1600}%
\special{pa 2250 1590}%
\special{fp}%
\special{sh 1}%
\special{pa 2250 1590}%
\special{pa 2184 1572}%
\special{pa 2198 1592}%
\special{pa 2184 1612}%
\special{pa 2250 1590}%
\special{fp}%
% SPLINE 2 0 3 0
% 8 2250 1590 2250 1500 2280 1310 2380 1270 2490 1280 2570 1370 2610 1450 2650 1570
% 
\special{pn 8}%
\special{pa 2250 1590}%
\special{pa 2250 1560}%
\special{pa 2252 1528}%
\special{pa 2250 1494}%
\special{pa 2248 1460}%
\special{pa 2248 1424}%
\special{pa 2250 1390}%
\special{pa 2256 1358}%
\special{pa 2266 1330}%
\special{pa 2284 1306}%
\special{pa 2310 1290}%
\special{pa 2340 1278}%
\special{pa 2374 1272}%
\special{pa 2408 1268}%
\special{pa 2442 1268}%
\special{pa 2472 1274}%
\special{pa 2500 1286}%
\special{pa 2524 1306}%
\special{pa 2546 1332}%
\special{pa 2564 1358}%
\special{pa 2580 1388}%
\special{pa 2594 1416}%
\special{pa 2608 1444}%
\special{pa 2620 1474}%
\special{pa 2630 1504}%
\special{pa 2640 1534}%
\special{pa 2650 1566}%
\special{pa 2650 1570}%
\special{sp}%
% VECTOR 2 0 3 0
% 2 2650 1570 3750 1570
% 
\special{pn 8}%
\special{pa 2650 1570}%
\special{pa 3750 1570}%
\special{fp}%
\special{sh 1}%
\special{pa 3750 1570}%
\special{pa 3684 1550}%
\special{pa 3698 1570}%
\special{pa 3684 1590}%
\special{pa 3750 1570}%
\special{fp}%
% VECTOR 2 0 3 0
% 2 4560 1650 4870 1650
% 
\special{pn 8}%
\special{pa 4560 1650}%
\special{pa 4870 1650}%
\special{fp}%
\special{sh 1}%
\special{pa 4870 1650}%
\special{pa 4804 1630}%
\special{pa 4818 1650}%
\special{pa 4804 1670}%
\special{pa 4870 1650}%
\special{fp}%
% LINE 2 2 3 0
% 2 2410 1410 2410 1410
% 
\special{pn 8}%
\special{pa 2410 1410}%
\special{pa 2410 1410}%
\special{dt 0.045}%
% LINE 2 2 3 0
% 2 2400 1820 3600 2210
% 
\special{pn 8}%
\special{pa 2400 1820}%
\special{pa 3600 2210}%
\special{dt 0.045}%
% STR 2 0 3 0
% 3 3140 2780 3140 2880 2 0
% {\bf Figure 6}
\put(31.4000,-28.8000){\makebox(0,0)[lb]{{\bf Figure 6}}}%
% SPLINE 2 0 3 0
% 15 3770 1570 3930 1580 4070 1700 4100 1880 3890 1980 3680 1980 3410 1990 3320 2150 3350 2310 3660 2410 3920 2320 4180 2120 4360 1870 4510 1650 4570 1640
% 
\special{pn 8}%
\special{pa 3770 1570}%
\special{pa 3804 1568}%
\special{pa 3836 1568}%
\special{pa 3868 1570}%
\special{pa 3900 1574}%
\special{pa 3930 1580}%
\special{pa 3960 1592}%
\special{pa 3986 1608}%
\special{pa 4012 1628}%
\special{pa 4034 1652}%
\special{pa 4056 1678}%
\special{pa 4074 1708}%
\special{pa 4090 1740}%
\special{pa 4102 1772}%
\special{pa 4108 1806}%
\special{pa 4110 1838}%
\special{pa 4106 1868}%
\special{pa 4094 1894}%
\special{pa 4074 1916}%
\special{pa 4048 1934}%
\special{pa 4020 1948}%
\special{pa 3986 1960}%
\special{pa 3952 1970}%
\special{pa 3916 1976}%
\special{pa 3882 1982}%
\special{pa 3848 1986}%
\special{pa 3816 1988}%
\special{pa 3786 1988}%
\special{pa 3754 1986}%
\special{pa 3722 1984}%
\special{pa 3690 1982}%
\special{pa 3658 1978}%
\special{pa 3624 1972}%
\special{pa 3590 1968}%
\special{pa 3556 1964}%
\special{pa 3522 1964}%
\special{pa 3490 1966}%
\special{pa 3458 1970}%
\special{pa 3430 1980}%
\special{pa 3404 1994}%
\special{pa 3382 2016}%
\special{pa 3362 2040}%
\special{pa 3346 2070}%
\special{pa 3332 2100}%
\special{pa 3324 2134}%
\special{pa 3318 2168}%
\special{pa 3318 2202}%
\special{pa 3322 2234}%
\special{pa 3330 2266}%
\special{pa 3342 2296}%
\special{pa 3358 2322}%
\special{pa 3380 2344}%
\special{pa 3404 2362}%
\special{pa 3432 2378}%
\special{pa 3462 2390}%
\special{pa 3496 2400}%
\special{pa 3530 2408}%
\special{pa 3564 2412}%
\special{pa 3600 2412}%
\special{pa 3634 2412}%
\special{pa 3668 2410}%
\special{pa 3702 2404}%
\special{pa 3734 2398}%
\special{pa 3766 2390}%
\special{pa 3796 2380}%
\special{pa 3826 2368}%
\special{pa 3854 2356}%
\special{pa 3882 2342}%
\special{pa 3910 2326}%
\special{pa 3938 2310}%
\special{pa 3966 2294}%
\special{pa 3992 2276}%
\special{pa 4018 2258}%
\special{pa 4044 2238}%
\special{pa 4070 2218}%
\special{pa 4096 2198}%
\special{pa 4120 2176}%
\special{pa 4144 2156}%
\special{pa 4166 2134}%
\special{pa 4190 2112}%
\special{pa 4212 2088}%
\special{pa 4232 2066}%
\special{pa 4252 2042}%
\special{pa 4272 2018}%
\special{pa 4290 1992}%
\special{pa 4308 1966}%
\special{pa 4326 1938}%
\special{pa 4342 1908}%
\special{pa 4358 1878}%
\special{pa 4372 1846}%
\special{pa 4386 1812}%
\special{pa 4400 1780}%
\special{pa 4414 1750}%
\special{pa 4430 1720}%
\special{pa 4450 1694}%
\special{pa 4470 1672}%
\special{pa 4496 1656}%
\special{pa 4524 1646}%
\special{pa 4556 1642}%
\special{pa 4570 1640}%
\special{sp}%
\end{picture}%
\end{center}
%%%%%%%%%%%%%%%%%%%%%%%%%%%%%%%%%%%%%%%
\vskip5mm

\noindent
Moreover, since $P_n^{+}$ has also zeroes at 
$\alpha_{a}=\beta_{n-1}, (a=1, \cdots , \ell-1)$, 
only the contour for $\alpha_{\ell}$ may be pinched.
Note that the singularity at $\alpha_{\ell}=\beta_{n-1}$ comes from $g_{M}$.
Hence, the pinch does not occur if $M \subset \{ 1, \cdots , n-2 \}$, 
and it suffices to consider the case of $M \cap \{ n-1, n\} \not= \phi$.
\newline

\noindent
(I). \quad $M=M'\cup \{ n-1 \}, (M' \subset \{ 1, \cdots , n-2 \})$ case

In the expansion (\ref{exp}), 
the integrand has a pole at $\alpha_{\ell}=\beta_{n-1}$ only when $\sigma(\ell)=\ell$.
For such terms, we deform the contour $C$ for $\alpha_{\ell}$
by taking the residue at $\alpha_{\ell}=\beta_{n-1}$, that is,
\bea
\int_{C}d\alpha_{\ell} ={\rm (regular \, term)}+(-2\pi i){\rm Res}_{\alpha_{\ell}=\beta_{n-1}}.
\ena
The residue above is given by
\bea
{\rm Res}_{\alpha_{\ell}=\beta_{n-1}}&=&
\int_{C^{\ell-1}}\prod_{a=1}^{\ell -1}d\alpha_a 
\prod_{a=1}^{\ell -1}\varphi(\alpha_{a}) 
\no 
\\
&\times&
\varphi'(\beta_{n-1}) \Gamma\left(-\frac{1}{2}\right)
e^{\frac{\beta_{n-1}-\beta_{n}}{2}} 
\Gamma\left( \frac{\beta_{n-1}-\beta_{n}+2\pi i}{2\pi i} \right)
\Gamma\left( \frac{\beta_{n}-\beta_{n-1}+\pi i}{-2\pi i} \right)
\no 
\\
&\times&
g_{M'}(\alpha_{\sigma(1)}, \cdots , \alpha_{\sigma(\ell-1)})
\prod_{j=1}^{n-2}\frac{\beta_{n-1}-\beta_{j}+\pi i}{\beta_{n-1}-\beta_{j}}
\prod_{a=1}^{\ell -1}(\alpha_{a}-\beta_{n-1}+\pi i) 
\no 
\\
&\times&
\prod_{p=1}^{\ell -1}(1-e^{-2(\alpha_{p}-\beta_{n-1})})
\overline{P}_{n}(e^{-\alpha}, e^{-\beta_{n-1}}).
\label{res1}
\ena
Here $\varphi'(\beta)$ is given by (\ref{varphi'}).

Using
\bea
&&
{\rm Res}_{\beta_{n}=\beta_{n-1}+\pi i}
\Gamma \left( \frac{\beta_{n}-\beta_{n-1}+\pi i}{-2\pi i} \right) d\beta_{n}
=2\pi i, \no \\
&&
\Gamma(-\frac{1}{2})\Gamma(\frac{1}{2})e^{-\frac{\pi i}{2}}=2\pi i, \no \\
&& 
\varphi(\alpha)|_{\beta_{n}=\beta_{n-1}+\pi i}=
\varphi'(\alpha) \times \frac{-(2\pi i)^{3}}
{(\alpha-\beta_{n-1}+\pi i)(1-e^{-2(\alpha -\beta_{n-1})})},
\ena
we get
\bea
{\rm Res}_{\beta_{n}=\beta_{n-1}+\pi i}I_{M}&=&
(-1)^{\ell}(2\pi i)^{3\ell}\varphi'(\beta_{n-1}) 
\prod_{j=1}^{n-2}\frac{\beta_{n-1}-\beta_{j}+\pi i}{\beta_{n-1}-\beta_{j}} \no \\
&\times&
\int_{C^{\ell-1}}\prod_{a=1}^{\ell -1}d\alpha_a 
\prod_{a=1}^{\ell -1}\varphi'(\alpha_{a}) w_{M'}(\alpha) 
\overline{P}_{n}(e^{-\alpha}, e^{-\beta_{n-1}}).
\label{res(I)}
\ena

\noindent
(II). $M=M'\cup\{ n \}, (M' \subset \{1, \cdots , n-2\})$ case

In a similar manner to (I), 
it sufficies to consider the terms of $\sigma(\ell)=\ell$.
The pole of the integrand at $\alpha_{\ell}=\beta_{n-1}$ stems from
$g_M$ and 
the residue of $g_M$ at $\alpha_{\ell}=\beta_{n-1}$ is given by
\bea
{\rm Res}_{\alpha_{\ell}=\beta_{n-1}}\, g_M
&=&
g_{M'}(\alpha_{\sigma(1)}, \cdots , \alpha_{\sigma(\ell-1)})
\frac{\pi i}{\beta_{n-1}-\beta_{n}}
\prod_{j=1}^{n-2}\frac{\beta_{n-1}-\beta_{j}+\pi i}{\beta_{n-1}-\beta_{j}}
\prod_{p=1}^{\ell -1}(\alpha_{p}-\beta_{n-1}+\pi i).
\no 
\ena
This differs from the case (I) by the term
$\frac{\pi i}{\beta_{n-1}-\beta_{n}}$ which becomes $-1$ 
if $\beta_{n} \to \beta_{n-1}+\pi i$.
Other calculations are the same as in the case (I).
Therefore the residue is given by minus of (\ref{res(I)}):
\bea
{\rm Res}_{\beta_{n}=\beta_{n-1}+\pi i}I_{M}&=&
(-1)^{\ell+1}(2\pi i)^{3\ell}\varphi'(\beta_{n-1}) 
\prod_{j=1}^{n-2}\frac{\beta_{n-1}-\beta_{j}+\pi i}{\beta_{n-1}-\beta_{j}} \no \\
&\times&
\int_{C}d\alpha \prod_{p=1}^{\ell -1}\varphi'(\alpha_{p}) w_{M'}(\alpha) 
\overline{P}_{n}(e^{-\alpha}, e^{-\beta_{n-1}}).
\label{res(II)}
\ena

\noindent
(III). $M=M'\cup\{ n-1, n\}, (M' \subset \{1, \cdots , n-2\})$ case

The function $g_{M}(\alpha_{\sigma(1)}, \cdots , \alpha_{\sigma(\ell)})$ has a pole
at $\alpha_{\ell}=\beta_{n-1}$ only if 
$\sigma(\ell -1)=\ell$ or $\sigma(\ell)=\ell$.

Let us consider the case $\sigma(\ell -1)=\ell$ first.
In the calculation of the residue at $\alpha_{\ell}=\beta_{n-1}$,
only the residue of $g_M$ at $\alpha_{\ell}=\beta_{n-1}$ can differ from
the calculation in (I).
We have
\bea
{\rm Res}_{\alpha_{\ell}=\beta_{n-1}}\, g_M
&=&
g_{M'}(\alpha_{\tau(1)}, \cdots , \alpha_{\tau(\ell-2)})
\prod_{j=1}^{n-2}\frac{\beta_{n-1}-\beta_{j}+\pi i}{\beta_{n-1}-\beta_{j}} 
\frac{1}{\alpha_{\tau(\ell-1)}-\beta_{n}}
\prod_{j=1}^{n-1}\frac{\alpha_{\tau(\ell-1)}-\beta_{j}+\pi i}{\alpha_{\tau(\ell-1)}-\beta_{j}}
\no \\
&\times&
\prod_{p=1}^{\ell -2}\left\{
(\alpha_{\tau(p)}-\beta_{n-1}+\pi i)(\alpha_{\tau(p)}-\alpha_{\tau(\ell-1)}+\pi i)\right\}
(\beta_{n-1}-\alpha_{\tau(\ell-1)}+\pi i)  
\no
\ena
where $\tau :=\sigma \cdot (\ell-1, \ell)$.
Then
\bea
&&
\sum_{\sigma\in S_\ell, \sigma(\ell-1)=\ell}
(\text{sgn}\,\sigma)
{\rm Res}_{\alpha_{\ell}=\beta_{n-1}}\, g_M
\vert_{\beta_n=\beta_{n-1}+\pi i}
\no
\\
&&
\qquad\quad
=
\prod_{j=1}^{n-2}\frac{\beta_{n-1}-\beta_{j}+\pi i}{\beta_{n-1}-\beta_{j}} 
\sum_{\tau\in S_{\ell-1}}
\text{sgn}\,\tau\,\,
g_{M'}(\alpha_{\tau(1)}, \cdots , \alpha_{\tau(\ell-2)})
\prod_{j=1}^{n-1}\frac{\alpha_{\tau(\ell-1)}-\beta_{j}+\pi i}{\alpha_{\tau(\ell-1)}-\beta_{j}}
\no 
\\
&&
\qquad\quad
\times
\prod_{p=1}^{\ell -2}\left\{
(\alpha_{\tau(p)}-\beta_{n-1}+\pi i)(\alpha_{\tau(p)}-\alpha_{\tau(\ell-1)}+\pi i)\right\}.
\no
\ena

For the case $\sigma(\ell)=\ell$, we find
\bea
&&
{\rm Res}_{\alpha_{\ell}=\beta_{n-1}}\, g_M
\no
\\
&&
=
g_{M'}(\alpha_{\sigma(1)}, \cdots , \alpha_{\sigma(\ell-2)})
\frac{\pi i}{\beta_{n-1}-\beta_{n}}
\prod_{j=1}^{n-2}\frac{\beta_{n-1}-\beta_{j}+\pi i}{\beta_{n-1}-\beta_{j}} 
\no \\
&&
\times
\frac{1}{\alpha_{\sigma(\ell-1)}-\beta_{n-1}}
\prod_{j=1}^{n-2}\frac{\alpha_{\sigma(\ell-1)}-\beta_{j}+\pi i}{\alpha_{\sigma(\ell-1)}-\beta_{j}}
\prod_{p=1}^{\ell -1}(\alpha_{\sigma(p)}-\beta_{n-1}+\pi i)
\prod_{p=1}^{\ell -2}(\alpha_{\sigma(p)}-\alpha_{\sigma(\ell-1)}+\pi i).
\label{resgmIII}
\ena
Then
\bea
&&
\sum_{\sigma\in S_\ell, \sigma(\ell)=\ell}
(\text{sgn}\,\sigma)
{\rm Res}_{\alpha_{\ell}=\beta_{n-1}}\, g_M
\vert_{\beta_n=\beta_{n-1}+\pi i}
=-(\ref{resgmIII}).
\ena
Thus
\bea
&&
{\rm Res}_{\beta_n=\beta_{n-1}+\pi i}\, I_M=0.
\no
\ena

By (\ref{res(I)}) and (\ref{res(II)}) we get
\bea
&&
{\rm Res}_{\beta_{n}=\beta_{n-1}+\pi i}
\Psi_{P_{n}^{+}}=(2\pi i)^{-n\ell}(-1)^{\ell-1}(2\pi i)^{3\ell}
\varphi'(\beta_{n-1})
\prod_{j=1}^{n-2}\frac{\beta_{n-1}-\beta_{j}+\pi i}{\beta_{n-1}-\beta_{j}} \no \\
&& \quad {}\times \left(
\sum_{\# M'=\ell -1}v_{M'}\int_{C^{\ell-1}}
\prod_{a=1}^{\ell-1}d\alpha_{a} \prod_{p=1}^{\ell -1}\varphi'(\alpha_{p}) 
w_{M'}(\alpha) \overline{P}_{n}(e^{-\alpha}, e^{-\beta_{n-1}}) \right)
\otimes
(v_{+}\otimes v_{-}-v_{-}\otimes v_{+}) \no \\
&& {}=
(2 \pi i)^{-n+\ell+2}(-1)^{\ell-1}
\varphi'(\beta_{n-1})
\prod_{j=1}^{n-2}\frac{\beta_{n-1}-\beta_{j}+\pi i}{\beta_{n-1}-\beta_{j}}
\Psi_{\widehat{P}_{n}^{+}}\otimes\mathbf{e}_0 
\label{residue1}
\ena

Using (\ref{gamma1}), it is not difficult to show
\bea
(2 \pi i)^{-n+\ell+2}(-1)^{\ell-1}
\varphi'(\beta_{n-1})
\prod_{j=1}^{n-2}\frac{\beta_{n-1}-\beta_{j}+\pi i}{\beta_{n-1}-\beta_{j}}
=U_{n}.
\ena
This completes the proof of (\ref{Res+}).
\newline

Next we prove (\ref{Res-}).
Note that $\Psi_{P_{n}^{-}}$ satisfies (\ref{eqI'}) and (\ref{eqII'}).
Hence we have
\bea
\Psi_{P_{n}^{-}}(\beta_{1}, \cdots , \beta_{n})
&=&
\widehat{S}_{n-1,n-2}(\beta_{n-1}-\beta_{n-2}) \cdots 
\widehat{S}_{n-1,1}(\beta_{n-1}-\beta_{1}) \no \\
&\times&
P_{n-1,n}
\Psi_{P_{n}^{-}}(\beta_{1}, \cdots , \beta_{n-2}, \beta_{n}, \beta_{n-1}+2\pi i).
\label{Ss}
\ena
Here we should note that the cycle $P_{n}^{-}$ of 
$\Psi_{P_{n}^{-}}$ 
in the rhs of (\ref{Ss}) is given by
\bea
P_{n}^{-}=P_{n}^{-}(A_{1}, \cdots , A_{\ell}|B_{1}, \cdots , B_{n-2}|B_{n-1}),
\ena
which means that we do not change the order of $\beta$'s in $P_{n}^{-}$.

{}From (\ref{Ss}), it suffices to calculate 
\bea
{\rm Res}_{\beta_{n}=\beta_{n-1}+\pi i}\Psi_{P_{n}^{-}}
(\beta_{1}, \cdots , \beta_{n-2}, \beta_{n}, \beta_{n-1}+2\pi i).
\ena
The calculation of this residue is quite similar to that of 
${\rm Res}\Psi_{P_{n}^{+}}$.
The result is 
\bea
&&
P_{n-1, n}{\rm Res}_{\beta_{n}=\beta_{n-1}+\pi i}
\Psi_{P_{n}^{-}}(\beta_{1}, \cdots , \beta_{n-2}, \beta_{n}, \beta_{n-1}+2\pi i) \no \\
&&{}=
(2\pi i)^{-n+\ell+2}(-1)^{\ell-1}\varphi'(\beta_{n-1}+\pi i)
\prod_{j=1}^{n-2}
\frac{\beta_{n-1}-\beta_{j}+2\pi i}{\beta_{n-1}-\beta_{j}+\pi i}
\Psi_{\widehat{P}_{n}^{-}}\otimes \mathbf{e}_{0}.
\ena
Moreover it can be shown that
\bea
&&
(2\pi i)^{-n+\ell+2}(-1)^{\ell-1}\varphi'(\beta_{n-1}+\pi i)
\prod_{j=1}^{n-2}
\frac{\beta_{n-1}-\beta_{j}+2\pi i}{\beta_{n-1}-\beta_{j}+\pi i} \no \\
&& {}=
(-1)^{\frac{n}{2}-1}U_{n}
\prod_{j=1}^{n-2}S_{0}(\beta_{n-1}-\beta_{j}).
\ena
This completes the proof of (\ref{Res-}).
\newline

Thus if we define $f_P$ by
\bea
&&
f_{P}=
e^{\frac{n}{4}\sum_{j=1}^{n}\beta_{j}}
\prod_{1 \le j<j' \le n}\zeta(\beta_{j}-\beta_{j'})\Psi_{P},
\no
\ena
then
\bea
&&
{\rm Res}_{\beta_n=\beta_{n-1}+\pi i}f_{P_n}
=(-2\pi i)^{\ell+\frac{n}{2}}\frac{\zeta(-\pi i)}{2\pi}
B_{n-1}^{-(n-1)}  \no \\
&& 
{}\times
\left\{
f_{\widehat{P}_{n}^{+}}\otimes \mathbf{e}_0+(-1)^{\frac{n}{2}-1}
S_{n-1, n-2}(\beta_{n-1}-\beta_{n-2}) \cdots S_{n-1, 
1}(\beta_{n-1}-\beta_{1})
f_{\widehat{P}_{n}^{-}}  \otimes \mathbf{e}_0\right\}.
\label{resresult1}
\ena
Here we used (\ref{zetarel}).

If (\ref{cycletoshow2}) is satisfied, that is,
\bea
&&
\widehat{P}_{n}^{\pm}(A_{1}, \cdots , A_{\ell-1}|B_{1}, \cdots , B_{n-2}|B_{n-1})
=\pm B_{n-1}^{n-1} d_{n} P_{n-2}(A_{1}, \cdots , A_{\ell-1}|B_{1}, \cdots , B_{n-2})
\no
\ena
with
\bea
d_{n}:=\frac{2\pi}{\zeta(-\pi i)}(-2\pi i)^{-\ell-\frac{n}{2}},
\label{defd}
\ena
then
\bea
\text{RHS of (\ref{resresult1})}=
\left\{
I-(-1)^{\frac{n}{2}-1}
S_{n-1, n-2}(\beta_{n-1}-\beta_{n-2}) \cdots S_{n-1, 
1}(\beta_{n-1}-\beta_{1})
\right\}
f_{P_{n-2}} \otimes \mathbf{e}_0.
\no
\ena
Thus (a) is proved.
\vskip3mm

\noindent
{\it Proof of (b). }
It is easy to see that $\text{deg}_{A_a}\,P_{n}\leq n$ for all $a$.
We shall prove (ii).

Using
\bea
H_{k+1}^{\pm}(B_{j})-B^{\pm 2}H_{k}(B_{j})
=(B_{j}^{\pm 2}-B^{\pm 2})H_{k}^\pm(B_{j}),
\ena
we find
\bea
&&
D_{n}^{\pm}(A_{1}, \cdots , A_{r}|B_{1}, \cdots , B_{n-2}, B, -B) \no \\
&& {}=
(-1)^{\frac{n+m}{2}}\exp{( -2\sum_{k=-\infty}^{\infty}t_{2k}B^{2k})}
\prod_{a=1}^{r}(1-A_{a}^{2}B^{-2}) 
D_{n-2}^{\pm}(A_{1}, \cdots , A_{r}|B_{1}, \cdots , B_{n-2}).
\label{Prel1}
\ena

\begin{lemma}
For any integer $N$ and a pair of integers $(r,\ell)$ such that $r<\ell$
we have
\bea
&&
\text{Asym}\left(
\prod_{a=r+1}^{\ell}A_{a}^{-2a} \right) 
=
\text{Asym}\left(
(-1)^{\ell-r-1}
\prod_{a=r+1}^{\ell-1}(1-A_{a}^{2}B^{-2}) 
A_{\ell}^{-2(r+1)}
\prod_{a=r+1}^{\ell-1}A_{a}^{-2a-2} \right),
\no
\ena
where \text{Asym} is the anti-symmetrization with respect to 
$A_{r+1}, \cdots , A_{\ell}$.
\end{lemma}

\noindent
{\it Proof.} For two functions $P_1$ and $P_2$ of $A_{r+1}, \cdots , A_{\ell}$
we write $P_1\simeq P_2$ if $\text{Asym}(P_1)=\text{Asym}(P_2)$.
Then
\bea
\prod_{a=r+1}^{\ell}A_{a}^{-2a}
&\simeq&
(-1)^{\frac{1}{2}(\ell-r)(\ell-r-1)}
\big(\prod_{a=r+1}^{\ell}A_{a}^{-2\ell}\big) 
\prod_{a=r+1}^{\ell}A_{a}^{2(a-r-1)}
\no
\\
&\simeq&
(-1)^{\frac{1}{2}(\ell-r)(\ell-r-1)}
\big(\prod_{a=r+1}^{\ell}A_{a}^{-2\ell}\big) 
\prod_{a=r+1}^{\ell}A_{a}^{2(a-r-1)}
(1-A_{r+1}^2B^{-2})
\no
\\
&\vdots&
\no
\\
&\simeq&
(-1)^{\frac{1}{2}(\ell-r)(\ell-r-1)}
\big(\prod_{a=r+1}^{\ell}A_{a}^{-2\ell}\big) 
\prod_{a=r+1}^{\ell}A_{a}^{2(a-r-1)}
\prod_{a=r+1}^{\ell-1}(1-A_{a}^2B^{-2})
\no
\\
&=&
(-1)^{\frac{1}{2}(\ell-r)(\ell-r-1)}
\prod_{a=r+1}^{\ell-1}(1-A_{a}^{2}B^{-2})
A_{\ell}^{-2(r+1)}\prod_{a=r+1}^{\ell -1}A_{a}^{-2\ell+2(a-r-1)}
\no 
\\
&\simeq&
(-1)^{\ell-r-1}
\prod_{a=r+1}^{\ell-1}(1-A_{a}^2B^{-2})
A_{\ell}^{-2(r+1)}
\prod_{a=r+1}^{\ell-1}A_{a}^{-2-2a}.
\no
\ena
q.e.d.

By the lemma we have
\bea
&&
\text{Asym}\left(
\prod_{a=r+1}^{\ell_{n}}A_{a}^{n+1-2s+2r-2a} \right) \no \\
&& {}=
\text{Asym}\left(
(-1)^{\ell_{n}-r-1}
\prod_{a=r+1}^{\ell_{n}-1}(1-A_{a}^{2}B^{-2}) 
A_{\ell_{n}}^{n-1-2s}
\prod_{a=r+1}^{\ell_{n}-1}A_{a}^{n-1-2s+2r-2a} \right).
\label{Prel2}
\ena
It follows from (\ref{Prel1}) and (\ref{Prel2}) that $P_{n}$ 
given by (\ref{defnP}) satisfies (\ref{cycletoshow1}) with
\bea
\overline{P}_{n(\pm)}
=
(-1)^{s-1}\frac{c_{n}^{m, r, s}}{c_{n-2}^{m, r, s}}
B^{2s}A_{\ell_{n}}^{n-1-2s}
P_{n-2(\pm)}(A_{1}, \cdots , A_{\ell_{n}-1}| B_{1}, \cdots , B_{n-2}).
\label{Prel3}
\ena
Here we used $\ell_{n}=r+(n-m)/2$.

Note that $n$ is even.
{}From (\ref{defc}) and (\ref{defd}), we see that (\ref{Prel3}) implies (\ref{cycletoshow2}). 
This completes the proof of (b).

\section{New symmetry}
In the construction of \S4.3, the cycles of initial form factors for
$m$-minimal local operators with charge $m$, which means $r=0$, are
either of the following two cycles;
\bea
&&
E_m(t\vert B)(\prod_{j=1}^m B_j)^s 
\prod_{1\leq j<j'\leq m}(B_j^{\pm1}+B_{j'}^{\pm1}),
\quad
E_m^{odd}(t\vert B)(\prod_{j=1}^m B_j)^s.
\no
\ena
The form factors of some $m$-minimal local operators can not be obtained
{}from linear combinations of the expansion coefficients in $t_j$'s of these
two cycles.
Important examples of such operators are given by
the operators $\Lambda_{-1}(y)$ and compositions of it introduced by
Lukyanov \cite{luk1} (see the next section).
To improve this drawback we shall introduce some functions which create
new cycles from the ones in \S4.3 by multiplication.
The functions are extracted from
the integral formula of the form factors of $\Lambda_{-1}(y)$ \cite{npt}
(see \S7).

Consider $m$ and $r$ which satisfy $m-2r\geq 0$ and fix them.
Set
\bea
&&
Q_{n(+)}(y)=Q_{n(+)}(y\vert A_1,\cdots,A_{\ell_n}\vert  B_1,\cdots,B_n)
=\frac{\prod_{a=r+1}^{\ell_n}(1-A_a^2\text{e}^{2y})}
{\prod_{j=1}^n(1-B_j\text{e}^{y})},
\label{newsymmetry1}
\\
&&
Q_{n(-)}(y)=Q_{n(-)}(y\vert A_1,\cdots,A_{\ell_n}\vert  B_1,\cdots,B_n)
=
\prod_{j=1}^nB_j^{-1}
\frac{\prod_{a=r+1}^{\ell_n}(\text{e}^{-2y}-A_a^2)}
{\prod_{j=1}^n(1-B_j^{-1}\text{e}^{-y})},
\label{newsymmetry2}
\ena
where $\ell_n=(n-m)/2+r$ as before.
Notice that $Q_{n(-)}(y)=\text{e}^{my}Q_{n(+)}(y)$.

\begin{prop}\label{newsymmetry}
Consider a set of polynomials 
$P_n(A_1,\cdots,A_{\ell_n}\vert B_1,\cdots,B_n)$ in $A_j$'s
with the coefficients in symmetric Laurent polynomials of $B_j^{\pm1}$'s.
Suppose that 
there exist sets of polynomials of $A_j$'s, 
$\tilde{P}_{n}=$
$\tilde{P}_{n}$$(A_1,\cdots,A_{\ell_n}\vert B_1,\cdots,B_{n-2}| B)$ 
and $\overline{P}_{n}=$
$\overline{P}_{n}$$(A_1,\cdots,A_{\ell_n}\vert B_1,\cdots,B_{n-2}| B)$,
such that
\bea
&&
P_n(A_{1}, \cdots , A_{\ell_{n}}|B_{1}, \cdots , B_{n-2}, B, -B)
=
\prod_{a=1}^{r}(1-A_{a}^{2}B^{-2}) 
\tilde{P}_{n},
\label{symcond1}
\\
&&
\text{Asym}\left\{ 
\tilde{P}_n
\right\} 
=
\text{Asym}\left\{ \prod_{a=r+1}^{\ell_n -1}(1-A_{a}^{2}B^{-2}) 
\overline{P}_{n}
\right\}, 
\label{symcond2}
\\
&&
\overline{P}_{n}(A_{1}, \cdots \! , A_{\ell_n \! -1}, \pm B|B_{1}, \cdots \! , B_{n-2}|B)=
\pm B^{n-1}d_{n}P_{n-2}(A_{1}, \cdots \! , A_{\ell_n \! -1}|B_{1}, \cdots \! , B_{n-2}),
\label{symcond3}
\ena
where $d_{n}$ is a set of constants and Asym is the anti-symmetrization
with respect to $A_{r+1}$, ..., $A_{\ell_n}$.
Then $P_n^{new}=P_nQ_{n(+)}(y)$ is symmetric in $B_j$'s and
satisfies (\ref{symcond1}), (\ref{symcond2}) 
and (\ref{symcond3}) for appropriate $\tilde{P}_n^{new}$ and 
$\overline{P}_n^{new}$.
\end{prop}
\vskip3mm

\noindent
{\it Proof.} Since $Q_n(y)$ does not contain $A_1$, ..., $A_r$ and
is symmetric with respect to $A_{r+1}$, ..., $A_{\ell_n}$, 
one has to prove (\ref{symcond3}). This follows from
\bea
&&
Q_{n(+)}(y)\vert_{B_n=-B, B_{n-1}=B, 
A_{\ell_n}=\pm B}=Q_{n-2(+)}(y).
\no
\ena
q.e.d

In the proof of (b) of the proof of Theorem \ref{mainth} we have proved
that $P_{n(\pm)}$ given by (\ref{defnP}) actually satisfies 
(\ref{symcond1}), (\ref{symcond2}), (\ref{symcond3}). Thus we have

\begin{cor}
Suppose that $m$, $r$, $s$ satisfy $m-2r\geq 0$, $0\leq s \leq \frac{m}{2}$.
Let $P_{n(\pm)}$ be given by (\ref{defnP}) and $P_{n(\pm)}(y_1,\cdots,y_s)$
be defined by
\bea
&&
P_{n(\pm)}^\epsilon(y_1,\cdots,y_s)=P_{n(\pm)}\prod_{i=1}^{s}
Q_{n(\epsilon)}(y_i),
\label{newcycle1}
\ena
for $\epsilon=\pm$.
Define the set of functions
\bea
&&
f_{P_{n(\pm)}^\epsilon(y_1,\cdots,y_s)}=
e^{\frac{n}{4}\sum_{j=1}^{n}\beta_{j}}
\prod_{1 \le j<j' \le n}\zeta(\beta_{j}-\beta_{j'})\Psi_{P_{n(\pm)}^\epsilon
(y_1,\cdots,y_s)}.
\label{minimalff3}
\ena
If we expand (\ref{minimalff3}) into the series of $\exp(\epsilon y_i)$'s at
$\exp(\epsilon y_i)=0$ respectively, then each coefficient of them
satisfies (I), (II), (III) and defines an $m$-minimal local operator.
Moreover each function of (\ref{minimalff3}) is in 
$(V^{\otimes n})_{m-2r}^{sing}$.
\end{cor}

\noindent
{\it Proof.} By Proposition \ref{newsymmetry}, 
(\ref{newcycle1}) satisfies condition
(i) in the beginning of the proof of Theorem \ref{mainth}.
The condition (ii) on the degree of $A_a$ is easily verified.
q.e.d

\section{Form factors of Lukyanov's operators $\Lambda_{-1}(y)$ and $T(y)$}
In \cite{luk1} Lukyanov has introduced the operators $\Lambda_{k}(y)$,
$k=0,\pm1$ and $T(y)$ which produce the generating functions of form factors
of local operators. The $n$-particle form factors corresponding to 
$\Lambda_{\pm1}(y)$, $\Lambda_0(y)$ form
the 3-dimensional irreducible representation of $sl_2$ 
whose highest weight vector is the form factor 
corresponding to $\Lambda_{-1}(y)$.
The form factors corresponding to $T(y)$ belong to the trivial representation
of $sl_2$. 

The integral formulae of form factors
corresponding to those operators in \cite{luk1} take different form from those 
in this paper. In \cite{npt} the cycles of $\Lambda_{-1}(y)$ and $T(y)$
in the sense of this paper are found.
The cycles given in \cite{npt} are determined up to multiplying 
$2\pi i$-periodic and symmetric functions of $\beta_j$'s. 
This is because the solutions of 
the qKZ equation were mainly concerned there and the $2\pi i$-periodic
functions of $\beta_j$'s simply play the role of constants.
Here we shall give the cycles of $\Lambda_{-1}(y)$ and $T(y)$ including 
those functions of $\beta_j$'s. They are in fact special cases of 
(\ref{newcycle1}).

First of all we remark that all the local operators obtained from
$\Lambda_{-1}(y)$ and $T(y)$ are $2$-minimal.
\vskip2mm

\noindent
{\bf 7.1.} $\Lambda_{-1}(y)$.
\par
\noindent
The $2n$-particle form factor corresponding to $\Lambda_{-1}(y)$ is
described by the cycle
\bea
&&
P^{\Lambda_{-1}(y)}=c^{2,0,1}_{2n}(-1)^{\frac{1}{2}n(n-1)}
\prod_{j=1}^{2n}B_j\prod_{a=1}^{n-1}A_a^{2n-1-2a}
Q_{2n(+)}(y)
\label{lamda1}
\ena
up to overall multiple of a function of $y$, where $m=2$, $r=0$.
Define the expansion coefficients $P^{\Lambda_{-1}(y)}_{l(\pm)}$ by
\bea
&&
P^{\Lambda_{-1}(y)}=
\sum_{l=0}^\infty P^{\Lambda_{-1}(y)}_{l(+)}\exp(ly),
\quad
\hbox{e}^{2y}P^{\Lambda_{-1}(y)}=
\sum_{l=0}^\infty P^{\Lambda_{-1}(y)}_{l(-)}\exp(-ly).
\no
\ena
Let us calculate them. 
To this end we introduce the complete symmetric function $h_l(B)$ 
and $\bar{h}_l(B)$ by
\bea
&&
\frac{1}{\prod_{j=1}^{2n}(1-B_jx)}=
\sum_{l=0}^\infty h_l(B)x^l,
\quad
\frac{1}{\prod_{j=1}^{2n}(1-B_j^{-1}x)}=
\sum_{l=0}^\infty \bar{h}_l(B)x^l.
\no
\ena
We write $P\cong Q$ if $Asym(P)=Asym(Q)$ in the space (\ref{solsp1}) 
({\it cf.} \S4.6).
Then we have
\bea
P^{\Lambda_{-1}(y)}_{l(+)}
&\cong&
c^{2,0,1}_{2n}(-1)^{\frac{1}{2}n(n-1)}
\prod_{j=1}^{2n}B_j
\sum_{k=0}^{n-1}(-1)^k
\prod_{a=1}^{k}A_a^{2n+1-2a}
\prod_{a=k+1}^{n-1}A_a^{2n-1-2a}
h_{l-2k}(B)
\no
\\
P^{\Lambda_{-1}(y)}_{l(-)}
&\cong&
c^{2,0,0}_{2n}(-1)^{\frac{1}{2}n(n-1)}
\sum_{k=0}^{n-1}
(-1)^k
\prod_{a=1}^{n-k-1}A_a^{2n+1-2a}
\prod_{a=n-k}^{n-1}A_a^{2n-1-2a}
\bar{h}_{l-2k}(B).
\no
\ena
We can rewrite these cycles to forms close to those of $su(2)$ currents as
\bea
P^{\Lambda_{-1}(y)}_{l(+)}
&\cong&
c^{2,0,1}_{2n}(-1)^{\frac{1}{2}n(n-1)}
\frac{1}{e_1(B)}
\prod_{j=1}^{2n}B_j
\sum_{k=0}^{n-1}e_{2k+1}(B)h_{l-2k}(B)
\prod_{a=1}^{n-1}A_a^{2n-1-2a}
\no
\\
P^{\Lambda_{-1}(y)}_{l(-)}
&\cong&
c^{2,0,0}_{2n}(-1)^{\frac{1}{2}n(n-1)}
\frac{1}{\overline{e}_1(B)}
\sum_{k=0}^{n-1}\overline{e}_{2k+1}(B)\bar{h}_{l-2k}(B)
\prod_{a=1}^{n-1}A_a^{2n+1-2a}.
\no
\ena
The poles at $e_1(B)=0$ of $P^{\Lambda_{-1}(y)}_{l(+)}$ and 
$\overline{e}_1(B)=0$ of $P^{\Lambda_{-1}(y)}_{l(-)}$ are of course apparent.

For $l=0$, $P^{\Lambda_{-1}(y)}_{0(\pm)}$
are overall constant multiples of $P_{\pm}$ which describe
the $su(2)$ currents $j^{+}_{\pm}$.
It is obvious that the spins of the operators corresponding to 
$P^{\Lambda_{-1}(y)}_{l(\pm)}$ are $\mp (l+1)$.
\vskip3mm

\noindent
{\bf 7.2.} $T(y)$.
\par
\noindent
The $2n$-particle form factor corresponding to $T(y)$ is
described by the cycle
\bea
&&
P^{T(y)}=c^{2,1,1}_{2n}(-1)^{\frac{1}{2}n(n-1)}
\prod_{j=1}^{2n}B_j
\prod_{a=2}^{n}A_a^{2n+1-2a}
Q_{2n(+)}(y)
\label{ty1}
\ena
up to overall multiple of a function of $y$, where $m=2$, $r=1$.
Define the expansion coefficients by
\bea
&&
P^{T(y)}=
\sum_{l=0}^\infty P^{T(y)}_{l(+)}\exp(ly),
\quad
\hbox{e}^{2y}P^{T(y)}=
\sum_{l=0}^\infty P^{T(y)}_{l(-)}\exp(-ly).
\no
\ena
They are given by
\bea
P^{T(y)}_{l(+)}
&\cong&
c^{2,1,1}_{2n}(-1)^{\frac{1}{2}n(n-1)}
\prod_{j=1}^{2n}B_j
\sum_{k=0}^{n-1}(-1)^k
\prod_{a=2}^{k+1}A_a^{2n+3-2a}
\prod_{a=k+2}^{n}A_a^{2n+1-2a}
h_{l-2k}(B)
\no
\\
P^{T(y)}_{l(-)}
&\cong&
c^{2,1,0}_{2n}(-1)^{\frac{1}{2}n(n-1)}
\sum_{k=0}^{n-1}
(-1)^k
\prod_{a=2}^{n-k}A_a^{2n+3-2a}
\prod_{a=n-k+1}^{n}A_a^{2n+1-2a}
\bar{h}_{l-2k}(B).
\no
\ena
Again we can rewrite these cycles as
\bea
P^{T(y)}_{l(+)}
&\cong&
c^{2,1,1}_{2n}(-1)^{\frac{1}{2}n(n-1)}
\frac{1}{e_1(B)}
\prod_{j=1}^{2n}B_j
\sum_{k=0}^{n-1}e_{2k+1}(B)h_{l-2k}(B)
\prod_{a=2}^{n}A_a^{2n+1-2a}
\no
\\
P^{T(y)}_{l(-)}
&\cong&
c^{2,1,0}_{2n}(-1)^{\frac{1}{2}n(n-1)}
\frac{1}{\overline{e}_1(B)}
\sum_{k=0}^{n-1}\overline{e}_{2k+1}(B)\bar{h}_{l-2k}(B)
\prod_{a=2}^{n}A_a^{2n+3-2a}.
\no
\ena

For $l=1$, $P^{T(y)}_{l(+)}$ and $P^{T(y)}_{l(-)}$ are overall
constant multiples of $P_{\bar{z}}$ and $P_z$ respectively.
The spins of the operators corresponding to $P^{T(y)}_{l(\pm)}$
are $\mp(l+1)$.

We remark that $P^{T(y)}_{l(\pm)}$ for even $l$ do not correspond to 
local operators. They satisfy axioms (I) and (II).
Moreover they satisfy axiom (III) for $2n$ and $2n-2$ particle form factors 
with $n>2$. But they do not satisfy (III) for two particle form factors, 
that is, two particle form factors have a pole at $\beta_2=\beta_1+\pi i$.

\section{Another form factor formula for weight zero}

If the weight of $\psi_W$ is zero, that is $n=2\ell$, 
then $\psi_W=0$ for $W\in {\cal F}_q^{\otimes \ell}$ (see \cite{npt}).
Using this property one can rewrite the form factor formula.
This rewriting was done in \cite{npt} for the solutions of
qKZ equation. Again, there, the $2\pi i$-periodic functions of $\beta_j$'s
were not taken care of. Here we present the form factor formulae 
which satisfy the axioms for locality.
 
We use $2n$ instead of $n$ in this section since it is more convenient.
We recall some notations from \cite{npt}.

For $M\subset \{1,...,2n\}$ we set
\bea
&&
P_M^{+}(\alpha)=\prod_{j\in M}(\alpha-\beta_j+2\pi i),
\quad
P_M^{-}(\alpha)=\prod_{j\notin M}(\alpha-\beta_j+2\pi i).
\no
\ena
Define the operator $T_x$ by 
\bea
T_xf(\alpha)=f(\alpha)-f(\alpha+x).
\no
\ena
For a rational function $f(\alpha)$ we define the polynomial 
$[f(\alpha)]_{+}$ in such a way that
$f(\alpha)-[f(\alpha)]_{+}$ is regular at $\alpha=\infty$.
Using these notations we define polynomials $Q^{(a)}_M(\alpha)$ 
$1\le a \le n-1$ by
\bea
&&
Q_M^{(a)}(\alpha)\,=\,
P^+_M(\alpha-\pi i)\>\Bigl[\>T_{-\pi i}
\Bigl(\frac{P^-_M(\alpha)}{P^+_M(\alpha-\pi i)}\,
\Bigl[\frac{P^+_M(\alpha-\pi i)}{(\alpha-\pi i)^a}\Bigr]_+\Bigr)\Bigr]_+\;+\;
P^-_M(\alpha)\>\Bigl[\>T_{- \pi i}
\Bigl(\frac{P^+_M(\alpha)}{\alpha^a}\Bigr)\Bigr]_+\,.
\no
\ena

Let $v^S_M=v^S_M(\beta_1,\cdots,\beta_{2n})$, $\sharp M=n$ be the basis of 
$\left(V^{\otimes 2n}\right)_{0}$ which satisfies the conditions
\bea
&&
v^S_{\cdots,\ep_{k+1},\ep_k,\cdots}(\cdots,\beta_{k+1},\beta_k,\cdots)
=P_{k,k+1}\widehat{S}(\beta_k-\beta_{k+1})
v^S_{\cdots,\ep_{k},\ep_{k+1},\cdots}(\cdots,\beta_{k},\beta_{k+1},\cdots),
\label{basis1}
\\
&&
v_{M_{ext}}^S=v_{M_{ext}}+\sum_{M\neq M_{ext}}c_Mv_M,
\label{basis2}
\ena
where $M=\{k | \ep_{k}=- \}, M_{ext}=\{1,\cdots,n\}$ and $c_M$'s are some constants.
It is known that $v^S_M$ are uniquely determined by the conditions 
(\ref{basis1}) and (\ref{basis2}). 
For the explicit construction of $v^S_M$ see \cite{npt}.
We set
\bea
&&
\tilde{v}^S_M=
\frac{v^S_M}{\prod_{j\notin M,k\in M}(\beta_j-\beta_k-\pi i)}.
\no
\ena

Consider the cycle
\bea
&&
W=\prod_{k=1}^{n}W_k(\alpha_k),
\quad
W_k(\alpha)=\frac{P_k(\alpha)}{\prod_{j=1}^{2n}(1-AB_j^{-1})}.
\no
\ena
Suppose that $W_k\in {\cal F}_q$ for $k\neq n$ and 
$W_n\in {\hat{\cal F}}_q$.
Then it is proved in \cite{npt} that
\bea
\psi_W(\beta_1,\cdots,\beta_{2n})&=&
2^n(-2\pi i)\big(W_n(+\infty)-W_n(-\infty)\big)
\no
\\
&\times&
\sum_{\sharp M=n}\tilde{v}^S_M
\int_{C^{n-1}}
\prod_{a=1}^{n-1}d\alpha_a
\prod_{a=1}^{n-1}\phi(\alpha_a)
\det(Q^{(a)}_M(\alpha_b))_{a,b=1}^{n-1}
\frac{\prod_{k=1}^{n-1}P_k(\alpha_k)}
{\prod_{a=1}^{n-1}\prod_{j=1}^{2n}(1-AB_j^{-1})}.
\no
\ena
If $P_n=1$ then
\bea
&&
W_n(+\infty)=
\lim_{A\rightarrow 0}\frac{1}{\prod_{j=1}^{2n}(1-AB_j^{-1})}=1,
\quad
W_n(-\infty)=
\lim_{A\rightarrow \infty}\frac{1}{\prod_{j=1}^{2n}(1-AB_j^{-1})}=0.
\no
\ena

For a polynomial $P$ of $A_1$,...,$A_{n-1}$ satisfying 
$P|_{A_{a}=0}=0$ and $\text{deg}_{A_a}\, P\le 2n-1, (a=1, \cdots , n-1)$, 
set
\bea
&&
\tilde{\Psi}_P(\beta_1,\cdots,\beta_{2n})=
\sum_{\sharp M=n}\tilde{v}^S_M
\int_{C^{n-1}}
\prod_{a=1}^{n-1}d\alpha_a
\prod_{a=1}^{n-1}\phi(\alpha_a)
\det(Q^{(a)}_M(\alpha_b))_{a,b=1}^{n-1}
\frac{P}{\prod_{a=1}^{n-1}\prod_{j=1}^{2n}(1-AB_j^{-1})}.
\no
\ena
Let 
\bea
&&
P_{2m(\pm)}(A_{1}, \cdots , A_{m-1}|B_{1}, \cdots , B_{2m})=
E_{2m}(t\vert B)(\prod_{j=1}^{2m} B_j)^s
\prod_{1 \le j<j' \le 2m}(B_{j}^{\pm 1}+B_{j'}^{\pm 1}) 
\prod_{a=1}^{m-1}A_{a}^{k_{a}}
\no
\ena
be the cycle for initial form factor, where $1\leq k_a\leq 2m-1$ for all $a$
and $0\leq s\leq m$.
Set
\bea
&&
P_{2n(\pm)}(A_{1}, \cdots , A_{n-1}|B_{1}, \cdots , B_{2n})=
\tilde{c}_{2n}^{2m, m, s}
E_{2n}(t\vert B)(\prod_{j=1}^{2n} B_j)^s
\prod_{a=1}^{m-1}A_{a}^{k_{a}}
\prod_{a=m}^{n-1}A_{a}^{2n-1-2s+2m-2a}
\no 
\\
&&
\qquad\qquad\qquad\qquad\qquad\qquad\qquad
\times
\tilde{D}_{2n}^{\pm}(A_{1}, \cdots , A_{m-1}|B_{1}, \cdots , B_{2n}),
\label{defP}
\ena
where
\bea
\tilde{D}_{2n}^{\pm}=D_{2n}^{\pm}\big\vert_{A_m=0},
\quad
\tilde{c}_{2n}^{2m, m, s}=2^{n-m} c_{2n}^{2m, m, s}.
\no
\ena

Then by Theorem \ref{mainth} we have

\begin{theorem}\label{nextmainth}
The set of functions
\bea
&&
\tilde{f}_{P_{2n(\pm)}}=
e^{\frac{n}{2}\sum_{j=1}^{2n}\beta_{j}}
\prod_{1 \le j<j' \le 2n}\zeta(\beta_{j}-\beta_{j'})\tilde{\Psi}_{P_{2n(\pm)}}
\label{minimalff2}
\ena
satisfies (I), (II), (III) and is $2m$-minimal.
Each form factor of (\ref{minimalff2}) is in 
$(V^{\otimes n})_{0}^{sing}$.
\end{theorem}
\vskip5mm

\noindent
{\large \bf Acknowledgements}\par
\noindent
We are grateful to Fedor Smirnov for discussions.

\vskip3mm
\appendix
\section{Appendix}
Here we list cycles corresponding to important local operators.
The cycles are decribed modulo constant multiple.
In the part "Representation of $sl_2$", the dimension of 
the irreducible representation of $sl_2$ to which the operator
belongs is written. For example the form factors of $j^{+}_{+}$
is the highest weight vector of the 3-dimensional irreducible 
representation of $sl_2$.

As explained in \S7, $\Lambda(y)$ and $T(y)$ describe the generating
functions of form factors of local operators. More precisely in the 
expansion of the form factor corresponding to $T(y)$ only the
coefficients of the odd power of $\exp(\pm y)$ are the form factors
of local operators.
\vskip5mm

\begin{tabular}{|c|l|c|c|c|}\hline
{\it Operator}&{\it P-cycle ($2n$ particle)}
&{\it Spin}&{\it Minimality}&{\it Representation of $sl_2$}\\ \hline
$j^{+}_{+}$ 
& 
$\prod_{j=1}^{2n}B_j\prod_{a=1}^{n-1}A_a^{2a-1}$ 
&
$-1$ & $2$ & $3$
\\
$j^{+}_{-}$ 
& 
$\prod_{a=1}^{n-1}A_a^{2a+1}$ 
&
$1$ & $2$ & $3$
\\
$T_z$ 
& 
$(\sum_{j=1}^{2n}B_j^{-1})\prod_{a=1}^{n-1}A_a^{2a+1}$ 
&
$2$ & $2$ & $1$
\\
$T_{\bar{z}}$ 
& 
$(-1)^{n-1}\prod_{j=1}^{2n}B_j(\sum_{j=1}^{2n}B_j)\prod_{a=1}^{n-1}A_a^{2a-1}$ 
&
$-2$ & $2$ & $1$
\\
$\Theta$ 
& 
$(\sum_{j=1}^{2n}B_j)\prod_{a=1}^{n-1}A_a^{2a+1}$ 
&
$0$ & $2$ & $1$
\\
$\Lambda_{-1}(y)$
&
$\prod_{j=1}^{2n}B_j\prod_{a=1}^{n-1}A_a^{2a-1}
\frac{\prod_{a=1}^{n-1}(1-A_a^2\text{e}^{2y})}
{\prod_{j=1}^{2n}(1-B_j\text{e}^{y})}$
&
$\ast$ & $2$ & $3$
\\
$T(y)$
&
$\prod_{j=1}^{2n}B_j
\prod_{a=2}^{n}A_a^{2n+1-2a}
\frac{\prod_{a=2}^{n}(1-A_a^2\text{e}^{2y})}
{\prod_{j=1}^{2n}(1-B_j\text{e}^{y})}$
&
$\ast$ & $2$ & $1$
\\
\hline
\end{tabular}

\end{document}